\documentclass[fleqn,usenatbib]{mnras}

\usepackage{newtxtext,newtxmath}
\usepackage{CJK}
\usepackage[T1]{fontenc}

\DeclareRobustCommand{\VAN}[3]{#2}
\let\VANthebibliography\thebibliography
\def\thebibliography{\DeclareRobustCommand{\VAN}[3]{##3}\VANthebibliography}

\usepackage{graphicx}	
\usepackage{amsmath}	
\usepackage{newtxtext,newtxmath}
\usepackage{subfig}
\usepackage{float}
\usepackage{rotating}
\usepackage{multicol}
\usepackage{hyperref}
\usepackage{pdfpages}
\usepackage{color}

\title[I: white dwarf parameters and accreted planetary abundances]{Seven white dwarfs with circumstellar gas discs I: white dwarf parameters and accreted planetary abundances}

\author[L. K. Rogers]{L. K. Rogers$^{1}$\thanks{E-mail: laura.rogers@ast.cam.ac.uk},
A. Bonsor$^{1}$,
S. Xu \begin{CJK*}{UTF8}{gbsn}(许\CJKfamily{bsmi}偲\CJKfamily{gbsn}艺\end{CJK*}$)^{2}$,
P. Dufour$^{3}$,
B. L. Klein$^{4}$,
A. Buchan$^{1}$,
S. Hodgkin$^{1}$, 
\newauthor F. Hardy$^{3}$,
M. Kissler-Patig$^{5}$,
C. Melis$^{6}$,
A. J. Weinberger$^{7}$,
B. Zuckerman$^{4}$
\\
$^{1}$ Institute of Astronomy, University of Cambridge, Madingley Road, Cambridge CB3 0HA, UK \\
$^{2}$ Gemini Observatory/NSF's NOIRLab, 670 N. A'ohoku Place, Hilo, HI 96720, USA \\
$^{3}$ D\'epartement de Physique, Universit\'e de Montr\'eal, C.P. 6128, Succ. Centre-Ville, Montr\'eal, Qu\'ebec H3C 3J7, Canada \\
$^{4}$ Department of Physics and Astronomy, University of California, Los Angeles, CA 90095-1562, USA \\
$^{5}$European Space Agency - European Space Astronomy Centre, Camino Bajo del Castillo, s/n., 28692 Villanueva de la Canada, Madrid, Spain \\
$^{6}$Center for Astrophysics and Space Sciences, University of California, San Diego, CA 92093-0424, USA \\
$^{7}$Earth and Planets Laboratory, Carnegie Institution for Science, 5241 Broad Branch Rd NW, Washington, DC 20015, USA
}

\date{Accepted XXX. Received YYY; in original form ZZZ}

\pubyear{2022}

\begin{document}
\label{firstpage}
\pagerange{\pageref{firstpage}--\pageref{lastpage}}
\maketitle

\begin{abstract}

Observations of planetary material {\it polluting} the atmospheres of white dwarfs are an important probe of the bulk composition of exoplanetary material. Medium- and high-resolution optical and ultraviolet spectroscopy of seven white dwarfs with known circumstellar dust and gas emission are presented. Detections or meaningful upper limits for photospheric absorption lines are measured for: C, O, Na, S, P, Mg, Al, Si, Ca, Ti, Cr, Fe, and Ni. For 16 white dwarfs with known observable gaseous emission discs (and measured photospheric abundances), there is no evidence that their accretion rates differ, on average, from those without detectable gaseous emission. This suggests that, typically, accretion is not enhanced by gas drag. At the effective temperature range of the white dwarfs in this sample (16,000--25,000\,K) the abundance ratios of elements are more consistent than absolute abundances when comparing abundances derived from spectroscopic white dwarf parameters versus photometric white dwarf parameters. Crucially, this highlights that the uncertainties on white dwarf parameters do not prevent white dwarfs from being utilised to study planetary composition. The abundances of oxygen and silicon for the three hydrogen-dominated white dwarfs in the sample with both optical and ultraviolet spectra differ by 0.62 dex depending on if they are derived from the optical or ultraviolet spectra. This optical/ultraviolet discrepancy may be related to differences in the atmospheric depth of line formation; further investigations into the white dwarf atmospheric modelling are needed to understand this discrepancy.

\end{abstract}

\begin{keywords}
planets and satellites: composition -- stars: abundances --  white dwarfs
\end{keywords}



\section{Introduction}

Exoplanets are found to be ubiquitous across most stages of stellar evolution \citep[e.g.][]{mayor1995jupiter,vanderburg2020giant}. To constrain a planet's bulk composition, measurements of its mass and radius are compared to theoretical mass-radius relationships for various interior compositions and structures \citep[e.g.][]{seager2007mass, dorn2015can}. However, degeneracies arise because different compositions can produce similar mass-radius curves, thus introducing uncertainties in determination of bulk compositions. 

White dwarfs that have been `polluted' by the accretion of elements heavier than helium, directly sample the bulk elemental composition of exoplanetary material; this is not possible with other observational techniques. Because of the strong surface gravity of white dwarfs their outer layers should contain only hydrogen or helium or both \citep{fontaine1979diffusion}. However, contrary to this, observations have revealed that 25--50 per cent of single white dwarfs have atmospheres that are `polluted' with elements heavier than helium \citep{zuckerman2003metal, zuckerman2010ancient, koester2014frequency, wilson2019unbiased}. Due to the rapid gravitational settling times ($\sim$ days for hot H-dominated DA white dwarfs, and $\sim$ millions of years for cool He-dominated DBs) in comparison to the white dwarfs' cooling age, there must be ongoing accretion of material \citep{koester2009accretion}. This material is from remnant planetary systems that have survived to the white dwarf phase \citep{jura2003tidally, farihi2010rocky}. Planetesimals from outer belts can become destabilised and are perturbed on to eccentric star grazing orbits \citep[e.g.][]{debes2002there,bonsor2011dynamical,veras2014formation,mustill2018unstable}. There are several potential pathways that lead to the accretion of the planetary material: tidal disruption into dust, sublimation directly into gas, or direct collision with the white dwarf \citep{veras2014formation,brown2017deposition,bonsor2017infrared,steckloff2021sublimation,mcdonald2021white,brouwers2022road}. Spectroscopic observations of white dwarfs combined with atmospheric models reveal the chemical composition of the planetary material that has polluted each white dwarf. So far, 23 heavy elements have been discovered across all polluted white dwarfs (see table 1. in \citealp{klein2021discovery} for references). The polluted white dwarf GD 362 has absorption features from the most elements detected for a given white dwarf \citep[e.g.][]{zuckerman2007chemical,xu2013two}. 

In order to obtain absolute abundances of the polluting material, it is crucial to obtain accurate white dwarf parameters. These parameters are most often derived based on spectra, where the H and/or He lines are fitted with white dwarf models to infer the effective temperature ($T_\mathrm{eff}$) and $\log(g)$ of the white dwarf, or from photometry, where broad-band photometry is fitted to obtain the effective temperature, and the parallax is used to constrain $\log(g)$. \citet{genest2019comprehensive} find that the spectroscopically derived effective temperatures of DA stars greater than 14,000\,K, are higher than those derived by photometry by 10 per cent. This is thought to be due to the inaccurate treatment of Stark broadening. The selection of the photometric bands used in the fit for the photometric $T_\mathrm{eff}$ cause the largest disparity in results; for hotter white dwarfs the $u$-band is crucial to obtain accurate parameters \citep{bergeron2019measurement}. Recent work by \citet{izquierdo2023systematic} highlights that for DB white dwarfs, different spectral data can result in a large spread of derived white dwarf parameters: 524\,K in $T_\mathrm{eff}$, 0.27 dex in $\log(g)$, and 0.31 dex in log(H/He). Additionally, when deriving the parameters from photometric data, depending on the data used, a spread of 1210\,K and 0.13 dex in $T_\mathrm{eff}$ and $\log(g)$ respectively were found. 

Previous studies have highlighted that there appears to be an optical and ultraviolet discrepancy, where the abundances of the polluting material derived from optical data are significantly discrepant from those derived from ultraviolet data \citep{jura2012two,gansicke2012chemical,xu2019compositions}. \citet{gansicke2012chemical} consider that this could be due to uncertain atomic data, abundance stratification, or real variation. Given that the optical and ultraviolet abundances are most often obtained from multiple studies where different white dwarf parameters are implemented, a more thorough investigation ensuring consistency is key to helping solve this issue.

Dust debris from tidally disrupted planetesimals has been discovered via excess infrared emission around 1.5--4 per cent of white dwarfs \citep[e.g.][]{becklin2005dusty, kilic2006debris, jura2007externally,rebassa2019infrared, wilson2019unbiased, xu2020infrared}. 21 of the white dwarfs with dust debris also show evidence of circumstellar gas in emission near the same radius as the dust \citep{gaensicke2006gaseous, gansicke2007sdss, gansicke2008sdss, melis2010echoes, farihi2012trio, melis2012gaseous, brinkworth2012spitzer, debes2012detection, dennihy2020five,melis2020serendipitous,gentile2020white}. These systems are identified by their double peaked emission features, usually strongest at the Ca\,\textsc{ii} infrared triplet. Gaia\,J0611$-$6931 has the most elements detected in emission, with observations of Ca, O, Si, Mg, and Na \citep{dennihy2020five,melis2020serendipitous}. The gaseous systems show line profiles with Doppler broadened features consistent with the gas rotating as a Keplerian disc. A number of theories have been proposed to explain the production of gas. A proportion of the gas produced at the sublimation radius could viscously spread outwards causing an overlap in the location of the dust and gas \citep{rafikov2011runaway,metzger2012global}. This outwardly spreading gas causes drag on the dust particles and thus accelerates their accretion on to the white dwarf creating a runaway effect; this might explain the highest accretion rates observed in polluted white dwarfs. An alternative explanation for gas emission is collisional cascades of planetesimals within the Roche radius of the white dwarf \citep{jura2008pollution,kenyon2017numericalb,kenyon2017numericala}, observations of infrared variability in WD\,0145+234 appear consistent with simple collisional cascade models \citep{wang2019ongoing, swan2021collisions}.

Circumstellar dust and gas around white dwarfs tell us about the current, potentially violent accretion of planetary material. With $>$\,1000 polluted white dwarfs known, but only 21 systems with both detectable circumstellar dust and gas, this represents an intriguing subsample of polluted white dwarfs with different circumstellar environments. These systems are extreme examples of polluted white dwarfs, and as such they are perfect targets for studying pollution in their atmospheres and understanding how the planetary material ultimately ends up there. This work focuses on seven such systems, Paper I (this paper) focuses on the methods to obtain the abundances of the metals in the white dwarfs and the limitations involved, and Paper II (Rogers et al. in preparation) provides an in depth analysis of the composition of the planetary material accreted. This paper is structured as follows. Section \ref{Obs} describes the optical and ultraviolet spectra of these seven systems taken with VLT X-shooter, Keck HIRES, Magellan MIKE, and \textit{HST} COS. Section \ref{Mod} explains the methods to determine the white dwarf parameters and the abundances of the metals in the white dwarfs. The effect on the abundances of differing white dwarf parameters and spectral ranges is reported in the results section in Section \ref{Results}. Section \ref{disc} discusses the results and limitations of the methods with the conclusions presented in Section \ref{Conclusions}. 

\section{Observations and Data Reduction} \label{Obs}

\subsection{Targets}

The targets were selected as those with clear infrared excesses from a dust disc using data from \textit{WISE} \citep{xu2020infrared}, which were confirmed with \textit{Spitzer} photometry \citep{lai2021infrared}. \citet{dennihy2020five} and \citet{melis2020serendipitous} report that these seven white dwarfs all host circumstellar gaseous discs. These seven systems are listed in Table\,\ref{tab:WDs-prop}. 

\begin{table*}
	\centering
	\caption{Stellar parameters derived from the spectroscopic (spec) and photometric (phot) fitting methods, see Section \ref{par} for further details. Distances (\textit{D}) are inferred from \textit{Gaia} parallaxes.}
	\label{tab:WDs-prop}
	\begin{tabular}{lllcccccc} 
		\hline
		WD Name & \textit{Gaia} eDR3 Number & Coordinates & SpT & Spec $T_\mathrm{eff}$ & Spec $\log(g)$ & Phot $T_\mathrm{eff}$ & Phot $\log(g)$ & \textit{D} (pc) \\
		\hline	
		Gaia\,J0006+2858 & 2860923998433585664 & 00:06:34.71 +28:58:46.54 & DAZ & 23921 (335) & 8.04 (0.04) & 22840 (197) & 7.86 (0.02) & 152 \\
		Gaia\,J0347+1624 & 43629828277884160 & 03:47:36.69 +16:24:09.74 & DAZ & 21820 (305)$^*$ & 8.10 (0.04)$^*$ & 18850 (164) & 7.84 (0.03) & 141 \\
		Gaia\,J0510+2315 & 3415788525598117248 & 05:10:02.15 +23:15:41.42 & DAZ & 21700 (304)$^*$ & 8.22 (0.04)$^*$ & 20130 (145) & 8.13 (0.02) & 65  \\
		Gaia\,J0611$-$6931 & 5279484614703730944 & 06:11:31.70 $-$69:31:02.15 & DAZ & 17749 (248) & 8.14 (0.04) & 16530 (561) & 7.81 (0.03) & 143 \\
		Gaia\,J0644$-$0352 & 3105360521513256832 & 06:44:05.23 $-$03:52:06.42 & DBZA & 18350 (524) & 8.18 (0.27) & 17000 (327) & 7.98 (0.02) & 112 \\
		WD\,1622+587 & 1623866184737702912 & 16:22:59.64 +58:40:30.90 & DBZA & 23430 (524) & 7.90 (0.27) & 21530 (313) & 7.98 (0.03) & 183 \\
		Gaia\,J2100+2122 & 1837948790953103232 & 21:00:34.65 +21:22:56.89 & DAZ & 25565 (358) & 8.10 (0.04) & 22000 (399) & 7.92 (0.02) & 88 \\
		\hline
	\end{tabular}
	\vspace{-6mm}
	\small \begin{flushleft}
      \item \textbf{Notes:} 
      \item $^*$ Spectroscopic parameters from \citet{melis2020serendipitous}.
       \end{flushleft}
\end{table*}

\begin{table*}
	\centering
	\caption{Observations of the seven white dwarfs listing dates of observations, exposure times in seconds, and SNR. The SNR for X-shooter UVB, HIRESb and MIKE-blue  were calculated from the continuum around the Ca\,\textsc{ii} K line (3933.7\,\AA). The SNR for X-shooter VIS, HIRESr and MIKE-red were calculated from the continuum around 6600\,\AA.}
	\label{tab:WDs-obs}
	\begin{tabular}{llllllllllll} 
		\hline
		WD Name & X-shooter & Exp UVB & Exp VIS & SNR UVB & SNR VIS & HIRESb & Exp & SNR & HIRESr & Exp & SNR \\
		\hline	
		Gaia\,J0006+2858 & 15-08-2019 & 3400\,s &  3458\,s & 31 & 36 & 07-07-2019 & 3300\,s & 57 & 16-07-2019 & 5400\,s & 40 \\
		Gaia\,J0347+1624 & - & - & - & - & - & 05-12-2019 & 5000\,s & 23 & - & - & - \\
		Gaia\,J0510+2315 & - & - & - & - & - & 05-12-2019 & 2400\,s & 33 & 09-12-2019 & 4800\,s & 71 \\
		Gaia\,J0611$-$6931$^{\dagger}$ & 15-10-2019 & 3400\,s & 3458\,s & 31 & 38 & 27-08-2021$^{\dagger}$ & 3100\,s & 12 & 27-08-2021$^{\dagger}$ & 3100\,s & 17 \\
		Gaia\,J0644$-$0352 & 15-09-2019 & 3400\,s & 3458\,s & 99 & 48 & 13-09-2020 & 1740\,s & 58$^*$ &  09-12-2019 & 2700\,s & 35 \\
		 & - & - & - & - & - & 08-10-2020 & 3000\,s & 58$^*$ &  - & - & - \\
		WD\,1622+587 & - & - & - & - & - & 10-07-2019 & 5400\,s & 33 & 16-07-2019 & 3300s & 31  \\
		Gaia\,J2100+2122 & 13-07-2019 & 3400\,s & 3458\,s & 134 & 127 & 10-07-2019 & 3600\,s & 115$^*$ & 16-07-2019 & 3300\,s & 73 \\
		 & - & - & - & - & - & 07-07-2019 & 2700\,s & 115$^*$ & - & - & - \\
		\hline
	\end{tabular}
	\vspace{-6mm}
	\small \begin{flushleft}
      \item \textbf{Notes:} 
      \item $^*$ SNR is reported based on the stacked spectra.
      \item $^{\dagger}$ Gaia\,J0611$-$6931 MIKE blue and red data listed under HIRESb and HIRESr respectively.
      \end{flushleft}
\end{table*}

\begin{table*}
	\centering
	\caption{NUV and FUV observations of the six white dwarfs listing dates of observations, exposure times in seconds, and SNR. The SNR for the NUV was calculated at the continuum around 1860\,\AA~ and the FUV were calculated at the continuum around the carbon lines at 1334--1335\,\AA.}
	\label{tab:WDs-obs-uv}
	\begin{tabular}{llllllllllll} 
		\hline
		WD Name & FUV & Exp FUV & SNR FUV & NUV & Exp NUV & SNR NUV \\
		\hline	
		Gaia\,J0006+2858 & 19-01-2022 & 1439\,s & 24 & 10-08-2021 & 5092\,s & 44 \\
		Gaia\,J0347+1624 & - & - & - & 07-08-2021 & 5000\,s & 33 \\
		Gaia\,J0510+2315 & 23-09-2021  & 1944\,s & 46 & 06-08-2021 & 4894\,s & 62 \\
		Gaia\,J0611$-$6931 & 22-01-2022 & 2015\,s & 21$^*$ & 25-08-2021 & 5724\,s & 28 \\
        & 23-07-2022 & 4796 s & 21$^*$ & - & - & -\\
        WD\,1622+587 & 14-10-2021 & 5426\,s & 25 & 09-08-2021 & 8672\,s & 13 \\
		Gaia\,J2100+2122 & - & - & - & 14-09-2021 & 4914\,s & 36 \\
		\hline
	\end{tabular}
\vspace{-3mm}
	\small \begin{flushleft}
      \item \textbf{Notes:} 
      \item $^*$ SNR reported for the stacked data. 
      \end{flushleft}

\end{table*}

\begin{table*}
	\centering
	\footnotesize
	\caption{Number abundances (log n(Z)/n(H(e)) of the material polluting the white dwarfs calculated using the photometric and spectroscopic white dwarf parameters separately. For the derivation of the upper limits the spectroscopic solutions are used. A dash denotes that it was not possible to derive an abundance or upper limit for this element due to strong gaseous emission lines present, strong non-photometric features contaminating the spectrum, or no data available in the required wavelength range. 
	}
	\label{tab:WDs-ab}
	\begin{tabular}{cccccccccc} 
		\hline
		[X/H(e)] & Spec/Phot & UV/Op & {Gaia\,J0006} & {Gaia\,J0347} & {Gaia\,J0510} & {Gaia\,J0611}$^{\ddagger }$ & {Gaia\,J0644} & {WD\,1622} & {Gaia\,J2100} \\
		\hline	
		
		H & - &  - & - & -  & - & - & $-$5.16 & $-$3.37 & - \\ \\

        C & Spec & UV & $-$6.90\,$\pm$\,0.10 & - & $<-$8.27 & $-$7.15\,$\pm$\,0.10 & - & $-$4.75\,$\pm$\,0.11& - \\
        " & Phot & UV & $-$6.93\,$\pm$\,0.10 & - &  & $-$7.29\,$\pm$\,0.10 & - & $-$4.69\,$\pm$\,0.25& - \\ \\
		
		O & Spec & Op & $<-$3.92$^*$ & - & $-$4.23\,$\pm$\,0.11 & $-$3.75\,$\pm$\,0.16  &  $-$5.17\,$\pm$\,0.13 & $<-$4.46$^*$ &  $<-$4.10$^*$ \\
		" & Phot & Op & & - & $-$4.35\,$\pm$\,0.11 & $-$3.84\,$\pm$\,0.16 & $-$5.60\,$\pm$\,0.13 &  &  \\
  		" & Spec & UV & $-$4.48\,$\pm$\,0.10 & - & $-$4.98\,$\pm$\,0.12 & $-$4.28\,$\pm$\,0.10 & - & $-$5.39\,$\pm$\,0.10 & -\\
		" & Phot & UV & $-$4.54\,$\pm$\,0.10 & - & $-$5.07\,$\pm$\,0.12  & $-$4.33\,$\pm$\,0.10 & - & $-$5.80\,$\pm$\,0.10 & -\\ \\

        S & Spec & UV & $-$6.35\,$\pm$\,0.12 & - & $-$6.20\,$\pm$\,0.10 & $-$5.30\,$\pm$\,0.25 & - & $-$5.99\,$\pm$\,0.10 & -\\
        " & Phot & UV & $-$6.46\,$\pm$\,0.11 & - &$-$6.19\,$\pm$\,0.10 & $-$5.47\,$\pm$\,0.14 & - & $-$6.01\,$\pm$\,0.24 & -\\ \\

        P & Spec & UV & $-$7.39\,$\pm$\,0.22 & - & $<-$7.90 & $-$7.49\,$\pm$\,0.12 & - & $-$7.74\,$\pm$\,0.15 & -\\
        " & Phot & UV & $-$7.55\,$\pm$\,0.22 & - &   & $-$7.66\,$\pm$\,0.12 & - & $-$8.01\,$\pm$\,0.15 & -\\ \\
  
		Na & Spec & Op & $<-$5.11 & -  & $<-$5.44 & - & $<-$5.65 & $<-$4.67 & $<-$5.18 \\ \\
		
		Mg & Spec & Op & $-$4.95\,$\pm$\,0.10 & $-$5.78\,$\pm$\,0.16 & $-$5.23\,$\pm$\,0.10 & $-$4.61\,$\pm$\,0.11 & $-$5.73\,$\pm$\,0.10 & $-$4.91\,$\pm$\,0.10 & $-$5.08\,$\pm$\,0.10  \\
		" & Phot & Op & $-$5.03\,$\pm$\,0.10 & $-$6.05\,$\pm$\,0.16 & $-$5.35\,$\pm$\,0.10 & $-$4.68\,$\pm$\,0.11 &  $-$6.33\,$\pm$\,0.10 & $-$5.49\,$\pm$\,0.17 & $-$5.35\,$\pm$\,0.10 \\ 
        " & Spec & UV & - & - & - & - &- & $-$4.76\,$\pm$\,0.10 & $-$5.23\,$\pm$\,0.10 \\
		" & Phot & UV & - & - &- & - &- & $-$5.27\,$\pm$\,0.10 & $-$5.57\,$\pm$\,0.10 \\ \\
  
	    Al & Spec & Op & $<-$5.60 & $<-$5.20 & $<-$5.30 & $<-$4.50 & $-$6.76\,$\pm$\,0.11 & $<-$5.90 & $<-$5.80 \\
	    " & Phot & Op & & & & & $-$7.05\,$\pm$\,0.11 & & \\
        " & Spec & UV & $-$6.5\,$\pm$\,0.18 & $-$7.34\,$\pm$\,0.20 & $-$7.04\,$\pm$\,0.10 & $-$6.46\,$\pm$\,0.10 & - & $-$6.28\,$\pm$\,0.10 & $-$6.48\,$\pm$\,0.10 \\
        " & Phot & UV & $-$6.5\,$\pm$\,0.18 & $-$7.26\,$\pm$\,0.20 & $-$6.99\,$\pm$\,0.15 & $-$6.71\,$\pm$\,0.18 & - & $-$6.38\,$\pm$\,0.10 & $-$6.49\,$\pm$\,0.10 \\ \\

		Si & Spec & Op & $-$4.93\,$\pm$\,0.10 &   $<-$5.30 & $-$5.12\,$\pm$\,0.10 & $-$4.70\,$\pm$\,0.11  & $-$5.97\,$\pm$\,0.10 & $-$5.20\,$\pm$\,0.10 & $-$5.12\,$\pm$\,0.10 \\
		" & Phot & Op & $-$5.03\,$\pm$\,0.10 &  & $-$5.13\,$\pm$\,0.10 & $-$4.72\,$\pm$\,0.15 & $-$6.26\,$\pm$\,0.10 & $-$5.68\,$\pm$\,0.10 & $-$5.36\,$\pm$\,0.10 \\  
        " & Spec & UV & $-$5.48\,$\pm$\,0.10 & - & $-$5.85\,$\pm$\,0.17 & $-$5.22\,$\pm$\,0.10 & - & $-$5.20\,$\pm$\,0.10 & -\\
        " & Phot & UV & $-$5.50\,$\pm$\,0.10 & - & $-$5.88\,$\pm$\,0.11 & $-$5.24\,$\pm$\,0.10 & - & $-$5.18\,$\pm$\,0.17& -\\ \\
		
        Ca & Spec & Op & $-$6.17\,$\pm$\,0.10 & $<-$5.68\,$\pm$\,0.11$^{\dagger}$ & $-$6.31\,$\pm$\,0.10 & $-$6.08\,$\pm$\,0.15 & $-$6.70\,$\pm$\,0.10 & $-$5.85\,$\pm$\,0.10 & $-$6.21\,$\pm$\,0.11 \\
        " & Phot & Op & $-$6.32\,$\pm$\,0.10 & $<-$5.92\,$\pm$\,0.35$^{\dagger}$ & $-$6.80\,$\pm$\,0.10 & $-$6.37\,$\pm$\,0.14 & $-$7.41\,$\pm$\,0.10 & $-$7.02\,$\pm$\,0.10 & $-$6.65\,$\pm$\,0.10 \\ \\
        
        Ti & Spec & Op & $<-$6.30 & $<-$6.43 & $<-$5.86 & $<-$5.64 & $-$8.35\,$\pm$\,0.11 & $<-$6.77 & $<-$6.69\\
        " & Phot & Op & & & & & $-$9.13\,$\pm$\,0.10 & & \\ \\
		
		Cr & Spec & Op & $<-$5.71 & $<-$5.47 & $<-$4.53 & $<-$3.50 & $-$7.80\,$\pm$\,0.10 & $<-$5.74 & $<-$5.78\\
        " & Phot & Op & & & & & $-$8.56\,$\pm$\,0.10 & & \\ \\
		
		Fe & Spec & Op & $<-$4.66$^*$ & $<-$4.38$^*$ & $<-$4.10$^*$ & - & $-$6.51\,$\pm$\,0.10 & $<-$5.23$^*$ & $-$4.96\,$\pm$\,0.14 \\
        " & Phot & Op & & & & - & $-$7.15\,$\pm$\,0.10 & & $-$5.49\,$\pm$\,0.14 \\
        " & Spec & UV & $<-$5.00 & - & $<-$5.60 & $-$5.23\,$\pm$\,0.10 & & $-$5.26\,$\pm$\,0.10 & -\\
        " & Phot & UV &  & - &  & $-$5.45\,$\pm$\,0.10 & & $-$5.55\,$\pm$\,0.13 & -\\ \\
		
		Ni & Spec & Op & $<-$4.99 & $<-$4.67 & $<-$3.77 & $<-$3.00 & $<-$7.21 & $<-$5.10 & $<-$5.11 \\
        " & Spec & UV & $<-$6.50 & - & $<-$7.20 & $-$6.83\,$\pm$\,0.10 & - & $-$6.29\,$\pm$\,0.24 & -\\
        " & Phot & UV &  & - &  & $-$7.05\,$\pm$\,0.10 & - & $-$6.71\,$\pm$\,0.28 & -\\ \\
		
		\hline
	\end{tabular}
	\vspace{-3mm}
	\small \begin{flushleft}
    \item \textbf{Notes:} 
    \item $^*$Denotes gaseous emission present when derived upper limit.
    \item $^{\dagger}$Non-photospheric lines at approximately the same radial velocity as white dwarf photosphere, so this should be treated as an upper limit.
    \item $^{\ddagger }$ Cu measured for Gaia\,J0611$-$6931 to 2.8\,$\sigma$, abundances are: $-$7.87\,$\pm$\,0.18 and $-$7.94\,$\pm$\,0.18 for the spectroscopic and photometric parameters respectively, see Fig.\,D6 for the model fit. 
    \end{flushleft}
\end{table*}

\subsection{X-shooter}

Four of the white dwarfs were observed with the echelle spectrograph X-shooter \citep{vernet2011x} on Unit Telescope 3 (UT3) of the Very Large Telescope (VLT) at Paranal Observatory, Chile. X-shooter allows simultaneous observations in the 3 arms: UVB (3000--5595\,\AA), VIS (5595--10240\,\AA), and NIR (10240--24800\,\AA). The white dwarfs are too faint to have strong signals in the NIR arm, so the NIR data were excluded from this study. The observations were taken between 2019 July-October during runs 0103.C-0431(B) and 0104.C-0107(A). For all observations, stare mode was used, with a 1.0 and 0.9 arcsec slit width for the UVB and VIS arms, respectively, this gives a resolving power ($\lambda / \Delta \lambda $) of 5400 and 8900. Two exposures were taken lasting 1700 and 1729\,s each for the UVB and the VIS arms, respectively. The data reduction was performed using \textsc{esoreflex} (v 2.11.3) with the X-shooter pipeline version 2.9.1 \citep{2013A&A...559A..96F}. The standard reduction procedures were followed including minor alterations that improved the signal-to-noise ratio (SNR) of the output spectrum, and reduced the number of cosmic ray contaminants. The details of the X-shooter observations are listed in Table\,\ref{tab:WDs-obs}. The SNR at the continuum around the Ca\,\textsc{ii} K (3993\,\AA) line was 30--134 for the UVB arm and the SNR at the continuum around 6600\,\AA\ was 35--127 for the VIS arms depending on the flux of the white dwarf and observing conditions. 

\subsection{HIRES}

Six of the white dwarfs were observed with the High Resolution Echelle Spectrometer (HIRES) on the Keck I Telescope, Hawaii \citep{vogt1994hires}. This has 2 modes, HIRESb and HIRESr, with a wavelength coverage of approximately 3200--5750\,\AA \, and 4700--9000\,\AA \, respectively. The C5 decker was used, which has a slit width of 1.148 arcsec and a spectral resolution of 37,000. The observations were taken between 2019 July and 2020 October. 

Data reduction including bias subtraction, flat fielding, wavelength calibration, and spectral extraction were performed using \textsc{makee} following \citet{xu2016evidence}. The final spectra were continuum normalised using low order polynomials and combined using \textsc{iraf} functions \citep{klein2010chemical}. The details of the HIRES  observations are listed in Table\,\ref{tab:WDs-obs}. The SNR for one observation was 32--89 around the continuum at the Ca K line for HIRESb and 31--73 around the continuum at 6600\,\AA\, for HIRESr.

\subsection{MIKE}

Gaia\,J0611$-$6931 was observed with the Magellan Inamori Kyocera Echelle (MIKE) spectrograph \citep{bernstein2003mike} on the 6.5\,m Magellan Clay Telescope at Las Campanas Observatory on 2021 August 27 with one exposure of 1800\,s followed by a second of 1300\,s. Observations were taken at airmass 1.5 with the atmospheric dispersion corrector installed, but possibly not correcting the spectrum optimally, which would result in a lower SNR than expected in the blue. The SNR of the continuum was about 15 near the Ca infrared triplet. The seeing was 0.8 arcsec, and the employed 1 arcsec slit produces spectral resolution of R\,$\approx$\,28,000 on the blue side (3500--5060\,\AA) and 22,000 on the red side (5000--9400\,\AA). ThAr lamps taken before and after the exposures were used for wavelength calibration. Data reduction with the standard Carnegie Python MIKE pipeline included extraction, flat-fielding, and wavelength calibration using methods described in \citet{kelson2000evolution} and \citet{ kelson2003optimal}.

\subsection{HST COS far ultraviolet spectra}

Far ultraviolet (FUV) spectroscopic observations of four of the white dwarfs were conducted with the FUV channel of comic origins spectrograph (COS) on the \textit{Hubble Space Telescope} (\textit{HST}) (Programme ID: 16752), the observations are reported in Table \ref{tab:WDs-obs-uv}. The G130M grating was used with a central wavelength of 1291\,\AA, resulting in a wavelength coverage of 1150--1430\,\AA ~(20\,\AA~gap in between the two segments). The data were reduced with the \textsc{calcos} reduction pipeline. The data were obtained using the TIME-TAG mode allowing the data taken when the Sun is below the geometric horizon from the point of view of \textit{HST} (`night' data) to be separated from those taken when it is above the horizon (`day' data). Data taken during the day can contain geocoronal contributions of Lyman alpha and O\,\textsc{i} emission lines, and all four of the white dwarf spectra show this. There are photospheric lines of O\,\textsc{i} and Si\,\textsc{ii} that are blended with these emission features, therefore, for the three white dwarfs with night data, the STScI COS notebooks\footnote{\url{https://www.stsci.edu/hst/instrumentation/cos/documentation/notebooks}} were used to separate out the night and day data. The COS2025 strategy means only FP-POS 3 and 4 can be used, the exposures were split between these positions and the final data resulted in a median stack between these observations. There are small discrepancies in the resolutions of these lifetime positions, however, tests reveal this affects the final abundances by less than 0.05 dex. The SNR reported in Table \ref{tab:WDs-obs-uv} was calculated from the continuum around the C\,\textsc{ii} lines at 1334.530 and 1335.708\,\AA~using the STScI COS notebooks.

\subsection{HST COS near ultraviolet spectra}

Near ultraviolet (NUV) spectroscopic observations of six of the white dwarfs were conducted with the NUV channel of COS (Programme ID: 16204). The data were reduced with the \textsc{calcos} reduction pipeline. The G230L grating was used with a central wavelength of 2950\,\AA. Unlike the well-calibrated FUV wavelength scale, this grating has a zero point accuracy to within 175 km\,s$^{-1}$, therefore the radial velocities of the lines are offset in comparison to those from the optical and FUV, as seen in Supplementary Tables B1--B13, thus the radial velocities of the NUV data are not meaningful. The SNR reported in Table \ref{tab:WDs-obs-uv} was calculated from the continuum around the Al\,\textsc{iii} lines at 1854.716 and 1862.790\,\AA.

\section{Modelling Methods} \label{Mod}

White dwarf atmospheric models were used to derive stellar parameters, and measure the abundance of the polluting material \citep[][]{dufour2012detailed}.

\subsection{White Dwarf Parameters} \label{par}

Two methods were used to derive the stellar parameters for each white dwarf. The first fitted white dwarf models to broad-band photometry (the photometric method), and the second fitted white dwarf models to the pressure broadened hydrogen and helium spectral lines (the spectroscopic method). The derived values for the spectroscopic and photometric methods are reported in Table \ref{tab:WDs-prop}. These methods are discussed in more detail below.

\subsubsection{Photometric method}

For the photometric method, white dwarf models were fitted to SDSS, Pan-STARRS, SkyMapper, and \textit{GALEX} broad-band photometry with parallaxes from \textit{Gaia} to extract the best-fitting effective temperature and $\log(g)$. For the two DBs in the sample, fixed values of H/He were used that matched the values measured from the spectra, as described in the following section. SDSS to AB corrections were included as outlined in \citet{eisenstein2006catalog}. Reddening becomes important for objects $>$ 100\,pc, and five of the white dwarfs fall into this distance range. For these white dwarfs the observational data were de-reddened using the method as described in \citet{genest2019comprehensive} before fitting. When available, a combination of SDSS $ugriz$ and Pan-STARRS $grizy$ were used to constrain the white dwarf parameters\footnote{Photometry listed on: \url{https://montrealwhitedwarfdatabase.org}}. As discussed in \citet{genest2019comprehensive}, this provides the most accurate and consistent results. Otherwise, either the SDSS $ugriz$, Pan-STARRS $grizy$, or SkyMapper $ugrizy$ were used. 

\subsubsection{Spectroscopic method}

\begin{figure*}
	\includegraphics[width=0.9\textwidth]{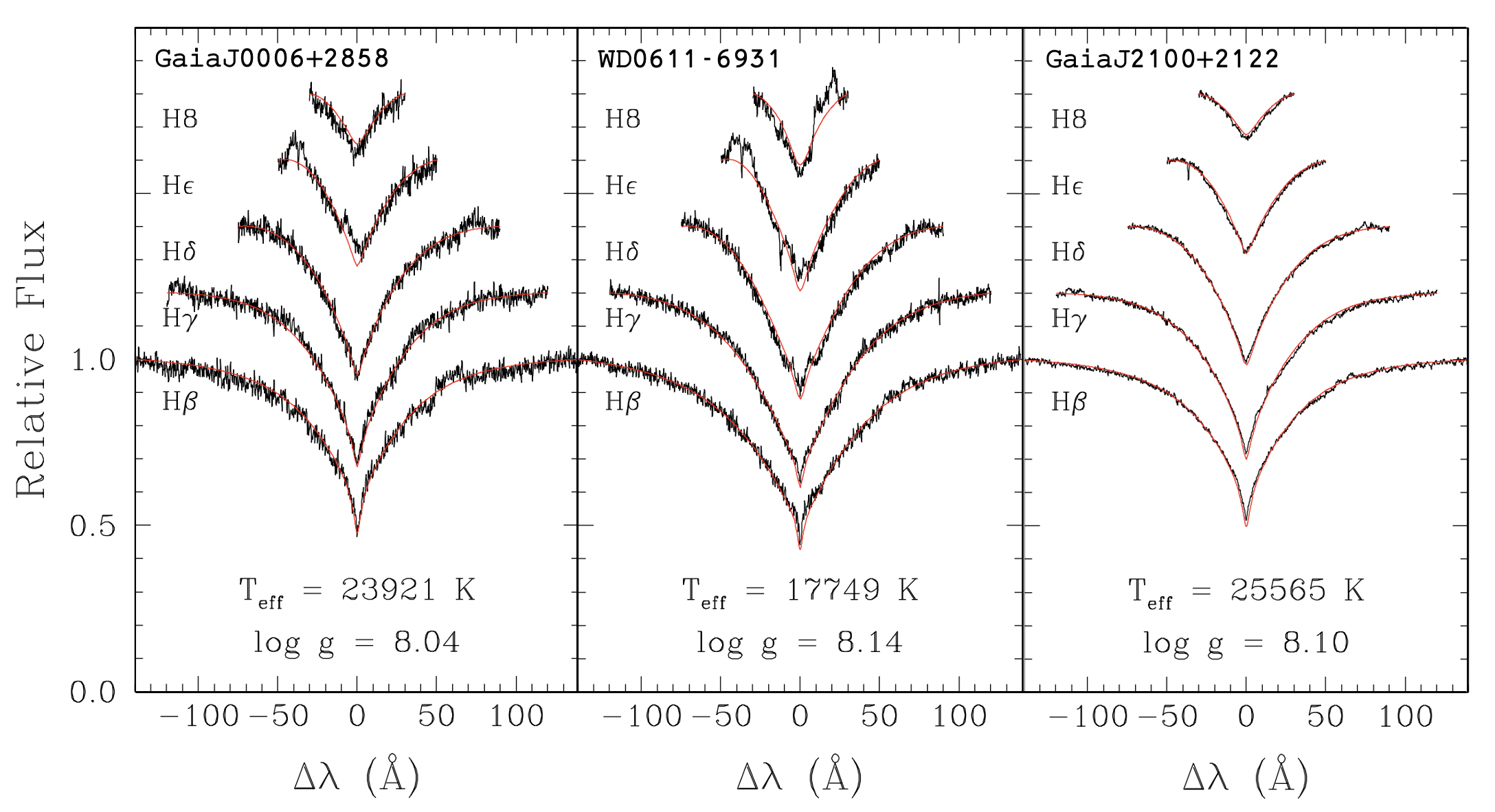}
    \caption{Model fits to the Hydrogen Balmer line profiles from the X-shooter spectra for Gaia\,J0006+2858, Gaia\,J0611$-$6931, and Gaia\,J2100+2122. The best-fitting model parameters are labelled on each panel. Double peaked emission lines originating from heavy elements ($>$\,He) in the circumstellar gas discs are visible in some of these Balmer profiles, most of these were identified in \citet{melis2020serendipitous} and \citet{dennihy2020five}. The emission feature near the core of H$\epsilon$ is from Ca\,\textsc{ii} at 3968.47\,\AA, not from hydrogen.} 
    \label{fig:spec-fit}
\end{figure*}

\begin{figure*}
\centering
\subfloat[Gaia\,J0644$-$0352]{
  \includegraphics[width= 0.45\textwidth]{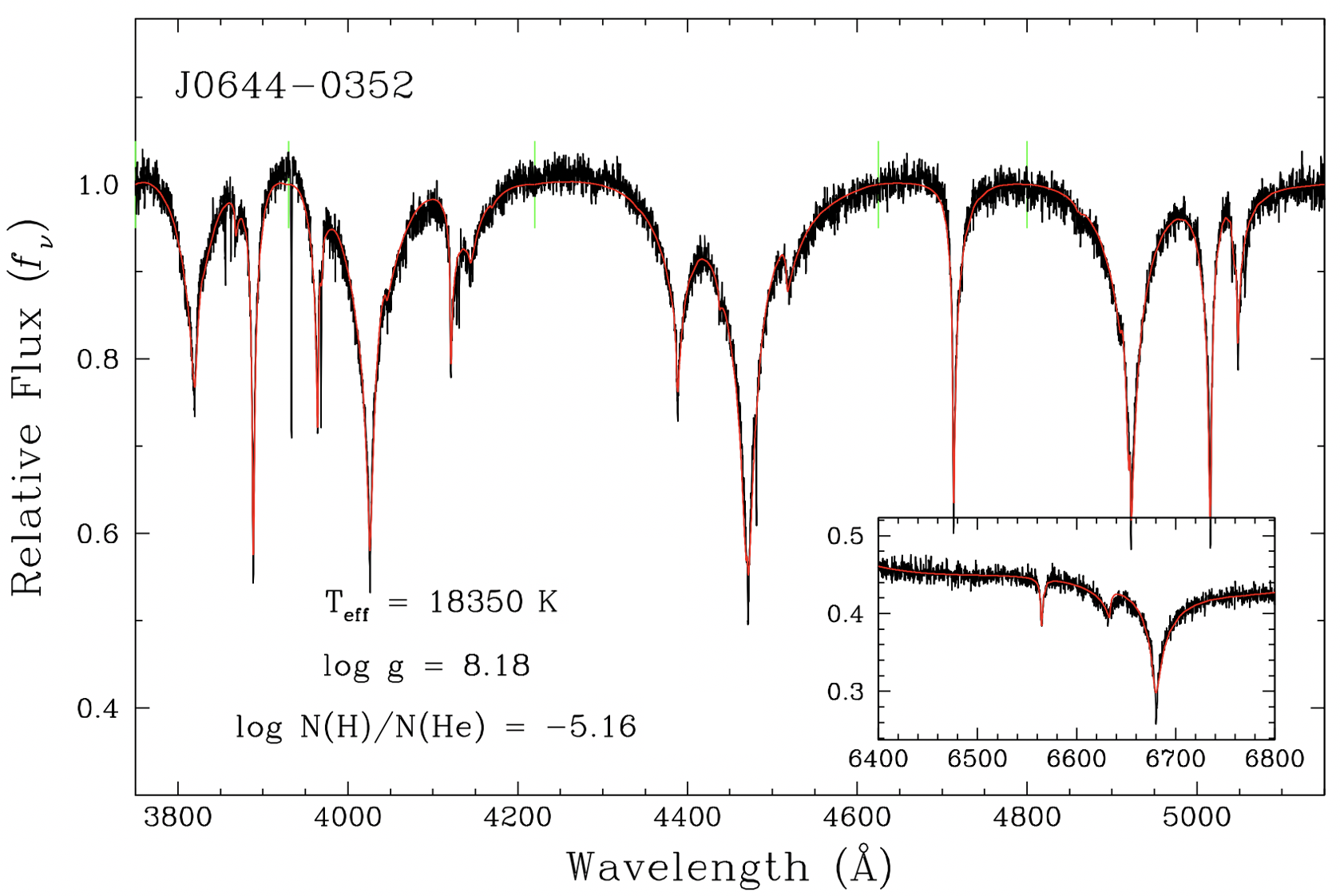}
  \label{fig:spec-fit-0644}
}
\subfloat[WD1622+587]{
  \includegraphics[width= 0.45\textwidth]{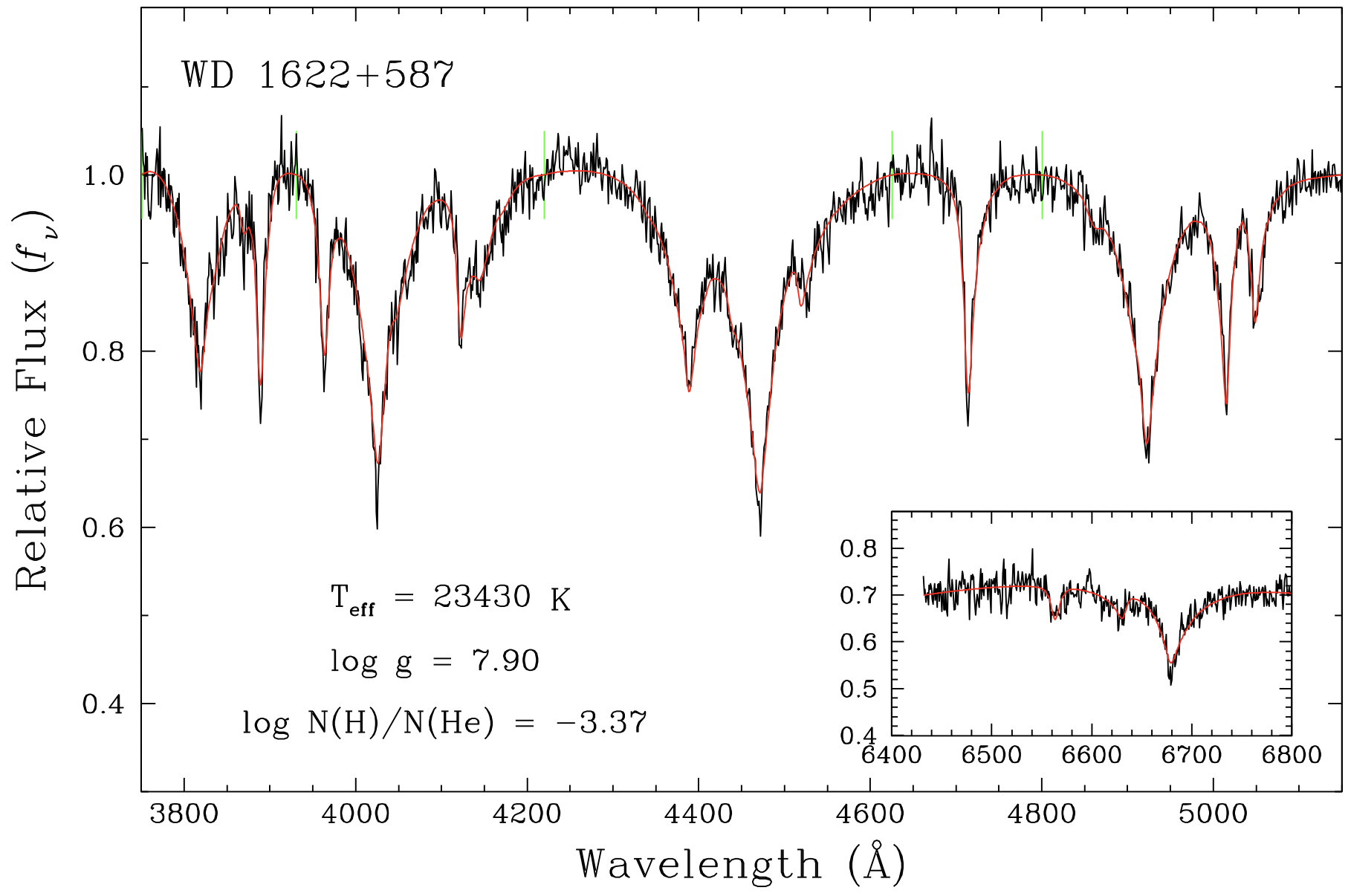}
  \label{fig:spec-fit-1622}
}
\caption{Model fits to the X-shooter Helium lines for Gaia\,J0644$-$0352 (left), and the KAST data from \citet{melis2020serendipitous} taken on UT 12-07-2019 (right). The best-fitting white dwarf parameters and hydrogen abundance are labelled.}
\end{figure*}

The spectroscopic method fits synthetic white dwarf model spectra to the hydrogen and helium optical absorption lines to extract the best-fitting effective temperature and $\log(g)$. Updated white dwarfs parameters were found for the four white dwarfs with X-shooter data presented here, and for WD\,1622+587 using the KAST data from \citet{melis2020serendipitous}. For the two remaining objects, the parameters derived in \citet{melis2020serendipitous} were used. The model fits to the Balmer lines for the three DA white dwarfs (Gaia\,J0006+2858, Gaia\,J0611$-$6931, and Gaia\,J2100+2122) are shown in Fig.\,\ref{fig:spec-fit}. Doubled peaked emission lines from the circumstellar gas discs are in the hydrogen lines, these features represent a small fraction of the frequency points and are not found to affect the derived parameters. For the DB white dwarfs, GaiaJ0644$-$0352 and WD\,1622+587, models were fitted to the Helium lines as shown in Figs.\,\ref{fig:spec-fit-0644} and \ref{fig:spec-fit-1622}, trace H is also present, and this abundance was also determined in this fit. Heavy elements and hydrogen in cool DBZ stars may affect the pressure/temperature structure and therefore affect the derived white dwarf parameters \citep{dufour2012detailed,coutu2019analysis}. It was tested whether including heavy elements in the models affected the derived parameters; as these white dwarfs are hot ($T_\mathrm{eff} > $ 20,000\,K), the inclusion of heavy elements had a negligible affect on the derived white dwarf parameters. 

The uncertainties in the spectroscopically derived effective temperature and $\log(g)$ for DA white dwarfs are from \citet{Liebert2005formation}, 1.4 per cent in $T_\mathrm{eff}$ and 0.042 dex in $\log(g)$. For the two DBA white dwarfs, the uncertainties derived in \citet{izquierdo2023systematic} are used 524\,K and in $T_\mathrm{eff}$ and 0.27 dex in $\log(g)$. These are  uncertainties on the fitting procedure and do not encapsulate uncertainties from the model atmospheres.

\subsection{Line identification} \label{Lines}

A multitude of lines from heavy elements were identified in the spectra of each white dwarf. Using the \textsc{iraf} task \textsc{splot} \citep{tody1986iraf} the equivalent width, line centre, and radial velocity of each spectral line was measured. The atomic data bases of Vienna Atomic Line Database (VALD)\footnote{\url{http://vald.astro.uu.se}}, National Institute of Standards and Technology (NIST)\footnote{\url{https://physics.nist.gov/PhysRefData/ASD/lines_form.html}}, and \citet{van2018recent}, as well as published line lists of polluted white dwarfs observed with Keck and \textit{HST} \citep[e.g.][]{klein2011rocky,jura2012two,gansicke2012chemical} were utilised to identify which element species are associated with the spectral features. For the equivalent width, a Voigt function was fitted to the profile of the line five times whilst changing the region used for the continuum fitting. From this the average equivalent width and standard deviation for each line was found, this was compared to a direct flux summation to ensure accuracy. The uncertainty on the equivalent width was calculated by combining in quadrature the standard deviation of the equivalent width measurements and the \textsc{splot} fitting error. The radial velocities of the lines were calculated using the core of the Voigt profile, the standard deviations of the radial velocities are reported in Table \ref{tab:RVs} and the variation may be due to Stark  shifts \citep{vennes2011pressure}; further investigation is beyond the scope of this work. Supplementary Tables B1--B13 list the spectral lines identified in both the ultraviolet and optical, the derived equivalent widths and errors, line centres, and radial velocities.

Non-photospheric absorption lines can be present in the spectra of polluted white dwarfs, and may be due to interstellar absorption or absorption from circumstellar material \citep{debes2012detection,vennes2013polluted,vanderbosch2021recurring}. This can cause additional uncertainties when deriving abundances of metals in the photospheres of white dwarfs. Table \ref{tab:RVs} shows the non-photospheric measurements of lines in the spectra of the seven white dwarfs in this study. These lines are usually offset from the velocity of the photospheric lines so can be distinguished. For the ultraviolet wavelengths, an interstellar medium model was used to fit Voigt profiles to the non-photospheric contributions, enabling abundance determination for the photospheric component. 

\begin{figure*}
	\includegraphics[width=0.7\textwidth]{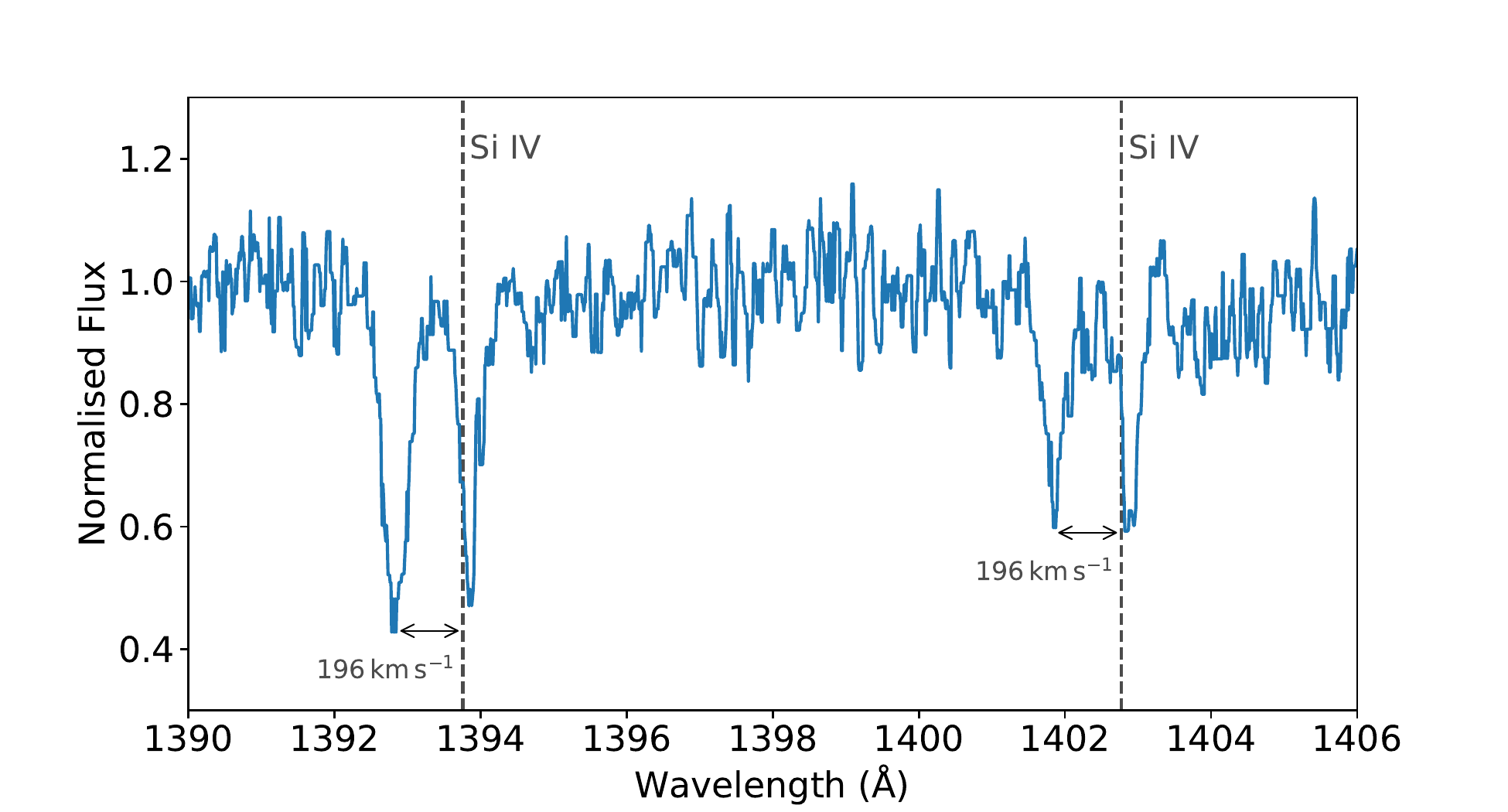}
    \caption{Si\,\textsc{iv} lines in the ultraviolet FUV spectrum of Gaia\,J0006+2858. The photospheric lines are redshifted compared to the rest  wavelength of the Si\,\textsc{iv} lines and an additional absorption component is present blueshifted from the rest wavelength of the Si\,\textsc{iv} lines of $-$196\,km\,s$^{-1}$.}
    \label{fig:Si4}
\end{figure*}

There are two Si\,\textsc{iv} lines observed in the photosphere of Gaia\,J0006+2858 at wavelengths of 1393.76 and 1402.77\,\AA. Both lines have a blueshifted component offset from the silicon rest frame wavelengths by $-$196\,km\,s$^{-1}$, with measured line centers of 1392.84 and 1401.85\,\AA~as shown in Fig.\,\ref{fig:Si4}. From the atomic data bases, absorption from other elemental species is ruled out. The relative shift between the line centers from the blueshifted component compared to the Si\,\textsc{iv} photospheric component are: 216\,km\,s$^{-1}$ and 222\,km\,s$^{-1}$ respectively. It is likely that these two absorption lines are blueshifted Si\,\textsc{iv} lines. Additional absorption components to these Si\,\textsc{iv} lines have been previously observed and are thought to be circumstellar absorption from close in hot gas, however, the velocity offset is much less extreme \citep{fortin2020modeling, gansicke2012chemical}. The width of these blueshifted lines seen in Gaia\,J0006+2858 is consistent with the range in velocities expected for a gas disc which occults the white dwarf. No circumstellar silicon gaseous \textit{emission} features are observed \citep{melis2020serendipitous}, so the radial extent of the silicon part of the gas disc cannot be compared. More detailed models of the gas disc are required to understand these observations. The other three white dwarfs observed in the FUV show no additional Si\,\textsc{iv} absorption components. 

\subsection{Abundance of polluting metals}

The abundances of metals in the atmosphere of each white dwarf were measured following \citet{dufour2012detailed}. The spectra were divided into panels which cover a region of 5--15\,\AA \,\,around each absorption line. The white dwarf effective temperature and $\log(g)$ was inputted, and the best-fitting abundance for that spectral line was found. The abundances of the lines were fitted using the effective temperature and $\log(g)$ from the photometric and spectroscopic methods separately. When more than one line of a particular element are present in the 5--15\,\AA \,region, the lines are fitted together.

\subsubsection{Absorption features in the presence of emission features}

Some absorption lines also have gaseous emission features present at the same wavelength which makes it difficult to disentangle the contribution from the photosphere from the circumstellar emission. In Supplementary Tables B1--B13 those lines which have photospheric absorption at the same wavelength as the gaseous emission features are noted. High order polynomials are fitted to the spectra to normalise out the broader gaseous emission features. In \citet{klein2010chemical} it is noted that through tests which varied the order of the normalisation polynomial, the effect of continuum normalisation on narrow absorption lines in the presence of bumpy features was $<$\,1 per cent. However, in this sample, the spectra contain both broad and sharp gaseous emission features which are less trivial to normalise out in order to obtain accurate equivalent widths of the photospheric features. To test how the sharp gaseous emission features affect the derived equivalent widths, tests were performed on the 3933\,\AA \,Ca\,\textsc{ii} line in Gaia\,J0006$+$2858 and 7771\,\AA \, O\,\textsc{i} line in Gaia\,J0510+2315. The equivalent widths were measured using \textsc{splot}, as explained in Section \ref{Lines}, using both the un-normalised and normalised HIRES spectra. The average equivalent width deviation was found to be 12 per cent. For those spectral lines with gaseous emission features, this additional error of 12 per cent was added in quadrature with the equivalent width error.  Gaia\,J0006$+$2858 has abundances determined from two calcium lines, 3179\,\AA \, and 3933\,\AA , where the later has a weak gaseous emission feature at the same wavelength. The calcium abundance derived from 3933\,\AA \, and that derived from 3179\,\AA \, are consistent within the errors of 0.1 dex, providing confidence that the derived abundances in the presence of gaseous emission features are accurate.  

\subsubsection{Upper limits} \label{lim}

Some important elements are not detected, and so an equivalent width upper limit that would have resulted in a 3$\sigma$ spectral line detection was calculated. Either a detection or upper limit was determined for these elements for each white dwarf: C, O, S, P, Na, Mg, Al, Si, Ca, Ti, Cr, Fe and Ni. Around the strongest line for a particular element, a spectral line was artificially inserted at decreasing values of equivalent width, corresponding to decreasing abundance. From this, the significance of the absorption feature was calculated, and repeated 10000 times. The equivalent width upper limit was taken to be the point at which 99.7 per cent of the lines were detected at 3$\sigma$ for a certain equivalent width. The equivalent width upper limits are reported in Table \ref{tab:WDs-EW-UL}. White dwarf models were used to convert from equivalent width to abundance assuming the spectroscopically derived $T_\mathrm{eff}$ and $\log(g)$; the abundance upper limits are reported in Table \ref{tab:WDs-ab}. As mentioned in Section \ref{sp_vs_ph}, the hotter the effective temperature of the white dwarf used in the models, the larger the abundance is for the same spectral line and equivalent width. Therefore, the abundance upper limits consider the abundance error associated with effective temperature and are applicable to both the spectroscopic and photometric abundances.

\section{Resulting Abundances} \label{Results}

\subsection{Abundances of the accreted material}

This work presents seven polluted white dwarfs with well characterised abundances of multiple elemental species. For the optical data, the average abundances from X-shooter and HIRES (or X-shooter and MIKE for Gaia\,J0611$-$6931) are reported in the Supplementary Tables B1--B13. For white dwarfs with observations from different instruments, the abundances are consistent within the uncertainties when measuring the abundance from the high resolution (R\,$\approx$\,40,000) and lower resolution (R\,$\approx$\,5,000--9,000) spectrographs separately. The higher resolution data provides more spectral lines and elemental species for abundances to be measured. For those white dwarfs with interstellar or circumstellar absorption features, lower resolution data is unable to distinguish these features and consequently gives a deceptively higher abundance. If the system is free from interstellar and circumstellar absorption, then low resolution and high resolution data give consistent abundances, but depending on the requirements, higher resolution ($>$\,40,000) spectra may be preferable. 

The abundances of the planetary material polluting these seven white dwarfs are reported in Table \ref{tab:WDs-ab}. For those white dwarfs observed with both high resolution and lower resolution instruments, the average abundance of these is used, weighted by the number of panels used to derive the abundance for that instrument. The reported uncertainties have two key contributions: the spread in abundances derived for a particular element, and the error associated with the measured equivalent widths. The spread error is taken as the standard error (${\sigma} / {\sqrt N}$) of the abundances derived from the panels, and the equivalent width error is calculated by propagating the individual equivalent width errors for each spectral line used in the abundance calculation. These two contributions are added in quadrature to give the error for each abundance. If only one line of a particular element is present, the average spread error based on the observations (0.08\,dex) was added in quadrature with the equivalent width error. Given this spread as well as additional unknown uncertainties, an uncertainty floor of 0.1\,dex is used. Examples of unknown uncertainties are: limits of the atomic data, uncertainties in white dwarf parameters, and uncertainties introduced from the white dwarf models. Supplementary Figs. D1 -- D11 show the abundances fit to the strongest spectral line for each element in each white dwarf.

\subsection{Spectroscopic versus photometric white dwarf parameters used to derive abundances} \label{sp_vs_ph}

\begin{figure*}
\centering
\subfloat[Absolute abundances]{
  \includegraphics[width=0.48\textwidth]{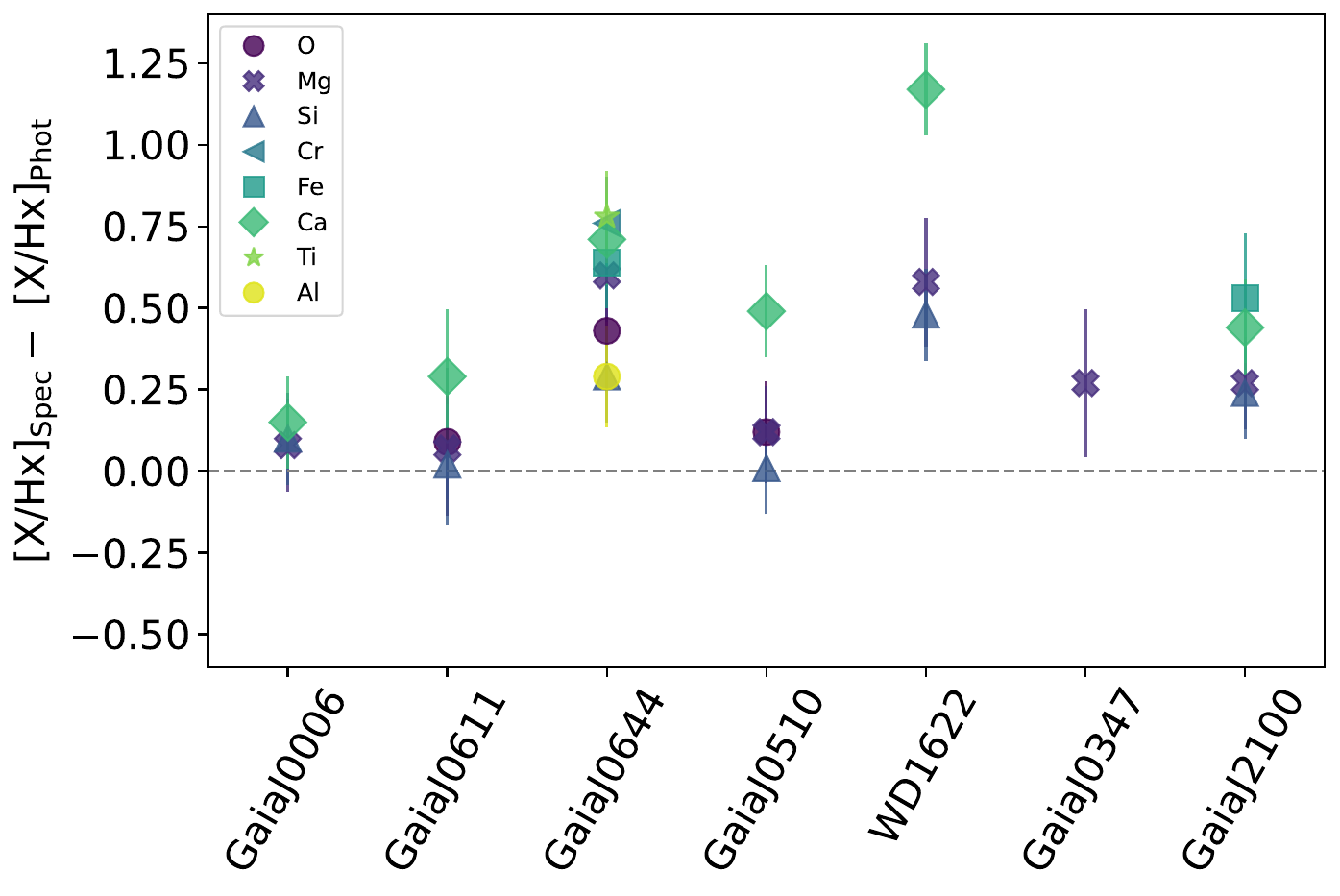}
  \label{fig:Ab-comp-absolute}
}
\subfloat[Abundance ratio to Mg]{
  \includegraphics[width=0.48\textwidth]{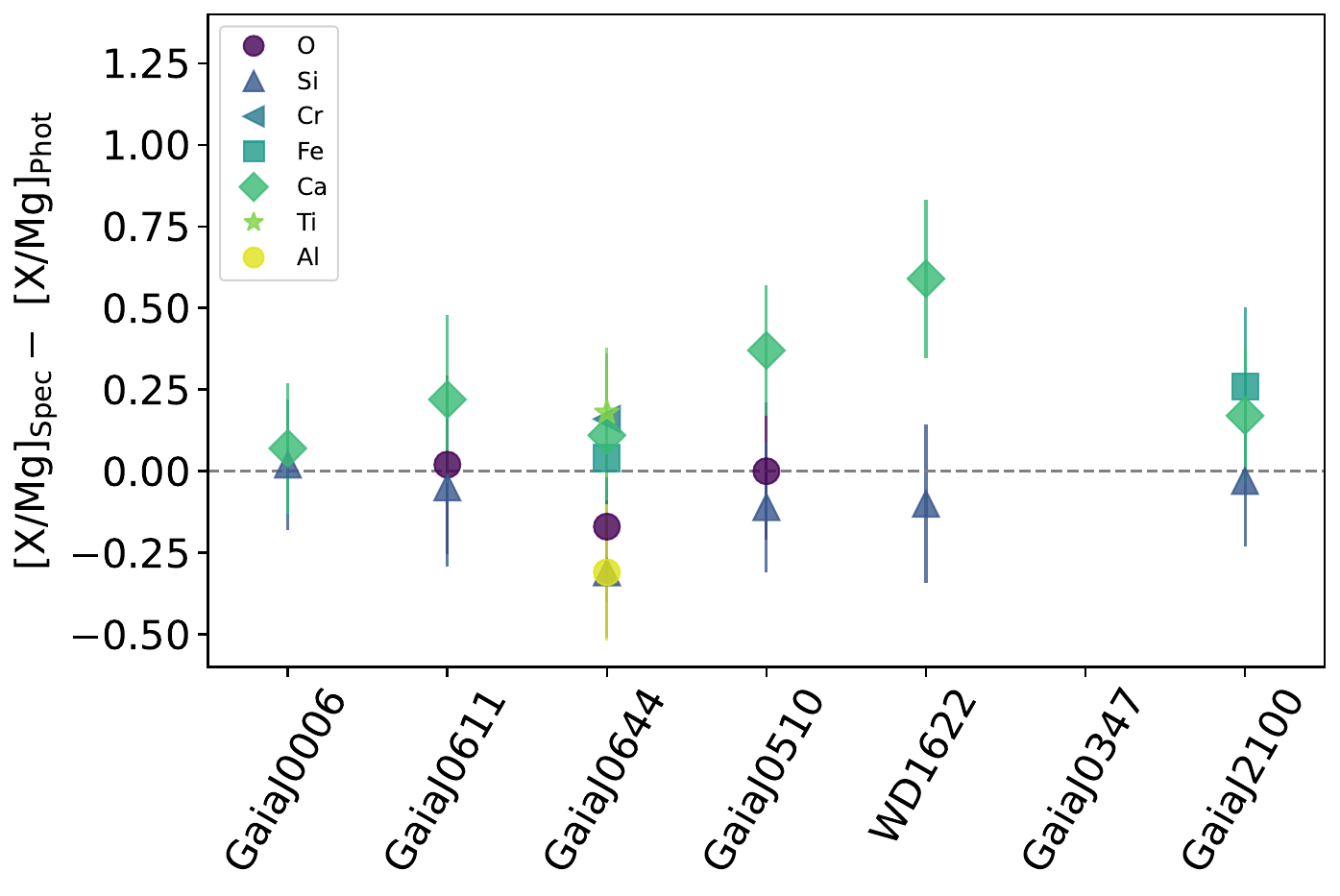}
  \label{fig:Ab-comp-ratio}
}

\caption{(a) Difference between the \textit{absolute abundance} of elements when derived from the spectroscopic white dwarf parameters versus the photometric parameters for the optical data. The white dwarfs are ordered from the smallest difference between the spectroscopic and photometric $T_\mathrm{eff}$ on the left to the largest on the right. (b) Difference between the \textit{abundance ratio} of elements with respect to Mg when derived from the spectroscopic white dwarf parameters versus the photometric parameters for the optical data. The errors on the abundance ratio assumes simple error propagation where the error on the abundance of [X/Hx] is added in quadrature with the error on the abundance of [Mg/Hx]. The $\tilde {\chi} ^{2}$ when testing the goodness-of-fit of a horizontal line at 0 is 10.24 for the absolute abundances versus 1.03 for the abundance ratios. The abundance ratios are less affected by the difference between the spectroscopic and photometric white dwarf parameters.}
\end{figure*}

The spectroscopic and the photometric methods for determining white dwarf parameters result in different sets of absolute abundances derived for the pollutant planetary material. All derived white dwarf effective temperatures are hotter for the spectroscopic method than for the photometric method. Derived heavy element abundance correlates with temperature, so the abundances of heavy elements when compared to the abundance of the principal element (H or He) as derived from the spectroscopic white dwarf parameters are larger. Figures \ref{fig:Ab-comp-absolute} and \ref{fig:Ab-comp-ratio} compare the absolute abundances compared to the abundance ratios, the $\tilde {\chi} ^{2}$ when testing the goodness-of-fit of a horizontal line at 0 is 10.24 for the absolute abundances versus 1.03 for the abundance ratios. Therefore, abundance ratios are less affected by differing white dwarf parameters for the effective temperature range of the white dwarfs in this sample. Figure \ref{fig:Ab-comp-ratio} shows that the Ca/Mg abundance ratio is most discrepant when comparing spectroscopic versus photometrically derived abundances. This is likely due to the ionisation levels of calcium being particularly sensitive to $T_\mathrm{eff}$.

\subsection{Optical versus ultraviolet spectra to derive abundances}

\begin{figure}
	\includegraphics[width=\columnwidth]{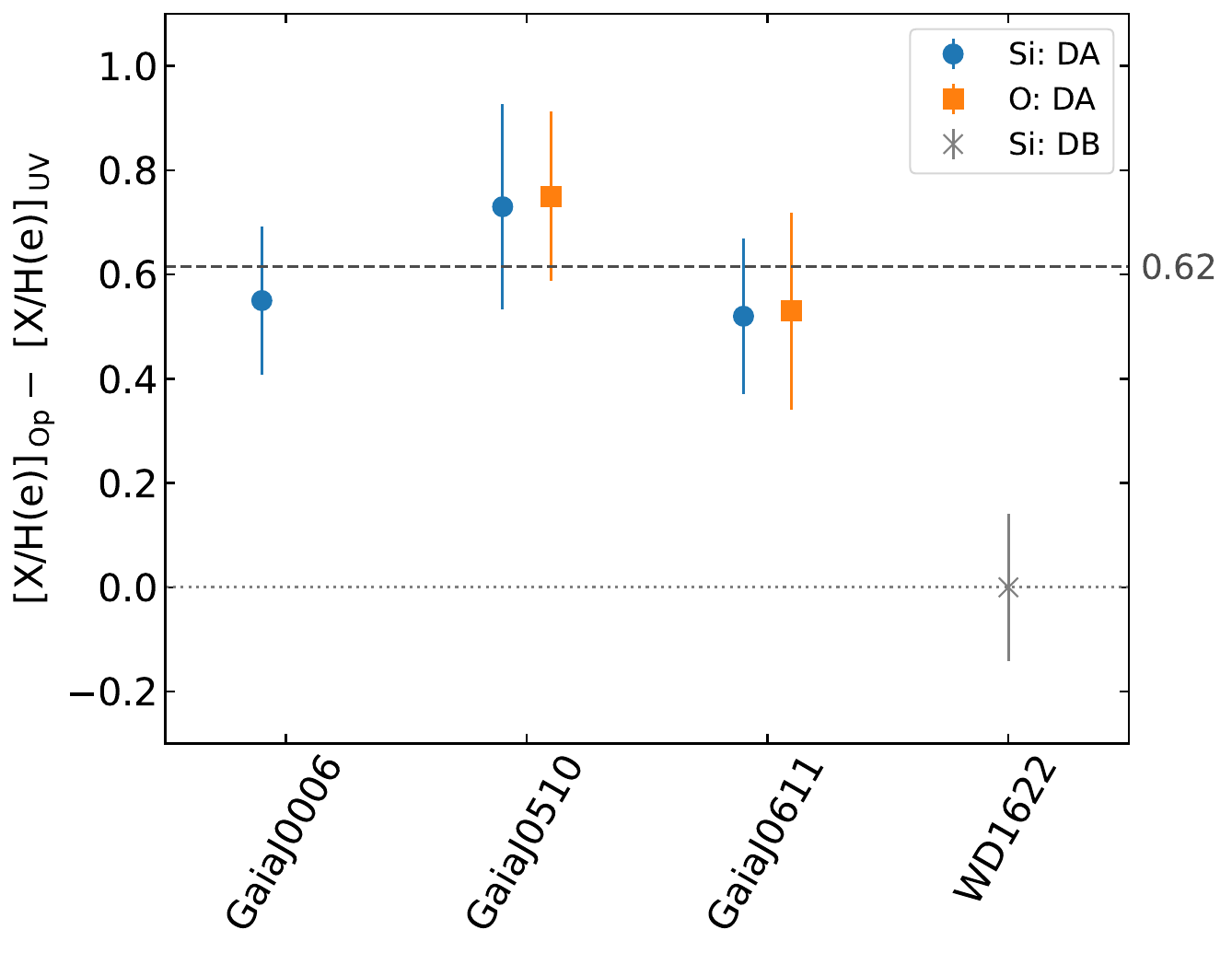}
    \caption{For the white dwarfs with both FUV and optically derived abundances, the difference between the optical and ultraviolet  abundances are shown. For the DAZ white dwarfs, the offset appears constant (mean offset of 0.62), and for the one DBZ there is no apparently offset.}
    \label{fig:uv-opt-ab}
\end{figure}

The abundances of the material polluting the white dwarfs were determined based on both optical and ultraviolet data. There are discrepancies between these abundances for the three DAZ white dwarfs, as shown in Fig.\,\ref{fig:uv-opt-ab}. The difference between the abundances appears constant with a mean offset of 0.62 dex. In this work, the one DBZ with FUV data does not show an apparent offset, however, previous studies \citep[e.g.][]{jura2012two,xu2019compositions} do observe a discrepancy.

\subsection{Accretion rates}

The white dwarfs in this sample represent seven of just 21 polluted white dwarf systems with detectable circumstellar gas and dust discs. Theory suggests that white dwarfs with circumstellar gas discs may have enhanced accretion from gas drag \citep{rafikov2011runaway} and so polluted white dwarfs with gas are hypothesised to accrete at higher rates than the general population of white dwarfs. A systematic approach is taken and investigates whether these systems are distinct in terms of their accretion rate properties. The mass accretion rates were calculated from the magnesium abundance, assuming magnesium makes up 15.8 per cent of the total mass, as in Bulk Earth \citep{allegre2001chemical}. \citet{xu2019compositions} demonstrated this as a more reliable and consistent way to measure and compare accretion rates for a sample of polluted white dwarfs, therefore only objects with magnesium abundance measurements are included. The derived mass accretion rates for white dwarfs with circumstellar gaseous emission discs with measured photospheric abundances are shown in Table \ref{tab:WDs-acc-rate},  and those without circumstellar gaseous emission discs are shown in Table \ref{tab:WDs-acc-rate-no-gas}. For the seven white dwarfs in this work, the accretion rates were found using the spectroscopic white dwarf parameters and the photometric parameters. The coolest white dwarf discovered with a circumstellar gaseous disc is WD\,0145+234, with a $T_\mathrm{eff}$ of 12,720\,K \citep{melis2020serendipitous}, thus, when comparing to the population of white dwarfs without circumstellar gas discs, the lower effective temperature cut off was set at 12,720\,K. There is no correlation between temperature (white dwarf cooling age) and accretion rate above 10,000\,K \citep{xu2019compositions, wyatt2014stochastic}.

\begin{figure}
	\includegraphics[width=\columnwidth]{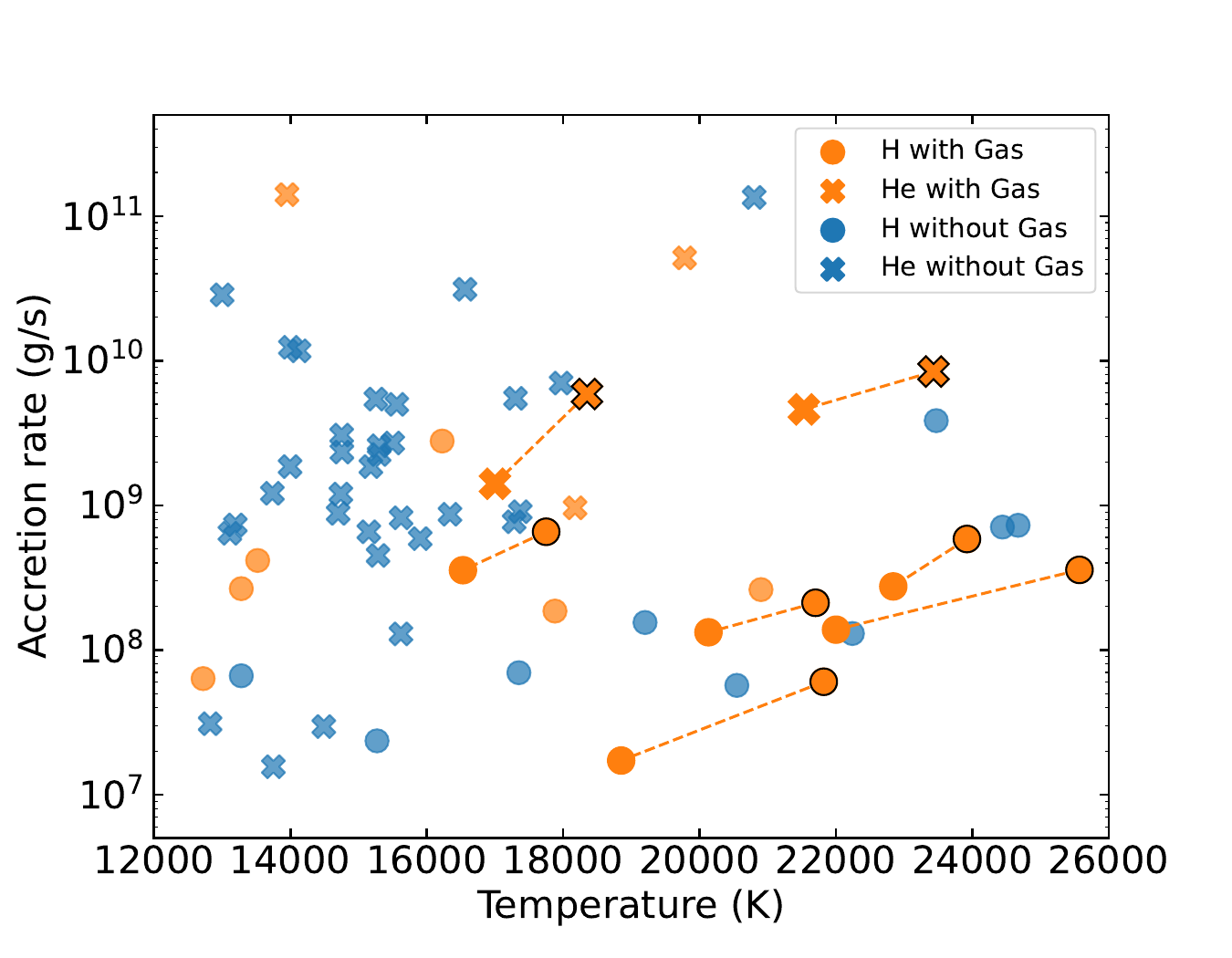}
    \caption{The accretion rates of white dwarfs with and without detectable gaseous discs in emission as a function of their effective temperatures. The data for the accretion rates of polluted white dwarfs without observable circumstellar gaseous discs are from Table \ref{tab:WDs-acc-rate-no-gas}, and the accretion rates for gaseous disc systems are from Table \ref{tab:WDs-acc-rate}. The seven systems reported in this work show both the spectroscopic and photometrically derived accretion rates and these are connected by dashed lines, where the higher effective temperatures are those derived from the spectroscopic method and have a black outline.}
    \label{fig:Acc-rates}
\end{figure}

The accretion rates as a function of white dwarf effective temperature are plotted in Fig.\,\ref{fig:Acc-rates}, which compares those white dwarfs with detectable circumstellar gas discs to the population of white dwarfs without circumstellar gas discs. The mass accretion rates differ by factors of 1.6--4.2 between the rates derived from the spectroscopic versus photometric white dwarf parameters; this demonstrates that accurate white dwarf parameters are important for determining accurate accretion rates. The white dwarfs with circumstellar gas have accretion rates that span the full range of accretion rates in the plot. A Kolmogorov-Smirnov (KS) test was used to test the null hypothesis that the two samples, the mass accretion rates of those white dwarfs with detectable gas discs versus those without, come from the same distribution. For both the accretion rates derived from the spectroscopic white dwarf parameters and the photometric parameters, the $p$-values are found to be large, and therefore, the null hypothesis cannot be rejected, and it remains plausible that the two samples came from the same distribution. Therefore, there is no evidence for enhanced accretion rates for white dwarfs with circumstellar gas discs compared to those without detectable circumstellar gas discs.

\begin{table*}
	\centering
	\caption{Total accretion rates based on Mg abundances for white dwarfs with an observable gaseous disc with emission features. This is calculated using $\dot{M} = (100/15.8) \times M_{\textrm{WD}} \times 10^q \times 10^{[\textrm{Mg}/\textrm{H(e)}]} \times A_{\textrm{Mg/H(e)}} / \tau_{\textrm{Mg}}$, where $q = \log _{10}(M_{\textrm{CVZ}}/ M_{\textrm{WD}})$, $A_{\textrm{Mg/H(e)}}$ is the atomic mass of Mg divided by the atomic mass of H or He, depending on the dominant atmospheric (atm) constituent, and $\tau_{\textrm{Mg}}$ is the sinking time of Mg. For the seven white dwarfs in this paper, the accretion rates are calculated for both the spectroscopic and photometric white dwarf parameters and abundances.}
	\label{tab:WDs-acc-rate}
	\begin{tabular}{lcccccccccc} 
		\hline
		WD Name & Atm & $T_\mathrm{eff}$ (K) & $\log(g)$ & $M_{\textrm{WD}}$ (M$_{\odot}$) & log(q) & log($\tau_{\textrm{Mg}}$) (yr) & log(Mg/H(e)) & $\dot{M}$ (g\,s$^{-1}$) & Reference \\
		\hline	
    Gaia\,J0006+2858 & H & 23920 & 8.04 & 0.66 & $-$15.5 & $-$1.43 & $-$4.95 & 5.86$\times 10^{8}$ & Spec, this work \\
    Gaia\,J0006+2858 & H & 22840 & 7.86 & 0.56 & $-$15.7 & $-$1.49 & $-$5.03 & 2.75$\times 10^{8}$ & Phot, this work \\
    Gaia\,J0347+1624 & H & 21820 & 8.10 & 0.69 & $-$16.2 & $-$1.93 & $-$5.78 & 6.04$\times 10^{7}$ & Spec, this work \\
    Gaia\,J0347+1624 & H & 18850 & 7.84 & 0.54 & $-$16.2 & $-$1.78 & $-$6.05 & 1.72$\times 10^{7}$ & Phot, this work \\
    Gaia\,J0510+2315 & H & 21700 & 8.22 & 0.76 & $-$16.6 & $-$2.26 & $-$5.23 & 2.12$\times 10^{8}$ & Spec, this work \\
    Gaia\,J0510+2315 & H & 20130 & 8.13 & 0.70 & $-$16.5 & $-$2.19 & $-$5.35 & 1.33$\times 10^{8}$ & Phot, this work \\
    Gaia\,J0611$-$6931 & H & 17750 & 8.14 & 0.70 & $-$16.7 & $-$2.32 & $-$4.61 & 6.57$\times 10^{8}$ & Spec, this work \\
    Gaia\,J0611$-$6931 & H & 16530 & 7.81 & 0.51 & $-$16.3 & $-$1.84 & $-$4.68 & 3.56$\times 10^{8}$ & Phot, this work \\
    Gaia\,J0644$-$0352 & He & 18350 & 8.18 & 0.70 & $-$6.4 & 5.29 & $-$5.73 & 5.89$\times 10^{9}$ & Spec, this work \\
    Gaia\,J0644$-$0352 & He & 17000 & 7.98 & 0.58 & $-$5.8 & 5.85 & $-$6.33 & 1.41$\times 10^{9}$ & Phot, this work \\
    WD\,1622+587 & He & 23430 & 7.80 & 0.50 & $-$7.3 & 4.92 & $-$4.91 & 8.42$\times 10^{9}$ & Spec, this work \\
    WD\,1622+587 & He & 21530 & 7.98 & 0.59 & $-$6.9 & 5.08 & $-$5.49 & 4.61$\times 10^{9}$ & Phot, this work \\
    Gaia\,J2100+2122 & H & 25570 & 8.10 & 0.69 & $-$15.5 & $-$1.47 & $-$5.23 & 3.58$\times 10^{8}$ & Spec, this work \\
    Gaia\,J2100+2122 & H & 22000 & 7.92 & 0.59 & $-$15.8 & $-$1.59 & $-$5.35 & 1.38$\times 10^{8}$ & Phot, this work \\

        \hline 
        WD\,0145+234 & H & 12720 & 8.10 & 0.67 & $-$15.6 & $-$1.55 & $-$5.90 & 6.35$\times 10^{7}$ & \citet{melis2020serendipitous} \\
        SDSS\,J0738+1835 & He & 13950 & 8.40 & 0.84 & $-$6.0 & 5.45 & $-$4.68 & 1.41$\times 10^{11}$ &  \citet{dufour2012detailed} \\
        WD\,0842+572 & H & 16225 & 8.00 & 0.62 & $-$16.6 & $-$2.16 & $-$3.90 & 2.78$\times 10^{9}$ & \citet{melis2020serendipitous} \\
		SDSS\,J0845+2257 & He & 19780 & 8.18 & 0.71 & $-$6.7 & 5.08 & $-$4.70 & 5.13$\times 10^{10}$ & \citet{wilson2015composition} \\
		SDSS\,J0959$-$0200 & H & 13280 & 8.06 & 0.64 & $-$15.8 & $-$1.59 & $-$5.20 & 2.66$\times 10^{8}$ & \citet{farihi2012trio} \\
		SDSS\,J1043+0855 & H & 17880 & 8.12 & 0.69 & $-$16.7 & $-$2.29 & $-$5.15 & 1.86$\times 10^{8}$ & \citet{manser2016another} \\
		SDSS\,J1228+1040 & H & 20900 & 8.15 & 0.71 & $-$16.5 & $-$2.15 & $-$5.10 & 2.62$\times 10^{8}$ & \citet{gansicke2012chemical} \\
		HE\,1349$-$2305 & He & 18170 & 8.13 & 0.67 & $-$6.3 & 5.40 & $-$6.50 & 9.62$\times 10^{8}$ & \citet{melis2012gaseous} \\
        SDSS\,J1617+1620 & H & 13520 & 8.11 & 0.68 & $-$15.9 & $-$1.70 & $-$5.02$^{*}$ & 4.16$\times 10^{8}$ & \citet{wilson2014variable} \\

        \hline
	\end{tabular}
 \flushleft$^{*}$Using the Convective at $\tau _{R} = 3.2$ case.
\end{table*}

\begin{table*}
	\centering
	\caption{Total accretion rates based on Mg abundances for white dwarfs without an observable gaseous disc with emission features, with effective temperature greater than 12,720\,K.}
	\label{tab:WDs-acc-rate-no-gas}
	\begin{tabular}{lcccccccccc} 
		\hline
		WD Name & Atm & $T_\mathrm{eff}$ (K) & $\log(g)$ & $M_{\textrm{WD}}$ (M$_{\odot}$) & log(q) & log($\tau_{\textrm{Mg}}$) (yr) & log(Mg/H(e)) & $\dot{M}$ (g\,s$^{-1}$) & Reference \\
		\hline	
        WD\,0002+729 & He & 13750 & 8.00 & 0.59 & $-$5.10 & 6.35 & $-$8.50 & 1.56$\times 10^{7}$ & \citet{Wolff2002element} \\
        WD\,0030+1526	&	He	&	15285	&	8.07	&	0.632	&	-5.57	&	5.96	&	-6.99	&	4.51$\times 10^{8}$ & \citet{izquierdo2023systematic}	\\
        WD\,0106$-$3253 & H & 17350 & 8.12 & 0.69 & $-$16.71 & $-$2.30 & $-$5.57 & 6.97$\times 10^{7}$ & \citet{xu2019compositions} \\
        Gaia\,J0218+3625 & He & 14700 & 7.86 & 0.51 &  $-$5.01 &  6.49 &$-$6.64 & 8.81$\times 10^{8}$ & \citet{doyle2023new} \\
        SDSS\,J0224+7503 & He & 16560 & 8.25 & 0.75 & $-$6.18 & 5.42 & $-$5.15 & 3.12$\times 10^{10}$ & \citet{izquierdo2021gd} \\
        WD\,0259$-$0721	&	He	&	14128	&	8.01	&	0.594	&	-5.19	&	6.28	&	-5.61	&	1.17$\times 10^{10} $ & \citet{izquierdo2023systematic}	\\
        WD\,0300$-$013 & He & 15300 & 8.00 & 0.59 & $-$5.44 & 6.10 & $-$6.20 & 2.58$\times 10^{9}$ & \citet{jura2012two} \\
        WD\,0408$-$041 & H & 15270 & 8.09 & 0.67 & $-$16.18 & $-$1.90 & $-5.55$ & 9.43$\times 10^{7}$ & \citet{xu2019compositions} \\
        WD\,0435+410 & He & 17280 & 8.20 & 0.72 & $-$6.24 & 5.41 & $-$6.69 & 7.70$\times 10^{8}$ & \citet{farihi2013evidence} \\
        PG\,0843+517 & H & 24670 & 7.93 & 0.60 & $-$15.30 & $-1.22$ & $-$4.82 & 7.29$\times 10^{8}$ & \citet{xu2019compositions} \\
        WD\,0859+1123	&	He	&	15253	&	8.09	&	0.644	&	-5.60	&	5.93	&	-5.92	&	5.43$\times 10^{9}$ & \citet{izquierdo2023systematic}	\\
        WD\,0930+0618	&	He	&	15560	&	8.01	&	0.597	&	-5.52	&	6.04	&	-5.9	&	4.99$\times 10^{9}$ & \citet{izquierdo2023systematic}	\\
        WD\,0944$-$0039	&	He	&	13113	&	8.15	&	0.678	&	-5.34	&	6.10	&	-6.96	&	6.42$\times 10^{8}$ & \citet{izquierdo2023systematic}	\\        
        PG\,1015+161 & H & 19200 & 8.22 & 0.75 & $-$16.75 & $-2.38$ & $-5.30$ & 1.55$\times 10^{8}$ & \citet{gansicke2012chemical} \\
        PG\,1018+411 & H & 24440 & 8.11 & 0.70 & $-$15.77 & $-$1.65 & $-$4.86 & 7.08$\times 10^{8}$ & \citet{xu2019compositions} \\
        WD\,1109+1318	&	He	&	15623	&	8.12	&	0.663	&	-5.75	&	5.81	&	-6.73	&	8.20$\times 10^{8}$ & \citet{izquierdo2023systematic}	\\
        SBSS\,1240+527 & He & 13000 & 8.00 & 0.59 & $-$4.98 & 6.45 & $-$5.26 & 2.86$\times 10^{10}$ & \citet{raddi2015likely} \\
        WD\,1244+498 & He & 15150 & 7.97 & 0.57 &  $-$5.34 & 6.19 & $-$6.79 & 6.57$\times 10^{8}$ & \citet{doyle2023new}\\        
        SDSS\,J1248+1005 & He & 15180 & 8.11 & 0.66 &  $-$5.63 &  5.90 &  $-$6.40 & 1.85$\times 10^{9}$ & \citet{doyle2023new}\\        
        WD\,1337+701 & H & 20546 & 7.95 & 0.60 & $-$16.16 & $-$1.82 & $-$5.66 & 5.70$\times 10^{7}$ & \citet{Johnson2022unusual} \\
        WD\,1359$-$0217	&	He	&	13995	&	7.78	&	0.47	&	-4.69	&	6.77	&	-6.32	&	1.86$\times 10^{9}$ & \citet{izquierdo2023systematic}	\\
        WD\,1415+234 & He & 17300 & 8.17 & 0.70 &  $-$6.20 & 5.46 & $-$5.82 & 5.49$\times 10^{9}$ & \citet{doyle2023new}\\
        WD\,1425+540 & He & 14490 & 7.95 & 0.56 & $-$5.15 & 6.35 & $-$8.16 & 2.96$\times 10^{7}$ & \citet{xu2017chemical} \\
        PG\,1457$-$086 & H & 22240 & 7.99 & 0.62 & $-$15.77 & $-$1.57 & $-$5.47 & 1.30$\times 10^{8}$ & \citet{xu2019compositions} \\
        WD\,1516$-$0040	&	He	&	13193	&	7.94	&	0.552	&	-4.88	&	6.55	&	-6.82	&	7.33$\times 10^{8}$ & \citet{izquierdo2023systematic}	\\
        WD\,1536+520 & He & 20800 & 7.96 & 0.58 & $-$6.72 & 5.23 & $-$4.06 & 1.35$\times 10^{11}$ & \citet{farihi2016solar} \\
        WD\,1551+175 & He & 14756 & 8.02 & 0.60 & $-$5.35 & 6.15 & $-$6.29 & 2.34$\times 10^{9}$ & \citet{xu2019compositions} \\
        WD\,1627+1723	&	He	&	15903	&	8.11	&	0.657	&	-5.79	&	5.79	&	-6.85	&	5.90$\times 10^{8}$ & \citet{izquierdo2023systematic}	\\
        SDSS\,J1734+6052 & He & 16340 & 8.04 & 0.62 &  $-$5.76 &  5.85 &  $-$6.62 & 8.67$\times 10^{8}$ & \citet{doyle2023new} \\        
        WD\,1822+410 & He & 15620 & 7.93 & 0.55 & $-$5.39 & 6.18 & $-$7.44 & 1.29$\times 10^{8}$ & \citet{klein2021discovery} \\
        Gaia\,J1922+4709 & He & 15500 & 7.95 & 0.56 &  $-$5.39 &  6.17 &  $-$6.14 & 2.70$\times 10^{9}$ & \citet{doyle2023new}\\        
        WD\,1929+012 & H & 23470 & 7.99 & 0.63 & $-$16.75 & $-1.36$ & $-4.10$ & 3.85$\times 10^{9}$ & \citet{melis2011accretion} \\
        WD\,J2047$-$1259 & He & 17970 & 8.04 & 0.62 & $-$6.14 & 5.58 & $-$5.60 & 7.00$\times 10^{9}$ & \citet{hoskin2020white} \\ 
        WD\,2207+121 & He & 14752 & 7.97 & 0.57 & $-$5.25 & 6.25 & $-$6.15 & 3.06$\times 10^{9}$ & \citet{xu2019compositions} \\
        EC\,22211$-$2525 & He & 14740 & 7.89 & 0.53 & $-$5.08 &  6.42 &  $-$6.52 & 1.20$\times 10^{9}$ & \citet{doyle2023new}\\
        WD\,2222+683 & He & 15300 & 8.00 & 0.59 & $-$5.44 & 6.10 & $-$6.26 & 2.25$\times 10^{9}$ & \citet{jura2012two} \\
        SDSS\,J2248+2632 & He & 17370 & 8.02 & 0.61 &  $-$5.97 &  5.72 &  $-$6.52 & 9.04$\times 10^{8}$ & \citet{doyle2023new}\\        
        WD\,2324$-$0018	&	He	&	12823	&	7.66	&	0.411	&	-4.21	&	7.20	&	-8.09	&	3.10$\times 10^{7}$ & \citet{izquierdo2023systematic}	\\
        Gaia\,J2339$-$0424 & He & 13735 & 7.93 & 0.55 & $-$4.95 & $-$6.50 & $-$6.58 & 1.21$\times 10^{9}$ & \citet{klein2021discovery} \\
        \hline
	\end{tabular}
\end{table*}

\section{Discussion} \label{disc}

This paper presents the abundances of the planetary material accreted by seven white dwarfs with circumstellar gas and dust. They represent seven of just 21 known polluted white dwarfs with circumstellar gas emission discs; it is crucial to understand the interplay between accretion, observed gas and atmospheric pollution. The abundances are derived from optical and ultraviolet spectra for two different sets of white dwarf parameters. The absolute abundances in the white dwarf photosphere are most affected by uncertainties in the derived stellar parameters, as well as the quality of the spectroscopic data and the white dwarf models used to analyse and obtain the abundances. Crucially, however, interpretation of the observed compositions, which is based on elemental ratios, are less affected by uncertainties in the stellar parameters.

In order to determine elemental abundances from the data, a number of systematics must be considered. Sometimes only one (or few) absorption line(s) of a particular element are present in the spectrum (see Supplementary Tables B1--B13 for details). This is especially important for the five DAZ white dwarfs where there are few absorption lines in the optical and they can be weak. Uncertain atomic data can also cause additional uncertainties to arise \citep{vennes2011pressure}. The abundances are also limited by the white dwarf models, for example, 3D effects such as convection may affect the derived abundances \citep{cunningham2019convective}. 

This work corroborates previous studies which identified the differences in deriving white dwarf parameters from spectral analysis compared to those derived from broad-band photometry. Above 14,000\,K spectroscopic $T_\mathrm{eff}$, and therefore also $\log(g)$, can exceed photometric $T_\mathrm{eff}$ by 5--10 per cent \citep{genest2019comprehensive}. All white dwarfs in this study fall into this range, and indeed this is reflected in the derived stellar parameters. The most accurate white dwarf parameters derived from broad-band photometry are those that include the SDSS $u$-band with additional optical photometry, e.g. Pan-STARRS  \citep{bergeron2019measurement}. For those white dwarfs with \textit{GALEX} photometry, the \textit{GALEX} $FUV$ and/or $NUV$ band magnitudes help to constrain the white dwarf parameters in a similar way to the $u$ band providing increased accuracy of the parameters. There are significant systematic uncertainties associated with the stellar parameters and a  hybrid approach similar to \citet{izquierdo2021gd} may result in improved parameters. 

The heavy element abundance uncertainties quoted in Table \ref{tab:WDs-ab} are measurement errors and do not include systematic errors arising from the different methods of obtaining white dwarf parameters. Instead the abundances based on the two sets of stellar parameters are considered. The difference in $T_\mathrm{eff}$ and $\log(g)$ between the two sets of stellar parameters are between 1000--3500\,K and 0.1--0.25 dex, respectively. In order to obtain reliable and accurate \textit{absolute} abundances of the accreted planetary material it is crucial to obtain accurate stellar parameters. However, as shown in Fig.\,\ref{fig:Ab-comp-ratio} the relative elemental ratios, which are used to interpret the observed compositions, when comparing abundances derived from the spectroscopic stellar parameters compared with the photometric stellar parameters. The ratios are less sensitive to the derived stellar parameters. Therefore, until the uncertainties on the absolute abundances can be reduced, the differences between the photometric and spectroscopic $T_\mathrm{eff}$ and $\log(g)$ will have little affect on the abundance analysis based on the relative ratios of elements. If absolute abundances are important, as when deriving accretion rates, the variations induced by differing white dwarf parameters must be considered. Table \ref{tab:WDs-acc-rate} and Fig.\,\ref{fig:Acc-rates} show the difference in derived accretion rates for the seven white dwarfs when using the spectroscopic parameters versus the photometric parameters. The accretion rates can change by up to a factor of 4.2 which would consequently affect inferences about the accretion and mass of the planetesimal currently present in the white dwarf photosphere. Therefore, when considering accretion rates, careful error propagation considering these systematic errors are crucial.

\begin{figure*}
\centering
\subfloat[]{
  \includegraphics[width=0.5\textwidth]{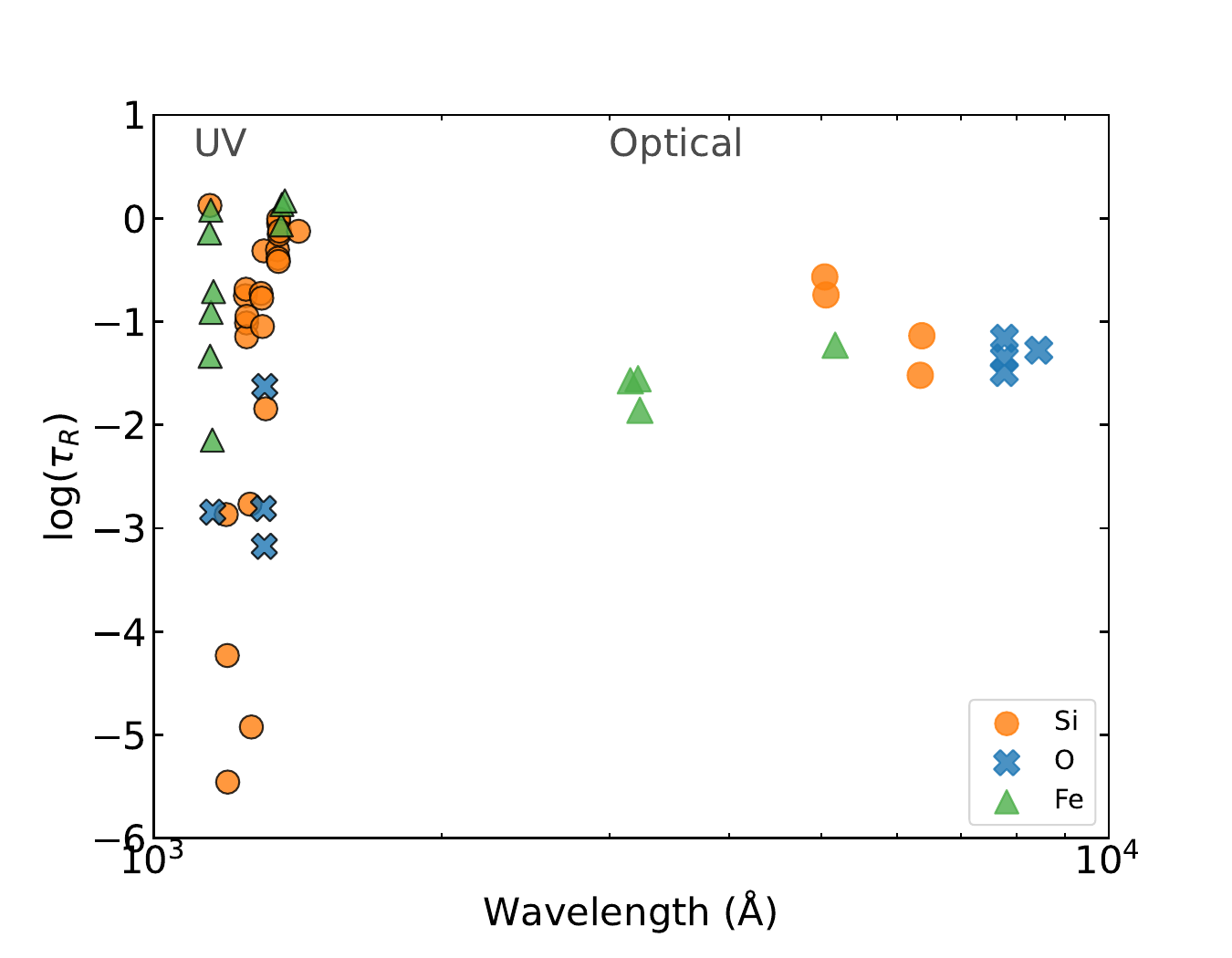}
  \label{fig:lines-uv-op}
}
\subfloat[]{
  \includegraphics[width=0.5\textwidth]{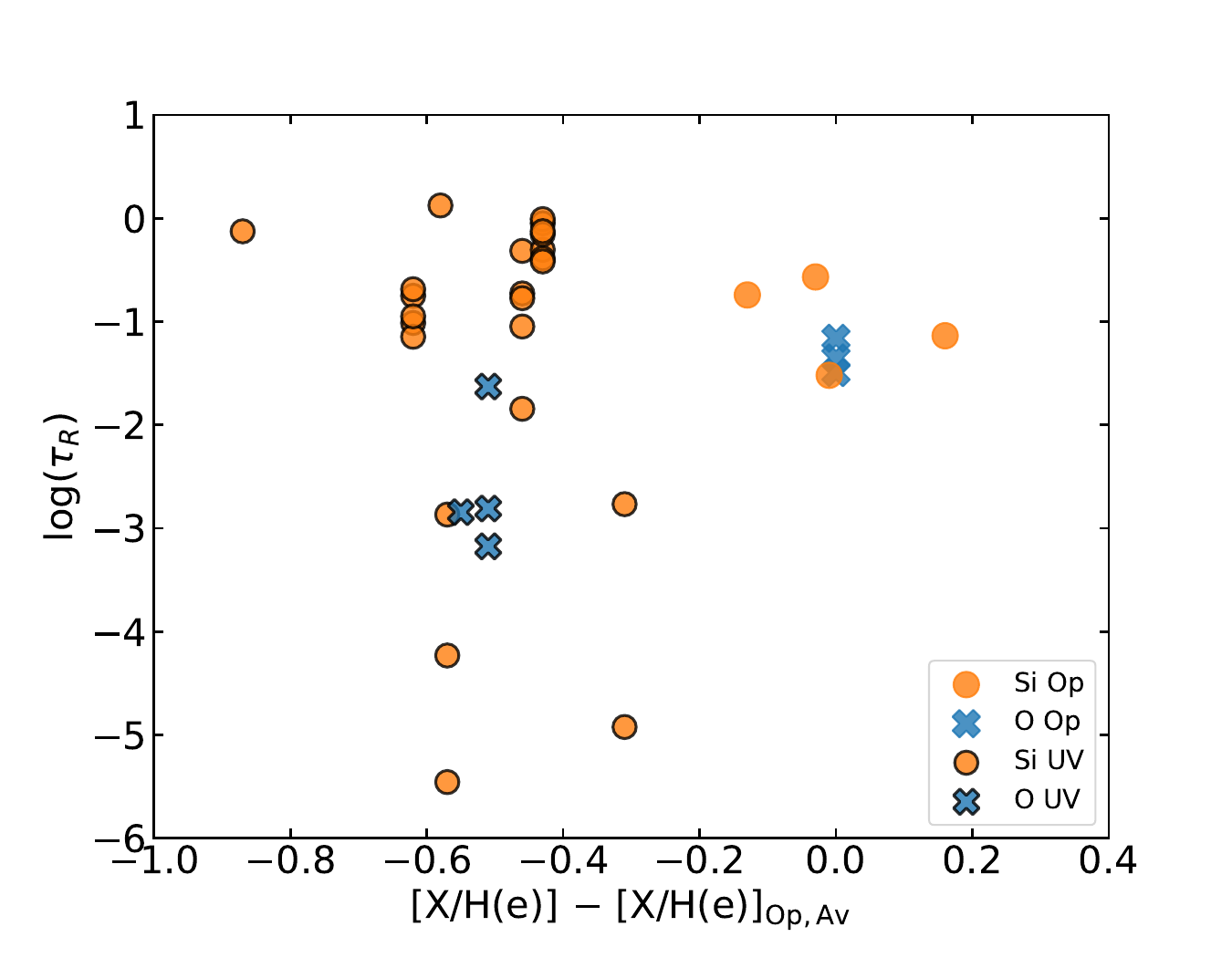}
  \label{fig:Si-uv-op}
}

\caption{(a) The depth of formation of spectral lines versus the wavelength of the lines. The UV lines have a black outline to the markers. The abundances derived using the optical data are all formed at a similar depth, compared to the ultraviolet which probes a variety of layers. (b) The $\tau _{R}$ as a function of abundance of silicon and oxygen plotted relative to the average optical abundances, [X/H(e)]$_\mathrm{Op,Av}$ from Table \ref{tab:WDs-ab}. The data points from the optical and UV are marked separately.}
\end{figure*}

It has been addressed in the literature that there exists an optical and ultraviolet discrepancy, where the abundances determined from optical data are offset from those derived from the ultraviolet data \citep{jura2012two,gansicke2012chemical,xu2019compositions}. This work uses a systematic approach, ensuring that the same white dwarf parameters are used when deriving abundances from the optical and ultraviolet. For the three DAZ white dwarfs in the sample, this discrepancy is observed, with an approximately constant offset of 0.62 observed between O and Si derived from the optical and ultraviolet. As is highlighted in previous works, atomic data uncertainties, accretion rate variation, or imperfect white dwarf atmosphere calculations may contribute to this effect.  Figure \ref{fig:lines-uv-op} shows the depth of formation of the lines in the ultraviolet and the optical. Depth of formation is the Rosseland optical depth, $\tau _{r}$, at which $\tau _{\nu} = 2/3$ for each line wavelength. The lines that form in the optical come from the same depth, whereas those from the ultraviolet form from a range of depths. Figure \ref{fig:Si-uv-op} shows that there is not an abundance dependence on the depth of formation, and lines that form at the same depth have different derived abundances. There may be issues with the white dwarf structure calculations, however, further investigations are outside the scope of this paper. Therefore, until the origin of this discrepancy is discovered, when analysing polluted white dwarf abundances, care must be given when combining results from the optical and UV. 

Previous work has found that detectable circumstellar gas is a rare phenomena in polluted white dwarfs \citep{manser2020frequency}. \citet{xu2019compositions} compared pollution levels for those white dwarfs with and without a detectable circumstellar \textit{dust} disc, compiling their work with data from the literature \citep{koester2011cool, koester2014frequency,koester2015db,hollands2018cool}. No strong difference in total mass accretion rate was found between those white dwarfs with detectable circumstellar dust versus those without. Here the work of \citet{xu2019compositions} is expanded to compare the pollution rate between white dwarfs with and without detectable circumstellar \textit{gaseous} discs, where without refers to all polluted white dwarfs without any evidence of circumstellar gas in emission. Previous studies investigating the link between the presence of a detectable gaseous disc and accretion rate based on a handful of systems found no correlation between accretion rate and the presence of circumstellar gas \citep{manser2016another,manser2020frequency}. However, gas drag may cause enhanced accretion rates above what may be expected from Poynting-Robertson drag alone. Combining white dwarfs with detectable circumstellar gas discs from the literature (which have reported abundances of at least Mg in their photosphere) with the seven in this work, the sample size is almost doubled. The white dwarfs that have circumstellar gaseous discs appear to trace the mass accretion rates of polluted white dwarfs without detectable gaseous discs, therefore, there is no evidence that accretion of material onto the white dwarfs is enhanced by gas drag. If instead it was found that gas discs systems accrete at higher rates, it would be difficult to confirm whether the presence of gas causes enhanced accretion, or vice-versa, whether the enhanced accretion results in gas. These conclusions are limited by both small number statistics and that the white dwarfs not studied in this paper have accretion rates derived from different methods of obtaining white dwarf parameters, therefore, the accretion rates for these could be inaccurate. 

Figure \ref{fig:Acc-rates} highlights how the accretion rates of He dominated white dwarfs are higher than those in H dominated atmospheres \citep{farihi2012scars, xu2019compositions}. This may be due to the orders of magnitude longer settling time-scales for He dominated white dwarfs and therefore the rates represent an average historical accretion rate. Direct impacts on to the white dwarf \citep{brown2017deposition, mcdonald2021white}, or the disruption of massive asteroids ($>$\,500\,km) which can collisionally evolve to produce enhanced accretion on short time-scales \citep{wyatt2014stochastic,brouwers2022road} can contribute to larger measured average accretion rates for DBZ white dwarfs.  However, for the sample of objects with dust and gas discs, direct impacts are unlikely to explain the enhanced accretion rates as direct impacts do not result in circumstellar discs. Thermohaline mixing may account for some differences between DA and DB accretion rates, but it should be stated the effect is debated \citep{koester2014thermohaline}. DBZ white dwarfs would experience less thermohaline mixing than DAZ white dwarfs \citep{bauer2019polluted}. When including thermohaline mixing into the white dwarf models, larger accretion rates are required in order to account for this instability \citep{bauer2018increases}. As it disproportionally affects DA white dwarfs, the accretion rates would need to be orders of magnitude larger for the DB white dwarfs to make DA and DB consistent using thermohaline mixing models.

\section{Conclusions} \label{Conclusions}

This paper presents VLT X-shooter, Keck HIRES, and Magellan MIKE optical spectroscopy and \textit{HST} COS ultraviolet spectroscopy of seven white dwarfs that host both detectable circumstellar gas and dust discs. All seven have accreted heavy elements; between three and 10 of these elements, C, O, S, P, Mg, Al, Si, Ca, Ti, Cr, Fe, and Ni, are detected in the atmosphere of each of the white dwarfs. White dwarfs with circumstellar gaseous discs are good targets for studying photospheric abundances as they reveal numerous elements allowing in depth compositional analysis of the polluting planetary material. 

All the white dwarfs show non-photospheric lines in their spectra. Most notably, Gaia\,J0006+2858 has a non-photospheric component for two Si\,\textsc{iv} photospheric lines in the FUV data blueshifted with a velocity of $-$196\,km\,s$^{-1}$. This may circumstellar gas absorption, however, the origin remains unsolved. 

Abundances of planetary material in the atmospheres of white dwarfs provide crucial constraints on the composition of exoplanetary material. This work shows that the ratio of abundances within the white dwarf atmosphere, for example, Fe/Mg, are less effected by uncertainties in the white dwarf parameters for the effective temperature range considered (16,000--25,500\,K) than absolute abundances (e.g. [Mg/H], [Fe/H]). Thus, it is preferable to use abundance ratios when interpreting planetary composition. This highlights the importance of considering the discrepancy between white dwarf parameters derived by the spectroscopic and photometric method when considering the total accretion onto white dwarfs, in this work the accretion rates differed by factors of 1.6--4.2.

A poorly understood discrepancy between abundances derived from optical or ultraviolet spectroscopy has previously been reported in the literature. This work derives the abundances in the optical and ultraviolet using a consistent approach and finds that there is an approximately constant offset between the optical and ultraviolet abundances of silicon and oxygen in three DAZ white dwarfs of 0.62 dex, and no offset is found for the one DBZ white dwarf analysed in this sample. This work speculates as to whether this discrepancy could be explained by vertical gradients in composition in the white dwarf atmosphere. The optical lines form at approximately the same depth, whereas, the ultraviolet lines form over a range of depths. Further work is needed to understand the origin of this discrepancy.

Combining the seven objects from this paper with nine white dwarfs from the literature with circumstellar gaseous discs and Mg abundance measurements of the polluting material, the mass accretion rates of systems with detectable circumstellar gaseous discs were compared to those without. This supports that polluted white dwarfs with circumstellar gaseous discs do not show enhanced accretion when compared to the population of polluted white dwarfs without gaseous discs, and so there is no evidence for enhanced accretion rates from gas drag.

The analysis of the abundances of the material that has accreted onto these seven white dwarfs is presented in the subsequent paper, Paper II, including in depth discussions on the composition and geological history of the planetesimals. 

\section*{Acknowledgements}

LKR is grateful for PhD funding from STFC and the Institute of Astronomy, University of Cambridge. AB and LKR acknowledge support of a Royal Society University Research Fellowship, URF\textbackslash R1\textbackslash 211421. LKR acknowledges support of an ESA Co-Sponsored Research Agreement No. 4000138341/22/NL/GLC/my = Tracing the Geology of Exoplanets. SX is supported by the international Gemini Observatory, a programme of NSF's NOIRLab, which is managed by the Association of Universities for Research in Astronomy (AURA) under a cooperative agreement with the National Science Foundation, on behalf of the Gemini partnership of Argentina, Brazil, Canada, Chile, the Republic of Korea, and the United States of America. BK acknowledges support from NASA/Keck research contracts 1654589, 1659075, and 1665572. AMB is grateful for the support of a PhD studentship funded by a Royal Society Enhancement Award, RGF\textbackslash EA\textbackslash 180174. STH is funded by the Science and Technology Facilities Council grant ST/S000623/1. C.M.\ and B.Z.\ acknowledge support from US National Science Foundation 
grants SPG-1826583 and SPG-1826550. The authors wish to acknowledge Marc Brouwers for useful discussions which helped shape the paper. Based on observations collected at the European Southern Observatory, Chile under ESO programmes: 0103.C-0431(B) and 0104.C-0107(A). This work was supported by a NASA Keck PI Data Award (\#1654589, \#1659075, and \#1665572), administered by the NASA Exoplanet Science Institute. Data presented herein were obtained at the W. M. Keck Observatory from telescope time allocated to the National Aeronautics and Space Administration through the agency's scientific partnership with the California Institute of Technology and the University of California. The Observatory was made possible by the generous financial support of the W. M. Keck Foundation. The authors wish to recognize and acknowledge the very significant cultural role and reverence that the summit of Maunakea has always had within the indigenous Hawaiian community. We are most fortunate to have the opportunity to conduct observations from this mountain. This paper includes data gathered with the 6.5 meter Magellan Telescopes located at Las Campanas Observatory, Chile. This research is based on observations made with the NASA/ESA Hubble Space Telescope obtained from the Space Telescope Science Institute, which is operated by the Association of Universities for Research in Astronomy, Inc., under NASA contract NAS 5-26555. These observations are associated with programmes 16204 and 16752.

\section*{Data Availability}

VLT X-shooter data available from the ESO archive (\href{http://archive.eso.org/eso/eso\_archive\_main.html}{http://archive.eso.org/eso/eso\_archive\_main.html}). Keck/HIRES data available from the Keck archive (\href{https://koa.ipac.caltech.edu/cgi-bin/KOA/nph-KOAlogin}{https://koa.ipac.caltech.edu}). Magellan MIKE data for Gaia\,J0611$-$6931 available on request. COS FUV (programme ID: 16752) and COS NUV (programme ID: 16204) data available from MAST (\href{https://mast.stsci.edu/search/ui/#/hst}{https://mast.stsci.edu/search/ui/\#/hst}).




\bibliographystyle{mnras}
\bibliography{Refs} 




\appendix

\section{Spectral Energy Distributions}

\begin{figure*}
\centering
\subfloat{
  \includegraphics[width=0.45\textwidth]{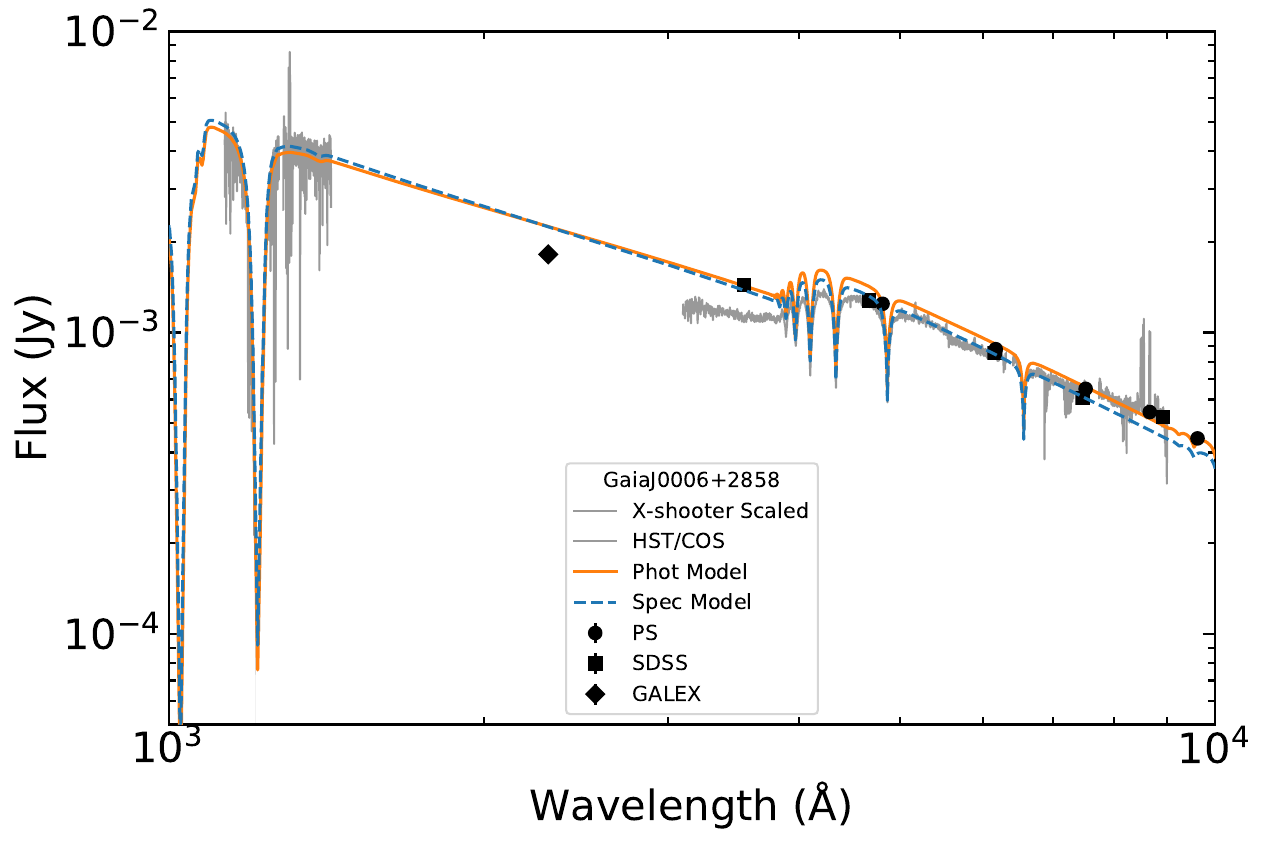}
}
\hspace{0.1cm}
\subfloat{
  \includegraphics[width=0.45\textwidth]{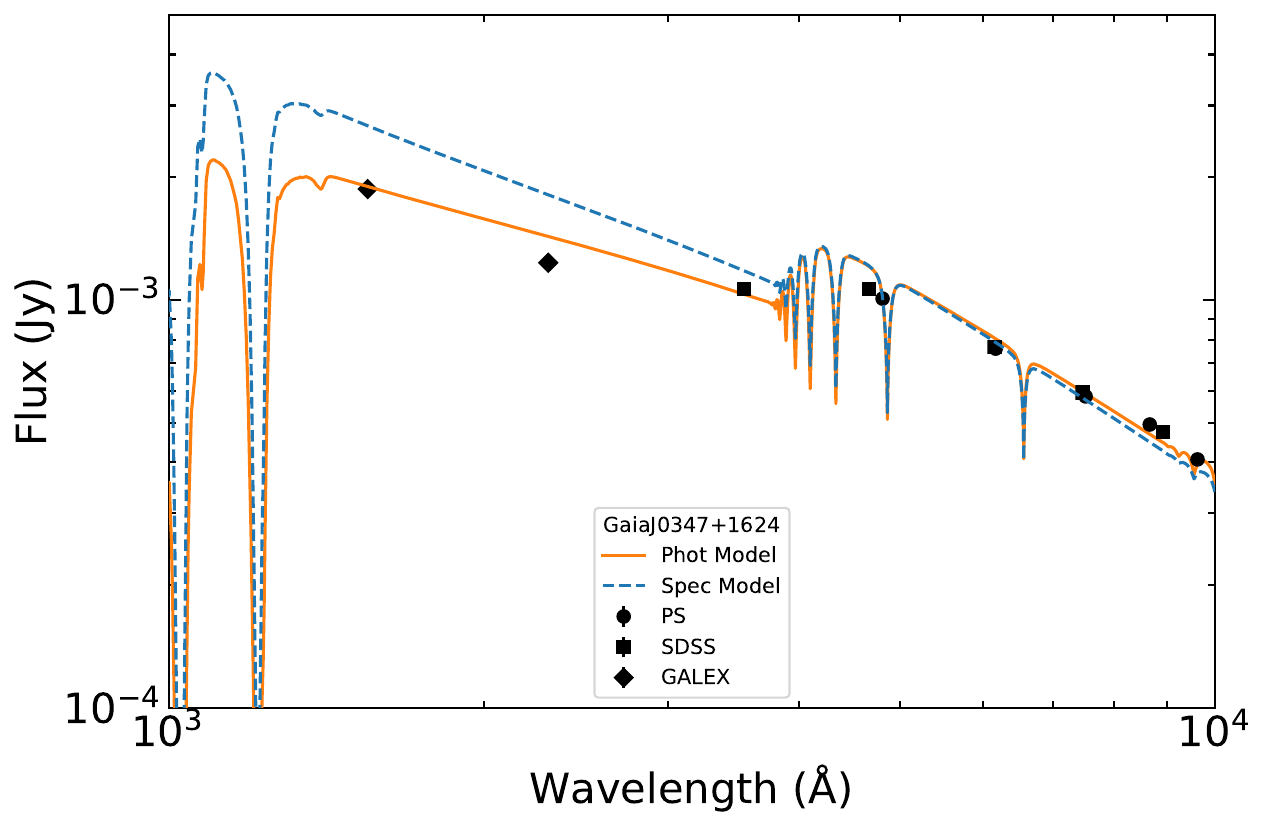}
}
\vspace{-0.5cm}
\subfloat{
  \includegraphics[width=0.45\textwidth]{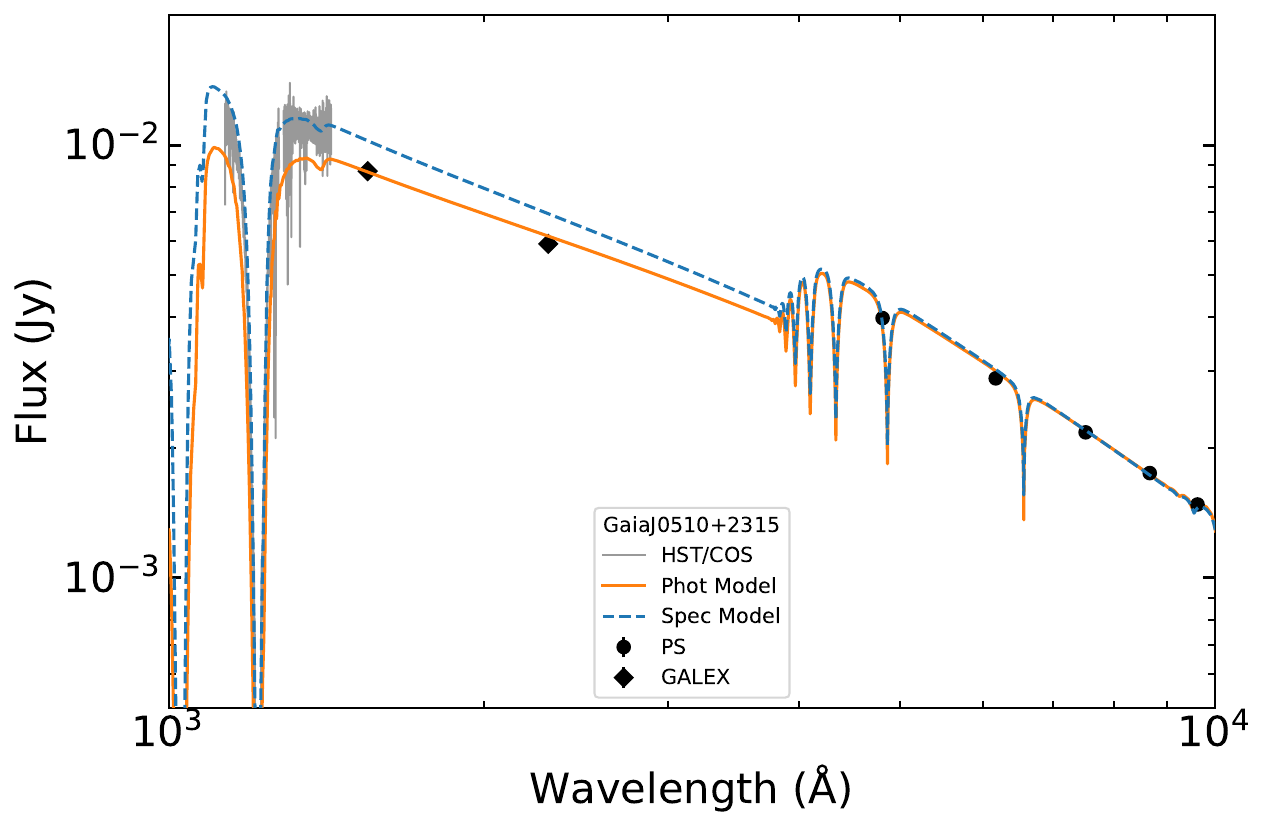}
}
\hspace{0.1cm}
\subfloat{
  \includegraphics[width=0.45\textwidth]{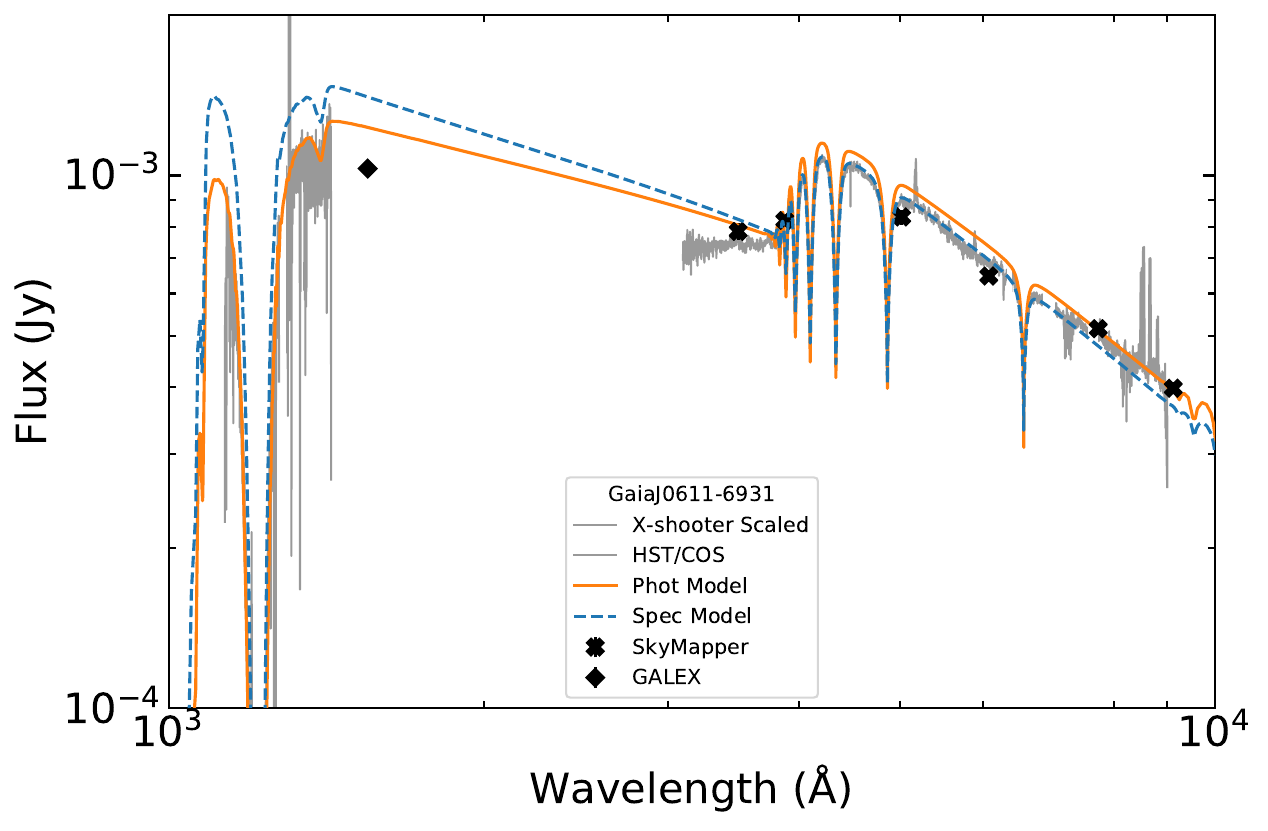}
}
\vspace{-0.5cm}
\subfloat{
  \includegraphics[width=0.45\textwidth]{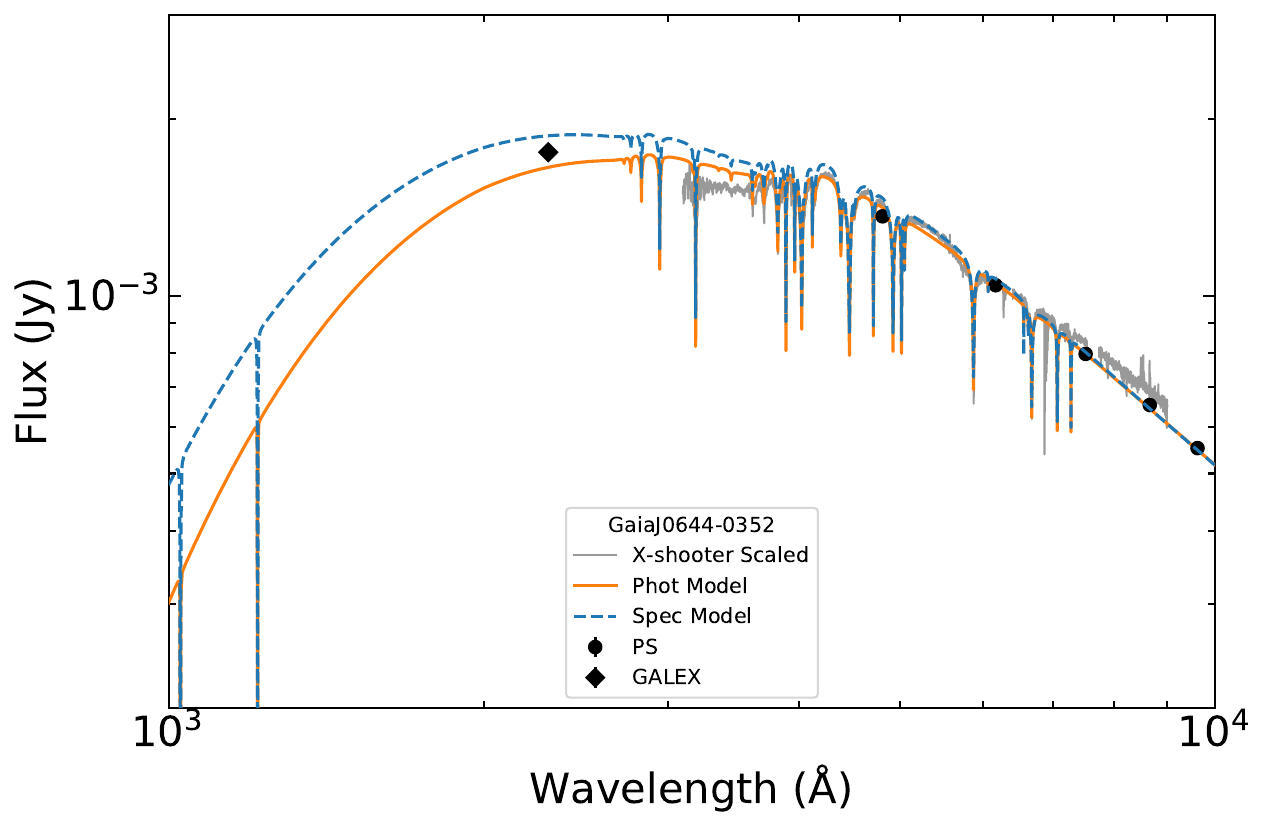}
}
\hspace{0.1cm}
\subfloat{
  \includegraphics[width=0.45\textwidth]{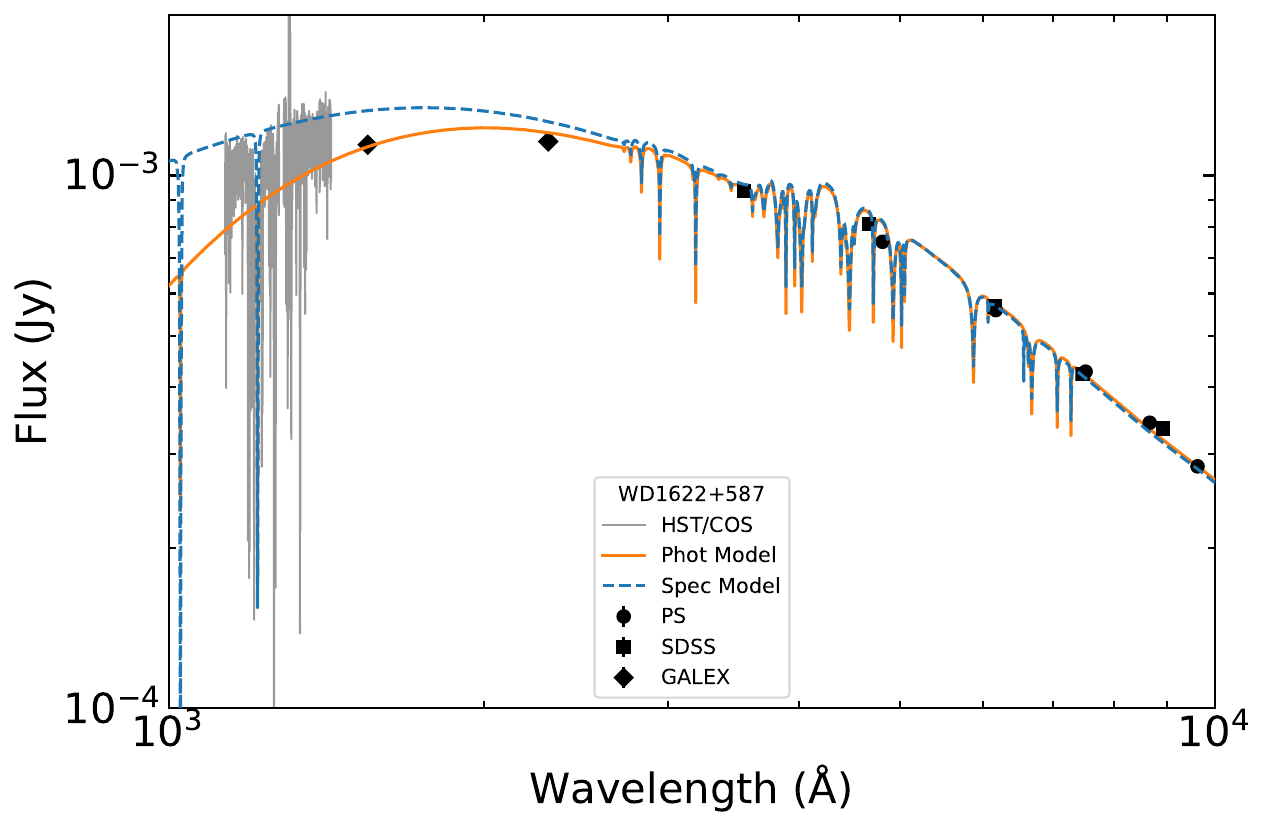}
}
\vspace{-0.5cm}
\subfloat{
  \includegraphics[width=0.45\textwidth]{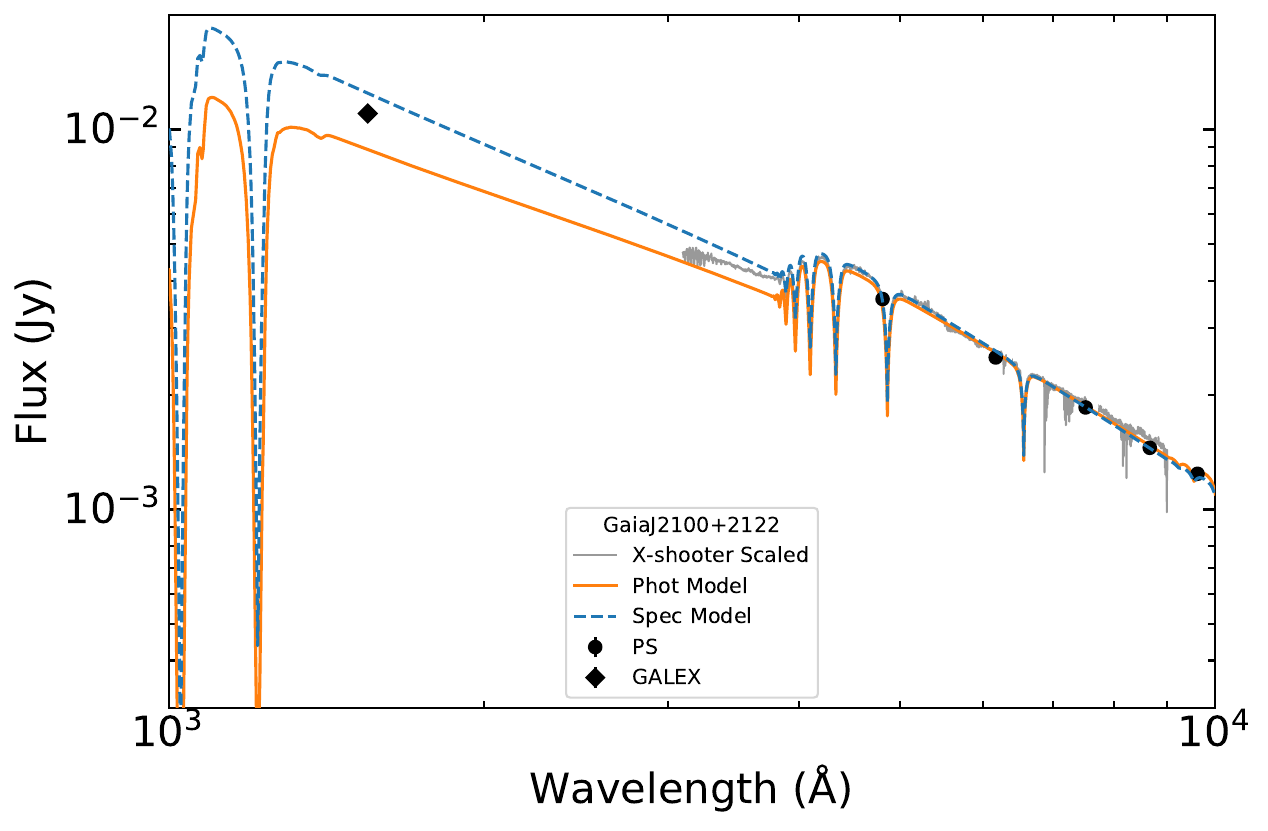}
  
}
\hspace{0.1cm}
\caption{Spectra (where available) and photometry for the seven white dwarfs. The X-shooter spectra for the four white dwarfs observed are shown in grey and scaled as the spectra suffer from flux loss due to non-ideal weather conditions and slit losses. There is a gap in the data between 6360\,\AA -- 6375\,\AA, and telluric absorption features are present in the reddest parts of the spectra. The \textit{Hubble} FUV data is also shown in grey for the four white dwarfs with FUV data. The black photometric data points are over-plotted with SDSS as square data points, Pan-STARRS (PS) as circular data points, and GALEX as diamonds; errors are plotted but are often smaller than the data points. The photometric and spectroscopic model fits are shown in orange (solid line) and blue (dashed line). Missing \textit{GALEX} FUV or NUV fluxes implies either a non-detection in that band, or the flux was flagged as it contained an artefact.
\label{fig:SEDs} }
\end{figure*}

Spectral energy distributions for the seven white dwarfs are shown in Fig.\,\ref{fig:SEDs} showing the best-fitting white dwarf models derived both spectroscopically and photometrically.

\section{Spectral Lines and abundances}

For each of the white dwarfs, the stellar parameters were determined spectroscopically and photometrically. For each set of parameters the abundances were obtained. Supplementary Tables B1 -- B13 show the spectral lines that were identified in the X-shooter, HIRES, MIKE and COS spectra for the seven white dwarfs, their associated equivalent width, radial velocity, and the abundance derived.

\section{Equivalent Width Upper Limits}

Using the method discussed in Section \ref{lim} the equivalent width upper limits were derived, as listed in Table \ref{tab:WDs-EW-UL} for the optical and Table \ref{tab:WDs-EW-UL-UV} for the ultraviolet. The strongest line in the wavelength range was used to derive the equivalent width upper limit. 

For the optical data, the HIRES data were used to obtain the equivalent width upper limits as the higher resolution allowed more stringent constraints to be used.

\begin{table*}
	\centering
	\footnotesize
	\caption{The upper limit equivalent widths in m\AA \, of the material polluting the seven white dwarfs in this study. $^*$ denotes when gaseous emission is present at this wavelength. When calculating abundance upper limit, add 12 per cent on to the EW quoted to be conservative.  
	}
	\label{tab:WDs-EW-UL}
	\begin{tabular}{ccccccccc} 
		\hline
		Element & Line (\AA) & {Gaia\,J0006} & {Gaia\,J0347} & {Gaia\,J0510} & {Gaia\,J0611} & {Gaia\,J0644} & {WD\,1622} & {Gaia\,J2100}  \\
		\hline	

		O\,\textsc{i} & 7771.9377 & 20.78$^*$ & - & - & - & - & 31.4$^*$ & 11.42$^*$ \\
		
		Na\,\textsc{ii} & 5889.9483 & 12.6 & - & 8.5 & n/a$^*$ & 25.1 & 16.5 & 8.7 \\
		
		Al\,\textsc{ii} & 3586.5564 & 7.0 & 16.6 & 14.3 & 92.1 & - & 13.0 & 3.4 \\
		
		Ti\,\textsc{ii} & 3349.0334 & 7.3 & 19.7 & 18.9 & 103.0 & - & 14.7 & 3.8 \\
		
		Cr\,\textsc{ii} & 3368.0416 & 7.2 & 19.1 & 18.4 & 121.8 & - & 14.8 & 3.8 \\
		
		Fe\,\textsc{ii} & 5169.0318 & 7.2$^*$ & 20.3$^*$ & 15.2$^*$ & - & - & 8.0$^*$ & - \\
		
		Ni\,\textsc{ii} & 3513.9871 & 7.0 & 16.9 & 14.5 & 91.2 & 7.3 & 12.7 & 3.4 \\
		
		\hline
	\end{tabular}
\end{table*}

\begin{table*}
	\centering
	\footnotesize
	\caption{The upper limit equivalent widths in m\AA \, of the material polluting two  white dwarfs in this study with COS FUV data.  
	}
	\label{tab:WDs-EW-UL-UV}
	\begin{tabular}{ccccccccc} 
		\hline
		Element & Line (\AA) & {Gaia\,J0006}  & {Gaia\,J0510}  \\
		\hline	
		Fe & 1144.938 & 24.0 & 13.8 \\
		Ni & 1370.132 & 17.5 & 10.0 \\
        C & 1335.708 & - & 14.5 \\
        P & 1153.995 & - & 11.7 \\
  
		\hline
	\end{tabular}
\end{table*}

\section{Model fits to spectral lines} \label{Mod-plots}

Supplementary Figs. D1 -- D11 show the model abundance fit to the strongest spectral line for each element in seven of the white dwarfs. 

\section{Non-photospheric absorption lines}

Table \ref{tab:RVs} shows the non-photospheric measurements of absorption lines in the spectra of the seven white dwarfs in this study.

\begin{table*}
	\centering
	\footnotesize
	\caption{Comparison between the mean radial velocities of the photospheric (phot) and non-photospheric (non-phot) spectral lines. Gaia\,J0006 has two Na non-photospheric lines, and Gaia\,J2100 has two Ca non-photospheric lines, these are listed on separate rows. As mentioned in \citet{doyle2023new}, if the difference between the photospheric and non-photospheric lines is of the order of the gravitational redshift of the white dwarf or less, it is possible that the non-photospheric lines are circumstellar in origin. However, particularly for the white dwarfs $>$\,100\,pc, absorption from the interstellar medium cannot be ruled out.}
	\label{tab:RVs}
	\begin{tabular}{ccccccccc}
\hline											
WD	&	RV$_{\textrm{phot}}$	&	$\sigma$ RV$_{\textrm{phot}}$	&	Element Species	&	RV$_{\textrm{non-phot}}$	&	Gravitation Redshift & Distance & RV$_{\textrm{phot}}$ $-$ RV$_{\textrm{non-phot}}$	\\
	&	(km s$^{-1}$)	&	(km s$^{-1}$)	&	non-phot	&	(km s$^{-1}$)	&	(km s$^{-1}$) &	(pc) & (km s$^{-1}$)	\\
\hline											
Gaia\,J0006	&	23.8	&	5.4	&	C, N, Na, Si, S, Ca, Fe?	&	$-$8.4	&	32.5 & 152 & 32.2	\\
 "	&	"	&	"	&	Na	&	$-$1.6	& "	& " & 25.4	\\
 "	&	"	&	"	&	Si	&	$-$196	& "	& " & 219.8	\\ 
Gaia\,J0347	&	16.0	&	$...$	&	Na, Ca	&	18.3	&	35.8 & 141 & $-$2.3	\\
Gaia\,J0510	&	26.1	&	4.8	&	C, N, Fe?	&	17.5	& 43.0	& 65 & 8.6	\\
Gaia\,J0611	&	58.7	&	8.1	&	C, O, Si, Fe	&	$-$2.0 & 37.8	&	143 & 60.7	\\
Gaia\,J0644	&	93.2	&	4.2	&	Ca	&	23.6	&	39.6 & 112 & 69.6	\\
WD\,1622	&	$-$23.0	&	6.0	&	C, N, O, Si, S	&	$-$21.8	&	21.5 & 183 & $-$1.2	\\
Gaia\,J2100	&	4.7	&	2.3	&	Ca	&	$-$11.4	& 35.8	& 88 & 16.1	\\
 "	&	"	&	"	&	Ca	&	$-$26.8	& " & " & 31.5	\\
\hline											
\end{tabular}
\end{table*}


\bsp	
\label{lastpage}

\clearpage


\section*{Supplementary material}

Table B1. Gaia J0006+2858 UV: The absorption lines in Gaia J0006+2858 from the FUV and NUV spectra. The notes column highlights those lines that have a significant ISM contribution (labelled 'i') and those where two spectral lines are blended ('b'). The NUV RV measurements are invalid and included in brackets, see Section 2.6 in the main body of the paper for further details.

| Line | $\lambda_{\rm vac}$ (Å) | log(gf) | $E_{low}$ (eV) | $\lambda_{\rm obs}$ (Å) | EW (Å) | RV (km/s) | [X/H]$_{\rm Spec}$ | [X/H]$_{\rm Phot}$ | Notes |
|---|---|---|---|---|---|---|---|---|---|
| C III | 1174.930 | −0.468 | 6.496 | 1175.01 | 0.038 (0.007) | 21.2 | −6.98 | −6.95 | |
| C III | 1175.260 | −0.565 | 6.493 | 1175.33 | 0.022 (0.005) | 18.6 | −6.98 | −6.95 | |
| C III | 1175.710 | 0.009 | 6.503 | 1175.77 | 0.067 (0.014) | 15.3 | −6.98 | −6.95 | |
| C III | 1176.370 | −0.468 | 6.503 | 1176.44 | 0.032 (0.010) | 18.1 | −6.98 | −6.95 | |
| C II | 1334.530 | −0.596 | 0.000 | 1334.55 | 0.225 (0.021)$^\beta$ | 4.0 | −6.83 | −6.92 | i |
| C II | 1335.708 | −0.341 | 0.008 | 1335.74 | 0.148 (0.023)$^\beta$ | 8.1 | −6.83 | −6.92 | i |
| **C Average:** | | | | | | | **−6.90** | **−6.93** | |
| O I | 1152.15 | −0.268 | 1.967 | 1152.24 | 0.054 (0.007) | 24.5 | −4.48 | −4.54 | |
| S II | 1253.811 | −1.4 | 0.000 | 1253.91 | 0.018 (0.007)$^\gamma$ | 23.2 | −6.29 | −6.41 | i |
| S II | 1259.519 | −1.32 | 0.000 | 1259.50 | 0.108 (0.024)$^\beta$ | −5.7 | −6.41 | −6.51 | i |
| **S Average:** | | | | | | | **−6.35** | **−6.46** | |
| Si III | 1141.579 | 0.47 | 16.098 | 1141.66 | 0.047 (0.015) | 21.8 | −5.57 | −5.56 | |
| Si III | 1154.998 | −0.35 | 16.081 | 1155.11 | 0.031 (0.009) | 29.1 | −5.37 | −5.38 | |
| Si III | 1155.959 | −0.35 | 16.098 | 1156.04 | 0.052 (0.017) | 21.0 | −5.37 | −5.38 | |
| Si III | 1158.101 | −0.26 | 16.098 | 1158.22 | 0.035 (0.011) | 30.0 | −5.37 | −5.38 | |
| Si III | 1160.252 | −0.26 | 16.130 | 1160.32 | 0.038 (0.015) | 18.3 | −5.37 | −5.38 | |
| Si III | 1161.579 | 0.22 | 16.130 | 1161.65 | 0.076 (0.016) | 18.3 | −5.37 | −5.38 | |
| Si II | 1190.416 | −0.245 | 0.000 | 1190.42 | 0.281 (0.033)$^\beta$ | 0.8 | −5.69 | −5.78 | i |
| Si II | 1193.290 | 0.075 | 0.000 | 1193.31 | 0.233 (0.033)$^\beta$ | 5.1 | −5.69 | −5.78 | i |
| Si II | 1194.500 | 0.487 | 0.036 | 1194.58 | 0.172 (0.026) | 19.5 | −5.69 | −5.78 | |
| Si II | 1197.394 | −0.202 | 0.036 | 1197.49 | 0.076 (0.018) | 22.8 | −5.69 | −5.78 | |
| Si II | 1227.604 | 0.528 | 5.345 | 1227.69 | 0.043 (0.020) | 21.0 | −5.38 | −5.47 | |
| Si II | 1228.739 | 0.651 | 5.323 | 1228.74 | 0.111 (0.020) | 0.4$^\alpha$ | −5.38 | −5.47 | b |
| Si II | 1229.383 | 0.882 | 5.345 | 1229.49 | 0.073 (0.021) | 25.3 | −5.38 | −5.47 | |
| Si II | 1246.740 | −0.236 | 5.309 | 1246.85 | 0.042 (0.012) | 26.5 | −5.53 | −5.63 | |
| Si II | 1248.426 | 0.066 | 5.323 | 1248.53 | 0.074 (0.012) | 23.8 | −5.53 | −5.63 | |
| Si II | 1250.091 | 0.228 | 6.857 | 1250.21 | 0.093 (0.012) | 27.5 | −5.53 | −5.63 | |
| Si II | 1250.436 | 0.423 | 6.859 | 1250.54 | 0.157 (0.013) | 25.9 | −5.53 | −5.63 | |
| Si II | 1251.164 | 0.24 | 5.345 | 1251.25 | 0.056 (0.008) | 20.2 | −5.53 | −5.63 | |
| Si II | 1260.422 | 0.462 | 0.000 | 1260.49 | 0.387 (0.046)$^\beta$ | 15.7 | −5.53 | −5.61 | i |
| Si II | 1264.738 | 0.71 | 0.036 | 1264.82 | 0.420 (0.048) | 19.5 | −5.53 | −5.61 | |
| Si II | 1265.002 | −0.273 | 0.036 | 1265.09 | 0.180 (0.035) | 21.6 | −5.53 | −5.61 | |
| Si III | 1294.545 | −0.037 | 6.553 | 1294.67 | 0.253 (0.045) | 28.3 | −5.64 | −5.60 | |
| Si III | 1296.726 | −0.127 | 6.537 | 1296.85 | 0.217 (0.031) | 28.9 | −5.64 | −5.60 | |
| Si III | 1298.892 | −0.257 | 6.553 | 1299.05 | 0.462 (0.049) | 37.4 | −5.64 | −5.60 | |
| Si II | 1309.276 | −0.448 | 0.036 | 1309.50 | 0.283 (0.052) | 51.8$^\alpha$ | −5.23 | −5.34 | b |
| Si III | 1312.591 | −0.84 | 10.276 | 1312.69 | 0.097 (0.022) | 21.5 | −5.23 | −5.34 | |
| Si II | 1346.884 | −0.144 | 5.323 | 1346.98 | 0.030 (0.010) | 22.3 | −5.20 | −5.29 | |
| Si II | 1348.543 | −0.186 | 5.309 | 1348.62 | 0.043 (0.011) | 17.5 | −5.20 | −5.29 | |
| Si II | 1350.072 | 0.216 | 5.345 | 1350.16 | 0.067 (0.020) | 20.3 | −5.20 | −5.29 | |
| Si II | 1353.721 | −0.158 | 5.345 | 1353.81 | 0.047 (0.013) | 20.4 | −5.20 | −5.29 | |
| Si IV | 1393.755 | 0.03 | 0.000 | 1393.85 | 0.213 (0.036) | 20.0 | −5.76 | −5.53 | |
| Si IV | 1402.770 | −0.28 | 0.000 | 1402.89 | 0.122 (0.045) | 25.6 | −6.05 | −5.83 | |
| Si III | 1417.237 | −0.106 | 10.276 | 1417.34 | 0.156 (0.021) | 22.2 | −5.42 | −5.35 | |
| **Si Average:** | | | | | | | **−5.48** | **−5.50** | |
| P II$^\delta$ | 1153.995 | 0.067 | 0.058 | 1154.10 | 0.025 (0.011) | 27.3 | −7.39 | −7.55 | |
| Al III | 1379.670 | −0.6 | 6.656 | 1379.78 | 0.051 (0.005) | 22.8 | −6.43 | −6.47 | |
| Al III | 1384.132 | −0.29 | 6.685 | 1384.23 | 0.093 (0.026) | 20.4 | −6.36 | −6.34 | |
| Al III | 1854.716 | 0.05 | 0.000 | 1854.93 | 0.165 (0.057) | (34.7) | −6.84 | −6.82 | |
| Al III | 1862.790 | −0.197 | 0.000 | 1863.05 | 0.097 (0.045) | (42.4) | −6.84 | −6.82 | |
| **Al Average:** | | | | | | | **−6.50** | **−6.50** | |

$^\alpha$ Merged with nearby line, RV measured with respect to the line with the strongest log(gf) as listed in the table.
$^\beta$ Significant ISM component merged with the photospheric line, the equivalent width and radial velocity are calculated for the combined line.
$^\gamma$ Significant ISM component but resolvable so the equivalent width is the estimated strength of the photospheric feature.
$^\delta$ Additional P III line observed at 1334.813Å, however, cannot be fit due to strong C lines.

**Table B2. Gaia J0006+2858 Optical:** The absorption lines in Gaia J0006+2858 from the optical. If two lines lie within $\sim 10\,\text{Å}$, equivalent widths and abundances were fitted simultaneously.

| Line | $\lambda_{\text{air}}$ (Å) | log(gf) | $E_{low}$ (eV) | $\lambda_{\text{obs}}$ (Å) | EW (Å) | RV (km/s) | [X/H]$_{\text{Spec}}$ HIRES | X-Sh | [X/H]$_{\text{Phot}}$ HIRES | X-Sh |
|---|---|---|---|---|---|---|---|---|---|---|
| Ca II | 3179.332 | 0.499 | 3.151 | 3179.66 | 0.013 (0.003) | 31.3 | −6.21 | | −6.33 | |
| Ca II$^{\beta}$ | 3933.663 | 0.105 | 0.000 | 3934.02 | 0.041 (0.006)$^{\alpha}$ | 27.3 | −6.13 | | −6.32 | |
| **Ca Average:** | | | | | | | **−6.17** | | **−6.32** | |
| Mg II | 4481.125 | 0.740 | 8.864 | 4481.59 | 0.208 (0.007) | 31.2 | −5.00 | −4.90 | −5.10 | −4.97 |
| Si II | 3856.018 | −0.370 | 6.859 | 3856.37 | 0.029 (0.003) | 27.5 | −4.79 | | −4.89 | |
| Si II | 3862.595 | −0.810 | 6.857 | 3862.94 | 0.015 (0.002) | 26.8 | −4.78 | | −4.90 | |
| Si II | 4128.054 | 0.410 | 9.836 | 4128.40 | 0.024 (0.003) | 24.8 | −4.99 | | −5.10 | |
| Si II | 4130.893 | 0.530 | 9.839 | 4131.26 | 0.024 (0.003) | 26.3 | −4.99 | | −5.10 | |
| Si II | 5041.023 | 0.150 | 10.066 | 5041.59 | 0.036 (0.006) | 33.8 | −5.10 | −4.97 | −5.20 | −5.05 |
| Si II | 5055.983 | 0.530 | 10.074 | 5056.53 | 0.055 (0.006) | 32.3 | −5.14 | −4.97 | −5.24 | −5.05 |
| Si II | 6347.106 | 0.170 | 8.121 | 6347.68 | 0.083 (0.008) | 27.0 | −4.95 | −4.86 | −5.04 | −4.94 |
| Si II | 6371.368 | −0.040 | 8.121 | 6371.99 | 0.046 (0.008) | 29.2 | −4.93 | | −5.03 | |
| **Si Average:** | | | | | | | **−4.94** | **−4.91** | **−5.04** | **−4.99** |

$^{\alpha}$ Gaseous emission from circumstellar gas at this wavelength, additional 12 percent error added in quadrature.
$^{\beta}$ Line has an additional resolvable absorption feature at the rest wavelength of line. This could be absorption from the ISM or the core of the double peaked circumstellar gas emission feature.

**Table B3. Gaia J0347+1624 UV:** The absorption line in Gaia J0347+1624 from the NUV spectrum. The NUV RV measurements are invalid and included in brackets, see Section 2.6 in the main body of the paper for further details.

| Line | $\lambda_{\text{vac}}$ (Å) | log(gf) | $E_{low}$ (eV) | $\lambda_{\text{obs}}$ (Å) | EW (Å) | RV (km/s) | [X/H]$_{\text{Spec}}$ | [X/H]$_{\text{Phot}}$ |
|---|---|---|---|---|---|---|---|---|
| Al III | 1854.716 | 0.05 | 0.000 | 1854.99 | 0.109 (0.055) | (43.9) | −7.34 | −7.26 |

**Table B4. Gaia J0347+1624 Optical:** The absorption lines measured in the Gaia J0347+1624 HIRES spectrum. There is likely a non-photospheric contribution to the Ca lines, and so the abundances should be taken as upper limits.

| Line | $\lambda_{\text{air}}$ (Å) | log(gf) | $E_{low}$ (eV) | $\lambda_{\text{obs}}$ (Å) | EW (Å) | RV (km/s) | [X/H]$_{\text{Spec}}$ | [X/H]$_{\text{Phot}}$ |
|---|---|---|---|---|---|---|---|---|
| Ca II | 3933.663 | 0.105 | 0.000 | 3933.90 | 0.062 (0.006) | 18.1 | −5.79 | −6.29 |
| Ca II | 3968.469 | −0.200 | 0.000 | 3968.71 | 0.022 (0.004) | 18.2 | −5.60 | −5.72 |
| **Ca Average:** | | | | | | | **−5.68** | **−5.92** |
| Mg II | 4481.1250* | 0.740 | 8.864 | 4481.36 | 0.040 (0.012) | 16.0 | −5.78 | −6.05 |

**Table B5. Gaia J0510+2315 UV:** The absorption lines in Gaia J0510+2315 from the FUV and NUV spectra. If two lines lie within $\sim$ 10 Å, equivalent widths and abundances were fitted simultaneously.

| Line | $\lambda_{\text{vac}}$ (Å) | log(gf) | $E_{low}$ (eV) | $\lambda_{\text{obs}}$ (Å) | EW (Å) | RV (km/s) | [X/H]$_{\text{Spec}}$ | [X/H]$_{\text{Phot}}$ |
|---|---|---|---|---|---|---|---|---|
| O I | 1152.150 | −0.268 | 1.967 | 1152.25 | 0.062 (0.010) | 26.5 | −4.98 | −5.07 |
| S II | 1253.811 | −1.4 | 0.000 | 1253.92 | 0.021 (0.006) | 25.0 | −6.17 | −6.13 |
| S II | 1259.519 | −1.32 | 0.000 | 1259.62 | 0.030 (0.003) | 23.1 | −6.23 | −6.26 |
| **S Average:** | | | | | | | **−6.20** | **−6.19** |
| Si III | 1141.579 | 0.47 | 16.098 | 1141.62 | 0.034 (0.004) | 11.3 | −5.96 | −5.87 |
| Si III | 1142.285 | −0.01 | 16.098 | 1142.37 | 0.032 (0.005) | 21.3 | −5.96 | −5.87 |
| Si III | 1144.309 | 0.74 | 16.130 | 1144.43 | 0.056 (0.011) | 31.2 | −5.96 | −5.87 |
| Si II | 1190.416 | −0.245 | 0.000 | 1190.51 | 0.091 (0.006) | 22.7 | −6.09 | −6.12 |
| Si II | 1193.290 | 0.075 | 0.000 | 1193.36 | 0.102 (0.009) | 17.6 | −6.09 | −6.12 |
| Si II | 1194.500 | 0.487 | 0.036 | 1194.59 | 0.096 (0.012) | 22.0 | −6.09 | −6.12 |
| Si II | 1197.394 | −0.202 | 0.036 | 1197.49 | 0.067 (0.011) | 23.6 | −6.09 | −6.12 |
| Si II | 1246.740 | −0.236 | 5.309 | 1246.85 | 0.027 (0.006) | 27.2 | −5.89 | −5.92 |
| Si II | 1248.426 | 0.066 | 5.323 | 1248.53 | 0.041 (0.008) | 24.6 | −5.89 | −5.92 |
| Si II | 1250.091 | 0.228 | 6.857 | 1250.20 | 0.036 (0.006) | 26.4 | −5.89 | −5.92 |
| Si II | 1250.436 | 0.423 | 6.859 | 1250.56 | 0.067 (0.008) | 29.2 | −5.89 | −5.92 |
| Si II | 1251.164 | 0.24 | 5.345 | 1251.26 | 0.044 (0.004) | 23.8 | −5.89 | −5.92 |
| Si II | 1260.422 | 0.462 | 0.000 | 1260.53 | 0.236 (0.007) | 25.6 | −5.90 | −6.02 |
| Si II | 1264.738 | 0.71 | 0.036 | 1264.86 | 0.271 (0.014) | 29.0 | −5.90 | −6.02 |
| Si II | 1265.002 | −0.273 | 0.036 | 1265.13 | 0.166 (0.013) | 31.2 | −5.90 | −6.02 |
| Si III | 1294.545 | −0.037 | 6.553 | 1294.66 | 0.098 (0.005) | 26.2 | −6.23 | −6.08 |
| Si III | 1296.726 | −0.127 | 6.537 | 1296.88 | 0.128 (0.009) | 34.4 | −6.23 | −6.08 |
| Si III | 1298.892 | −0.257 | 6.553 | 1299.07 | 0.210 (0.007) | 40.0 | −6.23 | −6.08 |
| Si III | 1301.149 | −0.127 | 6.553 | 1301.30 | 0.078 (0.009) | 34.4 | −6.23 | −6.08 |
| Si II | 1346.884 | −0.144 | 5.323 | 1347.00 | 0.039 (0.005) | 25.9 | −5.46 | −5.56 |
| Si II | 1348.543 | −0.186 | 5.309 | 1348.66 | 0.047 (0.005) | 25.6 | −5.46 | −5.56 |
| Si II | 1350.072 | 0.216 | 5.345 | 1350.17 | 0.070 (0.005) | 22.6 | −5.46 | −5.56 |
| Si II | 1352.635 | −0.193 | 5.323 | 1352.74 | 0.043 (0.005) | 22.1 | −5.46 | −5.56 |
| Si II | 1353.721 | −0.158 | 5.345 | 1353.84 | 0.050 (0.006) | 25.6 | −5.46 | −5.56 |
| **Si Average:** | | | | | | | **−5.85** | **−5.88** |
| Al III | 1379.670 | −0.6 | 6.656 | 1379.77 | 0.008 (0.002) | 21.9 | −7.08 | −7.13 |
| Al III | 1384.132 | −0.29 | 6.685 | 1384.23 | 0.031 (0.009) | 21.2 | −7.00 | −6.89 |
| **Al Average:** | | | | | | | **−7.04** | **−6.99** |

**Table B6. Gaia J0510+2315 Optical:** The absorption lines in the HIRES spectra.

| Line | $\lambda_{\text{air}}$ (Å) | log(gf) | $E_{low}$ (eV) | $\lambda_{\text{obs}}$ (Å) | EW (Å) | RV (km/s) | [X/H]$_{\text{Spec}}$ | [X/H]$_{\text{Phot}}$ |
|---|---|---|---|---|---|---|---|---|
| Ca II | 3933.663 | 0.499 | 3.151 | 3934.03 | 0.032 (0.004) | 27.6 | −6.31 | −6.80 |
| Mg II | 4481.125$^\alpha$ | 0.740 | 8.864 | 4481.53 | 0.114 (0.011) | 26.9 | −5.23 | −5.35 |
| Mg II | 4481.325$^\alpha$ | 0.590 | 8.864 | 4481.74 | 0.037 (0.007) | 27.8 | −5.23 | −5.35 |
| **Mg Average:** | | | | | | | **−5.23** | **−5.35** |
| O I | 7771.938 | 0.369 | 9.146 | 7772.63 | 0.023 (0.005)$^\beta$ | 26.6 | −4.23 | −4.35 |
| O I | 7774.156 | 0.223 | 9.146 | 7774.88 | 0.020 (0.004)$^\beta$ | 27.8 | −4.23 | −4.35 |
| O I | 7775.386 | 0.001 | 9.146 | 7776.04 | 0.005 (0.002)$^\beta$ | 25.3 | −4.23 | −4.35 |
| **O Average:** | | | | | | | **−4.23** | **−4.35** |
| Si II | 3856.018 | −0.370 | 6.859 | 3856.35 | 0.019 (0.003) | 26.0 | −5.00 | −5.02 |
| Si II | 4128.054 | 0.410 | 9.836 | 4128.44 | 0.015 (0.005) | 28.0 | −4.90 | −4.88 |
| Si II | 4130.893 | 0.530 | 9.839 | 4131.25 | 0.030 (0.006) | 25.8 | −4.90 | −4.88 |
| Si II | 5041.023 | 0.150 | 10.066 | 5041.52 | 0.014 (0.002) | 29.8 | −5.33 | −5.33 |
| Si II | 5055.983 | 0.530 | 10.074 | 5056.48 | 0.039 (0.003) | 29.2 | −5.36 | −5.36 |
| Si II | 6347.106 | 0.170 | 8.121 | 6347.65 | 0.058 (0.004) | 25.7 | −5.19 | −5.21 |
| Si II | 6371.368 | −0.040 | 8.121 | 6371.92 | 0.031 (0.004) | 26.1 | −5.15 | −5.16 |
| **Si Average:** | | | | | | | **−5.12** | **−5.13** |

$^\alpha$ Mg II doublet is resolved and fitted simultaneously for EW and RV.
$^\beta$ Gaseous emission from circumstellar gas at this wavelength, additional 12 percent error added in quadrature.

**Table B7. Gaia J0611−6931 UV:** The absorption lines in Gaia J0611−6931 from the FUV spectrum. If lines lie within ∼ 10 Å, equivalent widths and abundances were fitted simultaneously, and the strongest lines in the region are measured and reported here. The notes column labels the lines that have a significant ISM contribution as 'i', those where two spectral lines are blended as 'b', and those where the night data were used are labelled 'N'. The ISM components are offset by ∼60 km/s and are resolvable so the equivalent widths reported are those for the photospheric component.

| Line | $\lambda_{\rm vac}$ (Å) | log(gf) | $E_{low}$ (eV) | $\lambda_{\rm obs}$ (Å) | EW (Å) | RV (km/s) | [X/H]$_{\rm Spec}$ | [X/H]$_{\rm Phot}$ | Notes |
|---|---|---|---|---|---|---|---|---|---|
| C II | 1334.530 | −0.596 | 0.000 | 1334.79 | 0.052 (0.006) | 57.3 | −7.15 | −7.29 | i |
| C II | 1335.709 | −0.341 | 0.008 | 1335.96 | 0.077 (0.004) | 55.6 | −7.15 | −7.29 | |
| **C Average:** | | | | | | | **−7.15** | **−7.29** | |
| O I | 1152.150 | −0.268 | 1.967 | 1152.37 | 0.098 (0.010) | 57.6 | −4.30 | −4.27 | |
| O I | 1302.168 | −0.585 | 0.000 | 1302.47 | 0.156 (0.023) | 68.9 | −4.26 | −4.39 | i,N |
| O I | 1304.858 | −0.808 | 0.020 | 1305.13 | 0.117 (0.014) | 62.5 | −4.26 | −4.39 | N |
| O I | 1306.029 | −1.285 | 0.028 | 1306.29 | 0.098 (0.016) | 59.4 | −4.26 | −4.39 | N |
| **O Average:** | | | | | | | **−4.28** | **−4.33** | |
| Al III | 1379.670 | −0.600 | 6.656 | 1379.88 | 0.024 (0.007) | 45.2 | −6.42 | −6.89 | |
| Al III | 1384.132 | −0.290 | 6.685 | 1384.46 | 0.037 (0.009) | 71.0 | −6.51 | −6.59 | |
| **Al Average:** | | | | | | | **−6.46** | **−6.71** | |
| Si III | 1142.285 | −0.010 | 16.098 | 1142.50 | 0.058 (0.008) | 55.8 | −5.28 | −5.16 | |
| Si III | 1144.309 | 0.740 | 16.130 | 1144.47 | 0.095 (0.008) | 43.4 | −5.28 | −5.16 | |
| Si II | 1190.416 | −0.245 | 0.000 | 1190.65 | 0.150 (0.017) | 59.1 | −5.27 | −5.52 | i |
| Si II | 1193.290 | 0.075 | 0.000 | 1193.50 | 0.120 (0.014) | 51.7 | −5.27 | −5.52 | i |
| Si II | 1194.500 | 0.487 | 0.036 | 1194.75 | 0.235 (0.026) | 61.6 | −5.27 | −5.52 | |
| Si II | 1246.740 | −0.236 | 5.309 | 1246.99 | 0.077 (0.010) | 62.1 | −5.32 | −5.31 | |
| Si II | 1248.426 | 0.066 | 5.323 | 1248.69 | 0.089 (0.006) | 63.0 | −5.32 | −5.31 | |
| Si II | 1250.091 | 0.228 | 6.857 | 1250.37 | 0.110 (0.019) | 67.3 | −5.32 | −5.31 | |
| Si II | 1250.436 | 0.423 | 6.859 | 1250.71 | 0.220 (0.036) | 65.7 | −5.32 | −5.31 | |
| Si II | 1251.164 | 0.24 | 5.345 | 1251.45 | 0.069 (0.007) | 68.5 | −5.32 | −5.31 | |
| Si II | 1260.422 | 0.462 | 0.000 | 1260.62 | 1.065 (0.035) | 47.5 | −5.01 | −5.02 | i$^\beta$ |
| Si II | 1264.738 | 0.71 | 0.036 | 1265.09 | 1.390 (0.114) | 83.9$^\alpha$ | −5.01 | −5.02 | b |
| Si III | 1294.545 | −0.037 | 6.553 | 1294.77 | 0.112 (0.013) | 52.7 | −5.16 | −5.21 | N |
| Si III | 1296.726 | −0.127 | 6.537 | 1296.99 | 0.158 (0.028) | 60.3 | −5.16 | −5.21 | N |
| Si III | 1298.892 | −0.257 | 6.553 | 1299.20 | 0.332 (0.015) | 71.5 | −5.16 | −5.21 | N |
| Si III | 1303.323 | −0.037 | 6.585 | 1303.59 | 0.115 (0.022) | 61.0 | −5.16 | −5.21 | N |
| Si II | 1309.276 | −0.448 | 0.036 | 1309.65 | 0.555 (0.065) | 85.1$^\alpha$ | −5.16 | −5.21 | N,b |
| Si II | 1346.884 | −0.144 | 5.323 | 1347.14 | 0.110 (0.007) | 57.1 | −5.13 | −5.22 | |
| Si II | 1348.543 | −0.186 | 5.309 | 1348.77 | 0.109 (0.006) | 50.5 | −5.13 | −5.22 | |
| Si II | 1350.072 | 0.216 | 5.345 | 1350.32 | 0.188 (0.006) | 55.2 | −5.13 | −5.22 | |
| Si II | 1350.516 | −0.69 | 5.323 | 1350.76 | 0.068 (0.014) | 53.5 | −5.13 | −5.22 | |
| Si II | 1350.656 | −0.888 | 5.309 | 1350.91 | 0.087 (0.014) | 55.6 | −5.13 | −5.22 | |
| Si II | 1352.635 | −0.193 | 5.323 | 1352.88 | 0.118 (0.005) | 54.7 | −5.13 | −5.22 | |
| Si II | 1353.721 | −0.158 | 5.345 | 1353.96 | 0.137 (0.011) | 53.4 | −5.13 | −5.22 | |
| Si III | 1417.237 | −0.106 | 10.276 | 1417.51 | 0.077 (0.013) | 56.9 | −5.57 | −5.40 | |
| **Si Average:** | | | | | | | **−5.22** | **−5.24** | |
| P II | 1153.995 | 0.067 | 0.058 | 1154.20 | 0.029 (0.006) | 54.4 | −7.49 | −7.66 | |
| S II | 1253.811 | −1.4 | 0.000 | 1254.06 | 0.072 (0.008) | 59.3 | −5.13 | −5.36 | |
| S II | 1259.519 | −1.32 | 0.000 | 1259.75 | 0.065 (0.014) | 54.5 | −5.57 | −5.61 | |
| **S Average:** | | | | | | | **−5.30** | **−5.47** | |
| Fe II | 1143.226 | −0.716 | 0.000 | 1143.39 | 0.054 (0.010) | 42.2 | −5.29 | −5.43 | |
| Fe II | 1144.938 | 0.037 | 0.000 | 1145.14 | 0.078 (0.013) | 51.2 | −5.29 | −5.43 | |
| Fe II | 1146.832 | −1.177 | 0.107 | 1147.08 | 0.036 (0.007) | 64.1$^\alpha$ | −5.29 | −5.51 | b |
| Fe II | 1147.409 | −0.707 | 0.048 | 1147.64 | 0.036 (0.007) | 59.0 | −5.29 | −5.51 | |
| Fe II | 1151.146 | −0.451 | 0.083 | 1151.37 | 0.025 (0.007) | 57.2 | −5.29 | −5.51 | |
| Fe II | 1154.352 | −0.984 | 3.153 | 1154.58 | 0.015 (0.004) | 58.1 | −5.29 | −5.51 | |
| Fe II | 1358.937 | −0.193 | 3.245 | 1359.19 | 0.064 (0.009) | 56.2$^\alpha$ | −5.19 | −5.43 | b |
| Fe II | 1361.373 | −0.519 | 1.671 | 1361.64 | 0.020 (0.004) | 59.4 | −5.19 | −5.43 | |
| Fe II | 1371.022 | −0.229 | 2.635 | 1371.29 | 0.017 (0.003) | 57.5 | −5.15 | −5.44 | |
| **Fe Average:** | | | | | | | **−5.23** | **−5.45** | |
| Ni II | 1317.217 | −0.058 | 0.000 | 1317.45 | 0.038 (0.005) | 54.1 | −6.87 | −7.07 | |
| Ni II | 1370.132 | −0.105 | 0.000 | 1370.41 | 0.019 (0.004) | 60.0 | −6.82 | −7.05 | |
| Ni II | 1381.286 | −0.336 | 0.187 | 1381.54 | 0.019 (0.004) | 55.0 | −6.79 | −7.02 | |
| **Ni Average:** | | | | | | | **−6.83** | **−7.05** | |
| Cu II | 1358.773 | −0.268 | 0.000 | 1359.02 | 0.017 (0.006) | 54.5 | −7.87 | −7.94 | b |

$^\alpha$ Merged with nearby line, RV measured with respect to the line with the strongest log(gf) as listed in the table.
$^\beta$ Significant unresolved ISM component, EW is a combination of the ISM and WD component. Abundance calculated after subtracting the ISM component.

**Table B8. Gaia J0611−6931 Optical:** The absorption lines in Gaia J0611−6931. If two lines lie within ∼ 10 Å, equivalent widths and abundances were fitted simultaneously.

| Line | $\lambda_{\text{air}}$ (Å) | log(gf) | $E_{low}$ (eV) | $\lambda_{\text{obs}}$ (Å) | EW (Å) | RV (km/s) | [X/H]$_{\text{Spec}}$ X-Sh | [X/H]$_{\text{Spec}}$ MIKE | [X/H]$_{\text{Phot}}$ X-Sh | [X/H]$_{\text{Phot}}$ MIKE |
|---|---|---|---|---|---|---|---|---|---|---|
| Ca II | 3933.663 | 0.499 | 3.151 | 3934.51 | 0.118 (0.025)$^\alpha$ | 64.5 | −6.17 | −6.01 | −6.41 | −6.33 |
| Mg II | 4481.125 | 0.740 | 8.864 | 4482.14 | 0.546 (0.014) | 67.6 | −4.66 | −4.53 | −4.68 | −4.53 |
| Mg II | 7896.037 | 0.650 | 9.999 | 7898.08 | 0.080 (0.025) | 77.6 | −4.68 | −4.57 | −4.76 | −4.65 |
| **Mg Average:** | | | | | | | **−4.67** | **−4.55** | **−4.72** | **−4.59** |
| O I | 7771.938 | 0.369 | 9.146 | 7773.48 | 0.049 (0.014)$^\alpha$ | 59.4 | −3.81 | −3.69 | −3.90 | −3.78 |
| O I | 7774.156 | 0.223 | 9.146 | 7775.72 | 0.038 (0.014)$^\alpha$ | 60.3 | −3.81 | −3.69 | −3.90 | −3.78 |
| O I | 7775.386 | 0.001 | 9.146 | 7776.96 | 0.033 (0.017)$^\alpha$ | 60.6 | −3.81 | −3.69 | −3.90 | −3.78 |
| **O Average:** | | | | | | | **−3.81** | **−3.69** | **−3.90** | **−3.78** |
| Si II | 5055.983 | 0.530 | 10.074 | 5057.13* | 0.010 (0.003) | 68.1 | −4.83 | −4.68 | −4.87 | −4.71 |
| Si II | 6347.106 | 0.170 | 8.121 | 6348.40 | 0.117 (0.017) | 60.9 | −4.71 | −4.79 | −4.71 | −4.80 |
| Si II | 6371.368 | −0.040 | 8.121 | 6372.68 | 0.090 (0.0065) | 61.8 | | −4.54 | | −4.58 |
| **Si Average:** | | | | | | | **−4.77** | **−4.66** | **−4.78** | **−4.69** |

$^\alpha$ Gaseous emission from circumstellar gas at this wavelength, additional 12 percent error added in quadrature.

**Table B9. Gaia J0644−0352 Optical:** The absorption lines in Gaia J0644−0352. If two lines lie within ∼ 10 Å, equivalent widths and abundances were fitted simultaneously.

| Line | $\lambda_{\text{air}}$ (Å) | log(gf) | $E_{low}$ (eV) | $\lambda_{\text{obs}}$ (Å) | EW (Å) | RV (km/s) | [X/He]$_{\text{Spec}}$ HIRES | X-Sh | [X/He]$_{\text{Phot}}$ HIRES | X-Sh |
|---|---|---|---|---|---|---|---|---|---|---|
| Al II | 3586.556 | 0.637 | 11.847 | 3587.63 | 0.030 (0.005) | 89.9 | −6.76 | | −7.05 | |
| Al II | 3587.072 | 0.477 | 11.847 | 3588.18 | 0.021 (0.006) | 92.8 | −6.76 | | −7.05 | |
| Al II | 3587.445 | 0.307 | 11.847 | 3588.53 | 0.011 (0.004) | 90.5 | −6.76 | | −7.05 | |
| **Al Average:** | | | | | | | **−6.76** | | **−7.05** | |
| Ca II | 3158.870 | 0.241 | 3.123 | 3159.86 | 0.092 (0.004) | 93.5 | −6.97 | −6.91 | −7.60 | −7.52 |
| Ca II | 3179.332 | 0.499 | 3.151 | 3180.34 | 0.094 (0.009) | 95.1 | −6.97 | −6.72 | −7.58 | −7.32 |
| Ca II | 3181.275 | −0.455 | 3.151 | 3182.27 | 0.025 (0.004) | 93.4 | −6.97 | −6.72 | −7.58 | −7.32 |
| Ca II | 3706.024 | −0.48 | 3.123 | 3707.19 | 0.020 (0.006) | 94.2 | −6.94 | −6.57 | −7.66 | −7.44 |
| Ca II | 3736.903 | −0.173 | 3.151 | 3738.07 | 0.061 (0.003) | 93.5 | −6.89 | −6.86 | −7.54 | −7.50 |
| Ca II | 3933.663 | 0.105 | 0.000 | 3934.87 | 0.333 (0.004) | 91.7 | −6.67 | −6.70 | −7.50 | −7.52 |
| Ca II | 3968.469 | −0.200 | 0.000 | 3969.69 | 0.181 (0.013) | 92.4 | −6.84 | −6.84 | −7.69 | −7.68 |
| Ca II | 8498.018 | −1.416 | 1.692 | 8500.62 | 0.041 (0.015)$^\alpha$ | 91.8 | −6.60 | −6.76 | −7.21 | −7.40 |
| Ca II | 8542.086 | −0.463 | 1.700 | 8544.76 | 0.157 (0.028)$^\alpha$ | 93.8 | −6.42 | −6.56 | −7.14 | −7.26 |
| Ca II | 8662.135 | −0.723 | 1.692 | 8664.87 | 0.102 (0.022)$^\alpha$ | 94.7 | −6.63 | −6.45 | −7.30 | −7.14 |
| **Ca Average:** | | | | | | | **−6.73** | **−6.68** | **−7.42** | **−7.39** |
| Cr II | 3120.358 | 0.097 | 2.434 | 3121.30 | 0.022 (0.006) | 90.4 | −7.65 | | −8.41 | |
| Cr II | 3124.973 | 0.303 | 2.455 | 3125.95 | 0.011 (0.004) | 93.6 | −7.93 | | −8.70 | |
| Cr II | 3128.692 | −0.543 | 2.434 | 3129.61 | 0.009 (0.003) | 87.5 | −7.77 | | −8.58 | |
| Cr II | 3132.053 | 0.423 | 2.483 | 3133.04 | 0.023 (0.005) | 94.4 | −7.77 | | −8.58 | |
| Cr II | 3368.042 | −0.085 | 2.483 | 3369.06 | 0.009 (0.002) | 90.8 | −7.90 | | −8.61 | |
| **Cr Average:** | | | | | | | **−7.80** | | **−8.56** | |
| Fe II | 3154.193 | −0.513 | 3.768 | 3155.19 | 0.011 (0.003) | 94.5 | −6.69 | | −7.32 | |
| Fe II | 3213.309 | −1.388 | 1.695 | 3214.31 | 0.014 (0.003) | 93.0 | −6.49 | | −7.14 | |
| Fe II | 3227.742 | −1.178 | 1.671 | 3228.74 | 0.013 (0.002) | 92.3 | −6.44 | | −7.18 | |
| Fe II | 5169.032 | −1.25 | 2.891 | 5170.64 | 0.018 (0.005) | 93.1 | −6.45 | | −7.02 | |
| **Fe Average:** | | | | | | | **−6.51** | | **−7.15** | |
| Mg I | 3832.304 | 0.121 | 2.712 | 3833.45 | 0.015 (0.004) | 89.4 | −5.71 | | −6.46 | |
| Mg I | 3838.292 | 0.392 | 2.717 | 3839.43 | 0.021 (0.003) | 88.8 | −5.71 | | −6.46 | |
| Mg II | 4481.125 | 0.74 | 8.864 | 4482.59 | 0.229 (0.013) | 97.9 | −5.52 | −5.58 | −6.35 | −6.31 |
| Mg II | 7877.050 | 0.39 | 9.996 | 7879.88 | 0.147 (0.045) | 107.7 | −5.85 | −5.89 | −6.30 | −6.31 |
| Mg II | 7896.040 | 0.65 | 9.999 | 7898.94 | 0.179 (0.027) | 110.2 | −5.87 | −5.85 | −6.32 | −6.27 |
| **Mg Average:** | | | | | | | **−5.71** | **−5.75** | **−6.35** | **−6.30** |
| O I | 7771.938 | 0.369 | 9.146 | 7774.41 | 0.102 (0.011) | 95.2 | −5.08 | −5.10 | −5.50 | −5.55 |
| O I | 7774.156 | 0.223 | 9.146 | 7776.65 | 0.096 (0.016) | 96.0 | −5.08 | −5.10 | −5.50 | −5.55 |
| O I | 7775.386 | 0.001 | 9.146 | 7777.89 | 0.050 (0.017) | 96.6 | −5.08 | −5.10 | −5.50 | −5.55 |
| O I | 8446.355 | 0.236 | 9.521 | 8448.87 | 0.050 (0.027) | 89.1 | −5.38 | −5.20 | −5.79 | −5.62 |
| **O Average:** | | | | | | | **−5.20** | **−5.15** | **−5.62** | **−5.58** |
| Si II | 3853.665 | −1.33 | 6.857 | 3854.86 | 0.014 (0.003) | 93.2 | −5.80 | −5.93 | −6.31 | −6.40 |
| Si II | 3856.018 | −0.37 | 6.859 | 3857.21 | 0.099 (0.004) | 92.3 | −5.80 | −5.93 | −6.31 | −6.40 |
| Si II | 3862.595 | −0.81 | 6.857 | 3863.77 | 0.064 (0.003) | 91.0 | −5.80 | −5.93 | −6.31 | −6.40 |
| Si II | 4128.054 | 0.410 | 9.836 | 4129.34 | 0.050 (0.005) | 93.2 | −6.01 | −6.02 | −6.51 | −6.41 |
| Si II | 4130.893 | 0.530 | 9.839 | 4132.16 | 0.092 (0.005) | 91.7 | −6.01 | −6.02 | −6.51 | −6.41 |
| Si II | 5055.983 | 0.530 | 10.074 | 5057.73 | 0.092 (0.013) | 103.7 | −6.09 | | −6.47 | |
| Si II | 6347.106 | 0.170 | 8.121 | 6349.07 | 0.242 (0.011) | 92.6 | −6.00 | −5.98 | −6.06 | −6.01 |
| Si II | 6371.368 | −0.040 | 8.121 | 6373.34 | 0.145 (0.011) | 92.6 | −5.98 | | −6.24 | |
| **Si Average:** | | | | | | | **−5.96** | **−5.98** | **−6.29** | **−6.23** |
| Ti II | 3234.515 | 0.43 | 0.049 | 3235.52 | 0.013 (0.002) | 93.6 | −8.25 | | −9.05 | |
| Ti II | 3236.572 | 0.24 | 0.028 | 3237.54 | 0.013 (0.003) | 89.8 | −8.25 | | −9.05 | |
| Ti II | 3239.036 | 0.07 | 0.012 | 3239.99 | 0.008 (0.002) | 88.7 | −8.25 | | −9.05 | |
| Ti II | 3341.874 | 0.35 | 0.574 | 3342.89 | 0.007 (0.002) | 90.7 | −8.37 | | −9.15 | |
| Ti II | 3349.033$^\beta$ | 0.43 | 0.607 | 3350.07 | 0.015 (0.003) | 92.9 | −8.22 | | −9.04 | |
| Ti II | 3349.402$^\beta$ | 0.53 | 0.049 | 3350.42 | 0.011 (0.002) | 91.2 | −8.22 | | −9.04 | |
| Ti II | 3361.212$^\beta$ | 0.43 | 0.028 | 3362.24 | 0.007 (0.002) | 91.4 | −8.49 | | −9.28 | |
| Ti II | 3372.792 | 0.28 | 0.012 | 3373.79 | 0.010 (0.002) | 89.0 | −8.34 | | −9.13 | |
| Ti II | 3383.759 | 0.16 | 0.000 | 3384.79 | 0.008 (0.002) | 91.3 | −8.37 | | −9.10 | |
| Ti II | 3685.204$^\beta$ | 0.128 | 0.607 | 3686.31 | 0.004 (0.001) | 90.1 | −8.58 | | −9.29 | |
| Ti II | 3759.292 | 0.28 | 0.607 | 3760.45 | 0.007 (0.002) | 92.5 | −8.31 | | −9.04 | |
| **Ti Average:** | | | | | | | **−8.35** | | **−9.13** | |

$^\alpha$ Gaseous emission from circumstellar gas at this wavelength, additional 12 percent error added in quadrature.
$^\beta$ Merged with another photospheric absorption line.

Table B10. WD 1622+587 UV: The absorption lines in WD 1622+587 from the FUV and NUV. If two lines lie within $\sim 10\,\text{Å}$, equivalent widths and abundances were fitted simultaneously. The notes column labels the lines that have a significant ISM contribution as 'i', those where two spectral lines are blended as 'b', and those where the night data were used are labelled 'N'. The ISM components are blended with the photospheric components. The NUV RV measurements are invalid and included in brackets, see Section 2.6 in the main body of the paper for further details.

| Line | $\lambda_{\text{vac}}$ (Å) | log(gf) | $E_{low}$ (eV) | $\lambda_{\text{obs}}$ (Å) | EW (Å) | RV (km/s) | [X/He]$_{\text{Spec}}$ | [X/He]$_{\text{Phot}}$ | Notes |
|---|---|---|---|---|---|---|---|---|---|
| C III | 1174.930 | −0.468 | 6.496 | 1174.84 | 0.077 (0.007) | −22.5 | −4.95 | −4.35 | |
| C III | 1175.260 | −0.565 | 6.493 | 1175.18 | 0.043 (0.006) | −20.4 | −4.95 | −4.35 | |
| C III | 1175.710 | 0.009 | 6.503 | 1175.57 | 0.132 (0.009) | −35.7 | −4.95 | −4.35 | |
| C III | 1175.987 | −0.565 | 6.496 | 1175.90 | 0.076 (0.015) | −21.2 | −4.95 | −4.35 | |
| C III | 1176.370 | −0.468 | 6.503 | 1176.26 | 0.075 (0.017) | −27.5 | −4.95 | −4.35 | |
| C I | 1261.550 | −0.82 | 0.005 | 1261.50 | 0.049 (0.011) | −12.6 | −4.62 | −5.01 | |
| C II | 1323.950 | −0.15 | 9.290 | 1323.82 | 0.112 (0.016) | −29.2 | −4.97 | −5.09 | |
| C I | 1329.100$^{\alpha}$ | −1.108 | 0.002 | 1328.98 | 0.106 (0.025) | −26.8 | −4.97 | −5.09 | b |
| C I | 1329.577 | −0.624 | 0.005 | 1329.48 | 0.057 (0.008) | −21.9 | −4.97 | −5.09 | |
| C II | 1334.530 | −0.596 | 0.000 | 1334.43 | 0.488 (0.030) | −23.1 | −4.60 | −4.71 | |
| C II | 1335.708 | −0.341 | 0.008 | 1335.60 | 0.617 (0.034) | −25.1 | −4.60 | −4.71 | |
| **C Average:** | | | | | | | **−4.75** | **−4.69** | |
| O I | 1152.150 | −0.268 | 1.967 | 1152.06 | 0.060 (0.014) | −23.7 | −5.43 | −5.75 | |
| O I | 1304.858 | −0.808 | 0.020 | 1304.80 | 0.041 (0.013) | −13.3 | −5.36 | −5.85 | N |
| **O Average:** | | | | | | | **−5.39** | **−5.80** | |
| Al II | 1191.814 | −0.51 | 4.659 | 1191.77 | 0.053 (0.006) | −12.3 | −6.20 | −6.54 | |
| Al III | 1379.670 | −0.6 | 6.66 | 1379.57 | 0.048 (0.013) | −21.7 | −6.32 | −6.31 | b |
| Al III | 1384.132 | −0.29 | 6.685 | 1384.04 | 0.070 (0.009) | −20.6 | −6.30 | −6.25 | |
| Al III | 1854.716 | 0.05 | 0.000 | 1855.18 | 0.236 (0.081) | (74.9) | −6.31 | −6.50 | |
| Al III | 1862.790 | −0.197 | 0.000 | 1862.96 | 0.238 (0.060) | (27.9) | −6.31 | −6.50 | |
| **Al Average:** | | | | | | | **−6.28** | **−6.38** | |
| Si III | 1161.579 | 0.22 | 16.130 | 1161.48 | 0.036 (0.006) | −26.6 | −5.26 | −4.90 | |
| Si II | 1190.416 | −0.245 | 0.000 | 1190.30 | 0.394 (0.014) | −28.2 | −5.29 | −5.58 | i |
| Si II | 1193.290 | 0.075 | 0.000 | 1193.19 | 0.470 (0.019) | −25.0 | −5.29 | −5.58 | i |
| Si II | 1194.500 | 0.487 | 0.036 | 1194.40 | 0.711 (0.046) | −25.4 | −5.29 | −5.58 | |
| Si II | 1197.394 | −0.202 | 0.036 | 1197.31 | 0.264 (0.029) | −21.0 | −5.29 | −5.58 | |
| Si II | 1223.896 | −0.061 | 5.323 | 1223.82 | 0.082 (0.016) | −18.4 | −5.69 | −5.92 | |
| Si II | 1224.245 | 0.002 | 0.152 | 1224.12 | 0.105 (0.011) | −30.2 | −5.69 | −5.92 | |
| Si II | 1224.968 | −0.584 | 5.323 | 1224.88 | 0.040 (0.006) | −22.8 | −5.69 | −5.92 | |
| Si II | 1226.979 | 0.345 | 5.309 | 1226.81 | 0.275 (0.009) | −42.2 | −5.69 | −5.92 | |
| Si II | 1227.604 | 0.528 | 5.345 | 1227.51 | 0.175 (0.010) | −22.5 | −5.69 | −5.92 | |
| Si II | 1228.739 | 0.651 | 5.323 | 1228.58 | 0.341 (0.013) | −38.7 | −5.69 | −5.92 | |
| Si II | 1229.383 | 0.882 | 5.345 | 1229.30 | 0.263 (0.012) | −19.6 | −5.69 | −5.92 | |
| Si II | 1246.740 | −0.236 | 5.309 | 1246.66 | 0.091 (0.017) | −20.3 | −5.54 | −5.74 | |
| Si II | 1248.426 | 0.066 | 5.323 | 1248.34 | 0.112 (0.008) | −20.4 | −5.54 | −5.74 | |
| Si II | 1250.091 | 0.228 | 6.857 | 1250.01 | 0.188 (0.014) | −18.6 | −5.54 | −5.74 | |
| Si II | 1250.436 | 0.423 | 6.859 | 1250.38 | 0.232 (0.013) | −13.1 | −5.54 | −5.74 | |
| Si II | 1251.164 | 0.24 | 5.345 | 1251.07 | 0.153 (0.006) | −22.0 | −5.54 | −5.74 | |
| Si II | 1260.422 | 0.462 | 0.000 | 1260.34 | 1.247 (0.052) | −18.8 | −5.07 | −5.36 | i |
| Si II | 1264.738$^{\beta}$ | 0.71 | 0.036 | 1264.74 | 1.758 (0.045) | 1.0 | −5.07 | −5.36 | b |
| Si III | 1294.545 | −0.037 | 6.553 | 1294.46 | 0.182 (0.032) | −20.1 | −5.42 | −5.29 | |
| Si III | 1296.726 | −0.127 | 6.537 | 1296.65 | 0.206 (0.016) | −17.8 | −5.42 | −5.29 | |
| Si III | 1298.892 | −0.257 | 6.553 | 1298.84 | 0.385 (0.045) | −12.0 | −5.42 | −5.29 | |
| Si II | 1309.276 | −0.448 | 0.036 | 1309.29 | 0.710 (0.021) | 2.2 | −5.11 | −5.34 | |
| Si II | 1311.256 | 0.579 | 10.415 | 1311.16 | 0.103 (0.013) | −22.9 | −5.11 | −5.34 | |
| Si III | 1312.591 | −0.84 | 10.276 | 1312.50 | 0.068 (0.010) | −21.0 | −5.11 | −5.34 | |
| Si II | 1346.884 | −0.144 | 5.323 | 1346.78 | 0.135 (0.017) | −22.9 | −5.06 | −5.31 | |
| Si II | 1348.543 | −0.186 | 5.309 | 1348.43 | 0.143 (0.010) | −25.0 | −5.06 | −5.31 | |
| Si II | 1350.072 | 0.216 | 5.345 | 1349.96 | 0.168 (0.025) | −25.6 | −5.06 | −5.31 | |
| Si II | 1350.516 | −0.69 | 5.323 | 1350.46 | 0.157 (0.014) | −11.8 | −5.06 | −5.31 | b |
| Si II | 1352.635 | −0.193 | 5.323 | 1352.53 | 0.153 (0.018) | −24.0 | −5.06 | −5.31 | |
| Si II | 1353.721 | −0.158 | 5.345 | 1353.63 | 0.100 (0.017) | −20.8 | −5.06 | −5.31 | |
| Si IV | 1393.755 | 0.03 | 0.000 | 1393.65 | 0.094 (0.018) | −23.4 | −5.34 | −4.68 | |
| Si IV | 1402.770 | −0.28 | 0.000 | 1402.66 | 0.072 (0.020) | −23.7 | −5.17 | −5.32 | |
| Si III | 1417.237 | −0.106 | 10.276 | 1417.13 | 0.147 (0.012) | −23.7 | −5.13 | −5.00 | |
| Si II | 2072.700 | −0.29 | 6.86 | 2073.20 | 0.389 (0.100) | (72.3) | −4.91 | −5.22 | |
| **Si Average:** | | | | | | | **−5.20** | **−5.18** | |

**Table B10. WD 1622+587 UV:** Continued.

| Line | $\lambda_{\mathrm{vac}}$ (Å) | log(gf) | $E_{low}$ (eV) | $\lambda_{\mathrm{obs}}$ (Å) | EW (Å) | RV (km/s) | [X/He]$_{\mathrm{Spec}}$ | [X/He]$_{\mathrm{Phot}}$ | Notes |
|---|---|---|---|---|---|---|---|---|---|
| P II | 1153.995 | 0.067 | 0.058 | 1153.91 | 0.022 (0.006) | −22.4 | −7.74 | −8.01 | |
| S III | 1194.041 | −1.305 | 0.037 | 1193.95 | 0.030 (0.009) | −22.8 | −6.04 | −6.65 | |
| S III | 1200.956 | −1.03 | 0.103 | 1200.87 | 0.064 (0.019) | −21.5 | −6.02 | −5.80 | |
| S II | 1204.335$^\gamma$ | −1.121 | 1.845 | 1204.209 | 0.070 (0.010) | −31.4 | −5.92 | −5.94 | b |
| **S Average:** | | | | | | | **−5.99** | **−6.01** | |
| Fe II | 1144.938 | 0.037 | 0.000 | 1144.84 | 0.101 (0.016) | −24.9 | −5.14 | −5.29 | |
| Fe II | 1148.277 | −0.179 | 0.048 | 1148.10 | 0.082 (0.013) | −45.4 | −5.34 | −5.71 | b |
| Fe II | 1151.146 | −0.451 | 0.083 | 1151.05 | 0.022 (0.005) | −25.3 | −5.34 | −5.71 | |
| Fe II | 1358.937 | −0.193 | 3.245 | 1358.88 | 0.040 (0.007) | −12.1 | −5.18 | −5.53 | |
| Fe II | 1371.022 | −0.229 | 2.635 | 1370.88 | 0.020 (0.005) | −30.4 | −5.33 | −5.70 | |
| **Fe Average:** | | | | | | | **−5.26** | **−5.55** | |
| Ni II | 1317.217 | −0.058 | 0.000 | 1317.11 | 0.046 (0.008) | −25.3 | −6.56 | −7.01 | |
| Ni II | 1370.132 | −0.105 | 0.000 | 1369.99 | 0.034 (0.007) | −30.2 | −6.13 | −6.53 | |
| **Ni Average:** | | | | | | | **−6.29** | **−6.71** | |
| Mg II | 2790.775 | 0.273 | 4.420 | 2792.00 | 0.175 (0.051) | (131.6) | −4.76 | −5.27 | b |
| Mg II | 2795.528 | 0.085 | 0.000 | 2796.58 | 0.878 (0.106) | (112.9) | −4.76 | −5.27 | b |
| Mg II | 2802.704 | −0.218 | 0.000 | 2803.55 | 0.342 (0.085) | (90.8) | −4.76 | −5.27 | b |
| **Mg Average:** | | | | | | | **−4.76** | **−5.27** | |

$^\alpha$ Multiple merged C I lines (1328.83 Å, 1329.085 Å, 1329.1 Å, 1329.12 Å), RV with respect to line with highest log(gf).
$^\beta$ Two silicon lines merged, RV reported with respect to line with strongest log(gf).
$^\gamma$ Two S lines merged 1204.271 and 1204.324 Å, RV reported with respect to line with strongest log(gf).

**Table B11. WD 1622+587 Optical:** The absorption lines measured in the WD 1622+587 HIRES spectra. If two lines lie within ∼ 10 Å, equivalent widths and abundances were fitted simultaneously.

| Line | $\lambda_{\mathrm{air}}$ (Å) | log(gf) | $E_{low}$ (eV) | $\lambda_{\mathrm{obs}}$ (Å) | EW (Å) | RV (km/s) | [X/He]$_{\mathrm{Spec}}$ | [X/He]$_{\mathrm{Phot}}$ |
|---|---|---|---|---|---|---|---|---|
| Ca II | 3933.6631 | 0.105 | 0.000 | 3933.40 | 0.111 (0.010) | −20.3 | −5.85 | −7.02 |
| Mg II | 4481.125 | 0.74 | 8.864 | 4480.95 | 0.060 (0.009) | −11.6 | −4.92 | −5.84 |
| Mg II | 7877.050 | 0.39 | 9.996 | 7876.43 | 0.048 (0.016) | −23.5 | −4.80 | −5.29 |
| Mg II | 7896.362 | 0.65 | 9.999 | 7895.86 | 0.079 (0.018) | −19.0 | −5.03 | −5.50 |
| **Mg Average:** | | | | | | | **−4.91** | **−5.49** |
| Si II | 3856.018 | −0.37 | 6.859 | 3855.79 | 0.062 (0.009) | −17.9 | −5.23 | −5.74 |
| Si II | 3862.595 | −0.81 | 6.857 | 3862.31 | 0.052 (0.010) | −22.1 | −5.20 | −5.66 |
| Si II | 4128.054 | 0.410 | 9.836 | 4127.79 | 0.036 (0.008) | −19.4 | −5.24 | −5.71 |
| Si II | 4130.893 | 0.530 | 9.839 | 4130.52 | 0.067 (0.009) | −26.9 | −5.24 | −5.71 |
| Si II | 5055.983 | 0.530 | 10.074 | 5055.77 | 0.087 (0.011) | −12.8 | −5.36 | −5.81 |
| Si II | 6347.106 | 0.170 | 8.121 | 6346.69 | 0.207 (0.017) | −19.7 | −5.14 | −5.63 |
| Si II | 6371.368 | −0.040 | 8.121 | 6370.90 | 0.152 (0.018) | −22.4 | −5.06 | −5.57 |
| **Si Average:** | | | | | | | **−5.20** | **−5.68** |

**Table B12. Gaia J2100+2122 UV:** The absorption lines in Gaia J2100+2122 from the NUV. If two lines lie within ∼ 10 Å, equivalent widths and abundances were fitted simultaneously. The NUV RV measurements are invalid and included in brackets, see Section 2.6 in the main body of the paper for further details.

| Line | $\lambda_{\mathrm{vac}}$ (Å) | log(gf) | $E_{low}$ (eV) | $\lambda_{\mathrm{obs}}$ (Å) | EW (Å) | RV (km/s) | [X/H]$_{\mathrm{Spec}}$ | [X/H]$_{\mathrm{Phot}}$ |
|---|---|---|---|---|---|---|---|---|
| Al III | 1854.716 | 0.05 | 0.000 | 1855.39 | 0.242 (0.036) | (108.4) | −6.48 | −6.49 |
| Al III | 1862.790 | −0.197 | 0.000 | 1863.50 | 0.162 (0.036) | (113.5) | −6.48 | −6.49 |

**Table B13. Gaia J2100+2122 Optical:** The absorption lines measured in Gaia J2100+2122. If two lines lie within $\sim 10\,\text{Å}$, equivalent widths and abundances were fitted simultaneously.

| Line | $\lambda_{\text{air}}$ (Å) | log(gf) | $E_{low}$ (eV) | $\lambda_{\text{obs}}$ (Å) | EW (Å) | RV (km/s) | [X/H]$_{\text{Spec}}$ HIRES | [X/H]$_{\text{Spec}}$ X-Sh | [X/H]$_{\text{Phot}}$ HIRES | [X/H]$_{\text{Phot}}$ X-Sh |
|---|---|---|---|---|---|---|---|---|---|---|
| Ca II | 3179.331 | 0.499 | 3.151 | 3179.35 | 0.009 (0.002) | 2.1 | −6.32 | | −6.66 | |
| Ca II | 3933.663$^{\alpha}$ | 0.105 | 0.000 | 3933.72 | 0.022 (0.001) | 4.1 | −6.13 | −5.79 | −6.65 | −6.25 |
| **Ca Average:** | | | | | | | **−6.21** | **−5.79** | **−6.65** | **−6.25** |
| Fe II | 3227.742 | −1.178 | 1.671 | 3227.80 | 0.004 (0.001) | 5.4 | −4.96 | | −5.49 | |
| Mg II | 4481.125 | 0.74 | 8.864 | 4481.25 | 0.140 (0.005) | 8.4 | −5.12 | −5.04 | −5.40 | −5.30 |
| Si II | 3856.018 | −0.37 | 6.859 | 3856.07 | 0.015 (0.002) | 3.9 | −5.04 | | −5.32 | |
| Si II | 3862.595 | −0.81 | 6.857 | 3862.65 | 0.005 (0.001) | 4.1 | −5.10 | | −5.38 | |
| Si II | 4128.054 | 0.410 | 9.836 | 4128.10 | 0.017 (0.002) | 3.0 | −4.96 | | −5.16 | |
| Si II | 4130.893 | 0.530 | 9.839 | 4130.94 | 0.025 (0.002) | 3.4 | −5.12 | | −5.33 | |
| Si II | 5055.983 | 0.530 | 10.074 | 5056.15 | 0.027 (0.004) | 9.7 | −5.31 | −5.32 | −5.54 | −5.53 |
| Si II | 6347.106 | 0.170 | 8.121 | 6347.20 | 0.051 (0.003) | 4.2 | −5.14 | −5.01 | −5.41 | −5.26 |
| Si II | 6371.368 | −0.040 | 8.121 | 6371.44 | 0.019 (0.002) | 3.6 | −5.19 | | −5.46 | |
| **Si Average:** | | | | | | | **−5.11** | **−5.14** | **−5.36** | **−5.37** |

$^{\alpha}$ Ca II has extra absorption bluewards. This is not resolved in X-shooter and is resolved in HIRES. Therefore the abundance is higher in X-shooter and the final average abundance only includes the HIRES data.

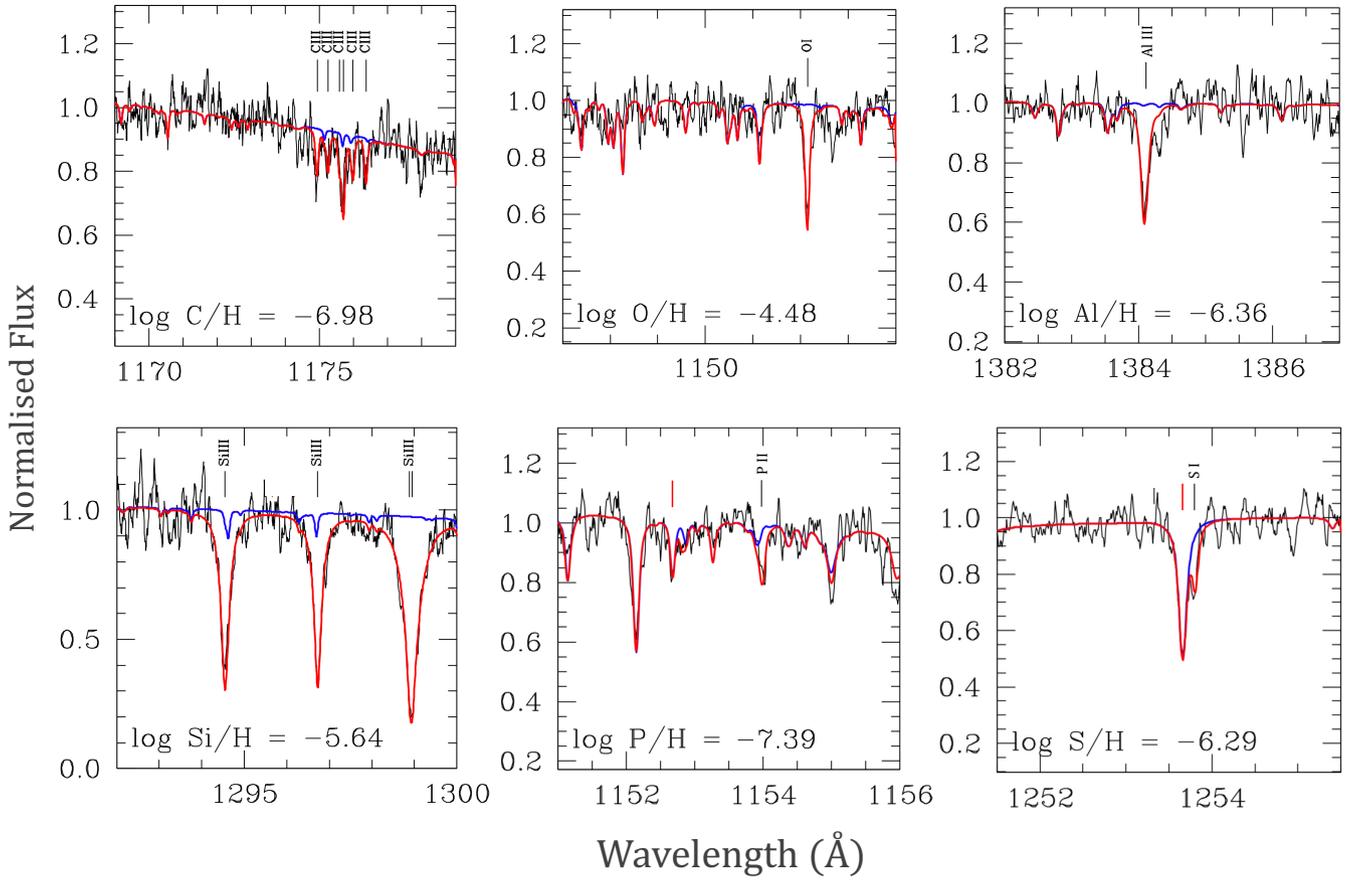

**Figure D1.** Model fits (red lines) to individual spectral lines in the *HST*/COS FUV spectra for Gaia J0006+2858, the spectra have a five point smoothing applied for clarity. Each panel correlates to a different element measured in the spectra (C, O, Al, Si, P, and S), and the lines fitted are denoted with a black dash and labelled. The blue line shows the model without the photospheric contribution of that element to the model. The blue model therefore encapsulates the photospheric components from other elements, as well as a model which uses a Voigt profile to match the non-photospheric absorption contribution to the spectral lines, these lines are marked with a red dash and are blueshifted in comparison to the photospheric component.

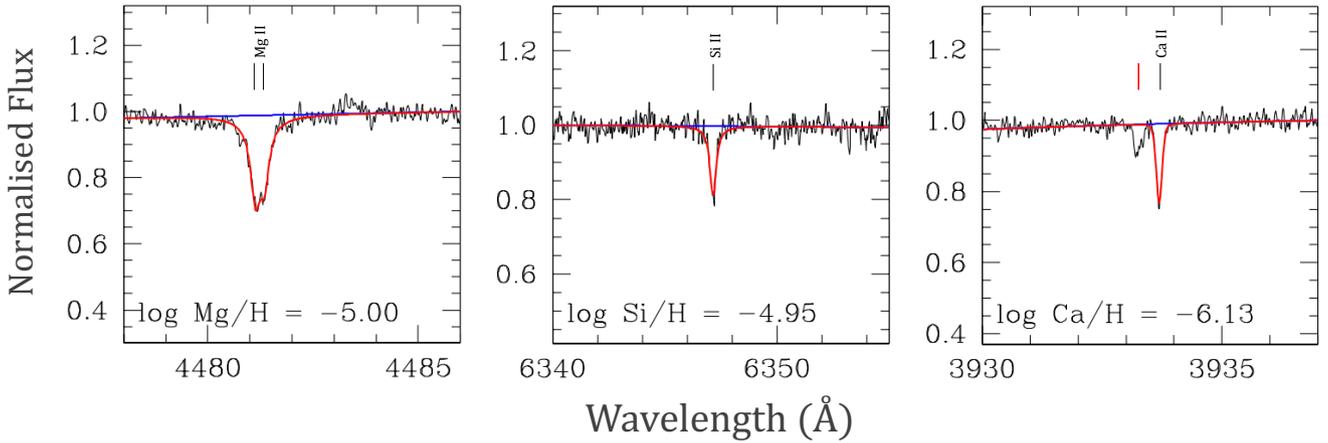

**Figure D2.** Model fits (red lines) to individual spectral lines in the HIRES optical spectra for Gaia J0006+2858, the spectra have a five point smoothing applied for clarity. Each panel correlates to a different element measured in the spectra (Mg, Si, and Ca), and the lines fitted are denoted with a black dash and labelled. The blue line shows the model without the photospheric contribution of that element to the model. The Ca II line has a non-photospheric absorption component blueshifted from the photospheric component, it is spatially separated and so is not modelled.

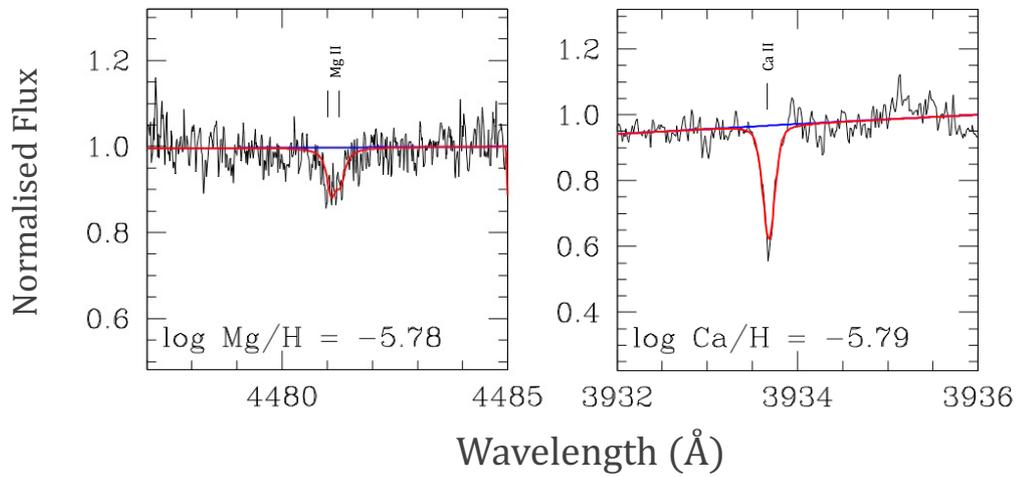

**Figure D3.** Model fits (red lines) to individual spectral lines in the HIRES optical spectra for Gaia J0347+1624. Each panel correlates to a different element measured in the spectra (Mg and Ca), and the lines fitted are denoted with a black dash and labelled. The blue line shows the model without the photospheric contribution of that element to the model. The calcium line has a merged non-photospheric contribution and is therefore an upper limit.

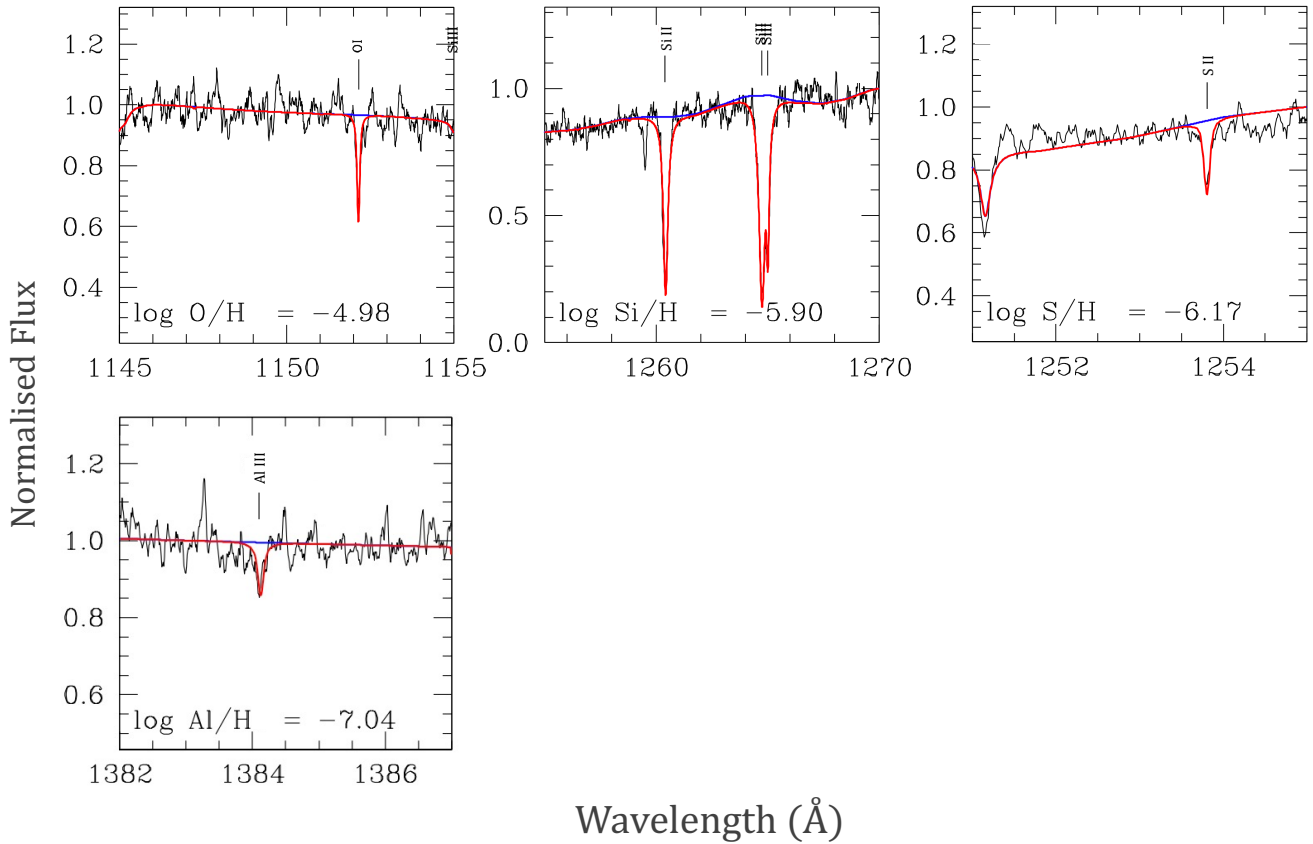

**Figure D4.** Model fits (red lines) to individual spectral lines in the *HST*/COS FUV spectra for Gaia J0510+2315, the spectra have a five point smoothing applied for clarity. Each panel correlates to a different element measured in the spectra (O, Si, S, and Al), and the lines fitted are denoted with a black dash and labelled. The blue line shows the model without the photospheric contribution of that element to the model.

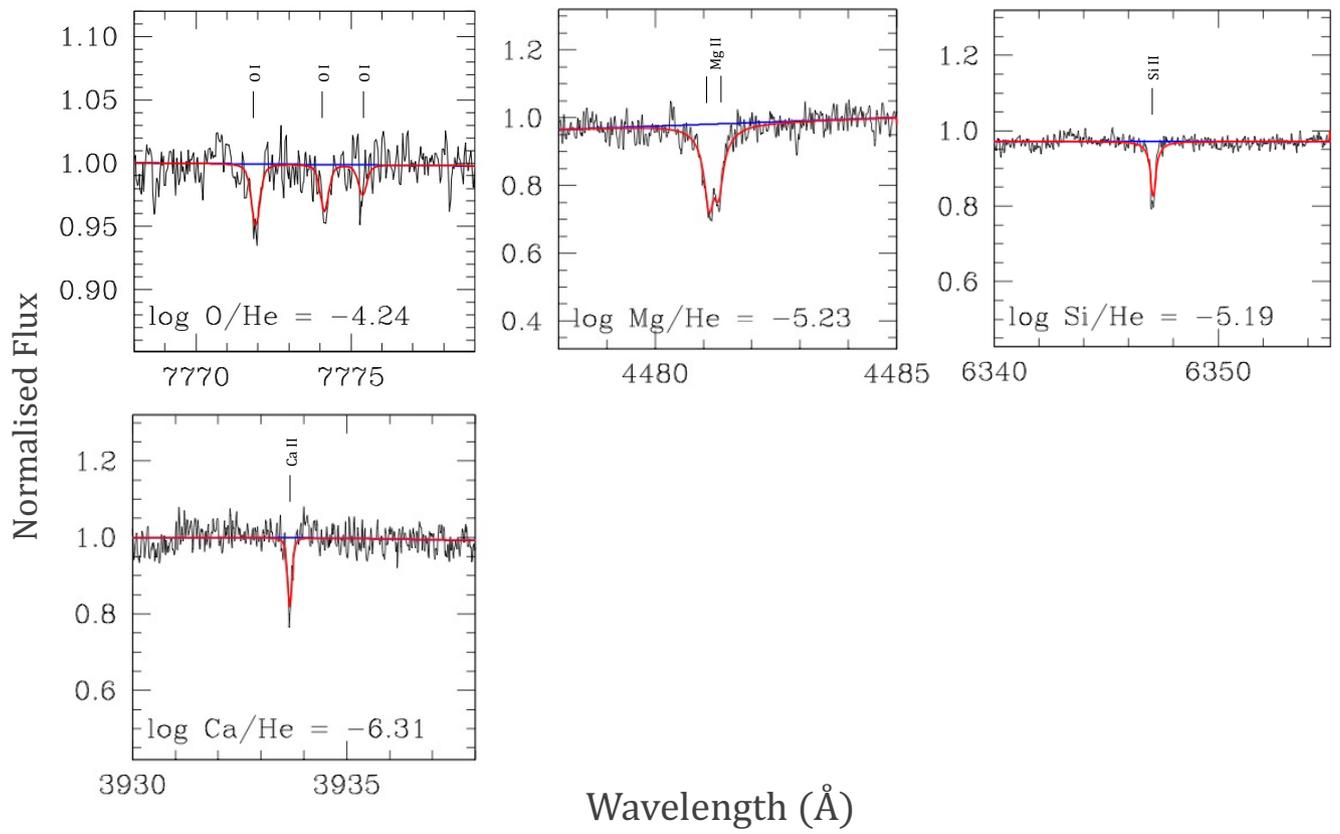

**Figure D5.** Model fits (red lines) to individual spectral lines in the HIRES optical spectra for Gaia J0510+2315, the spectra have a five point smoothing applied for clarity. Each panel correlates to a different element measured in the spectra (O, Mg, Si, and Ca), and the lines fitted are denoted with a black dash and labelled. The blue line shows the model without the photospheric contribution of that element to the model.

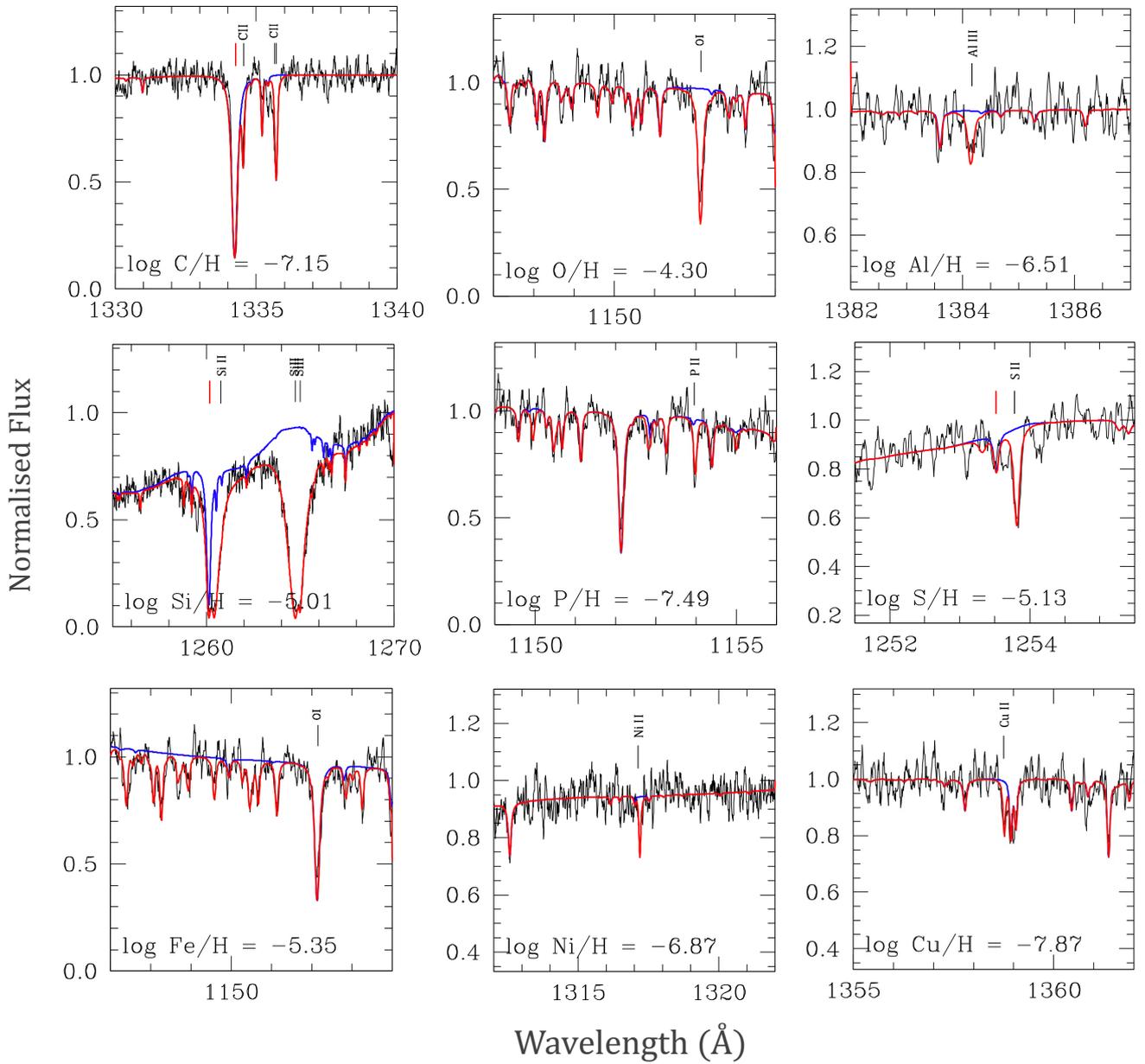

**Figure D6.** Model fits (red lines) to individual spectral lines in the *HST*/COS FUV spectra for Gaia J0611−6931, the spectra have a five point smoothing applied for clarity. Each panel correlates to a different element measured in the spectra (C, O, Al, Si, P, S, Fe, Ni, and Cu), and the lines fitted are denoted with a black dash and labelled. The blue line shows the model without the photospheric contribution of that element to the model. The blue model therefore encapsulates the photospheric components from other elements, as well as a model which uses a Voigt profile to match the non-photospheric absorption contribution to the spectral lines, these lines are marked with a red dash and are blueshifted in comparison to the photospheric component.

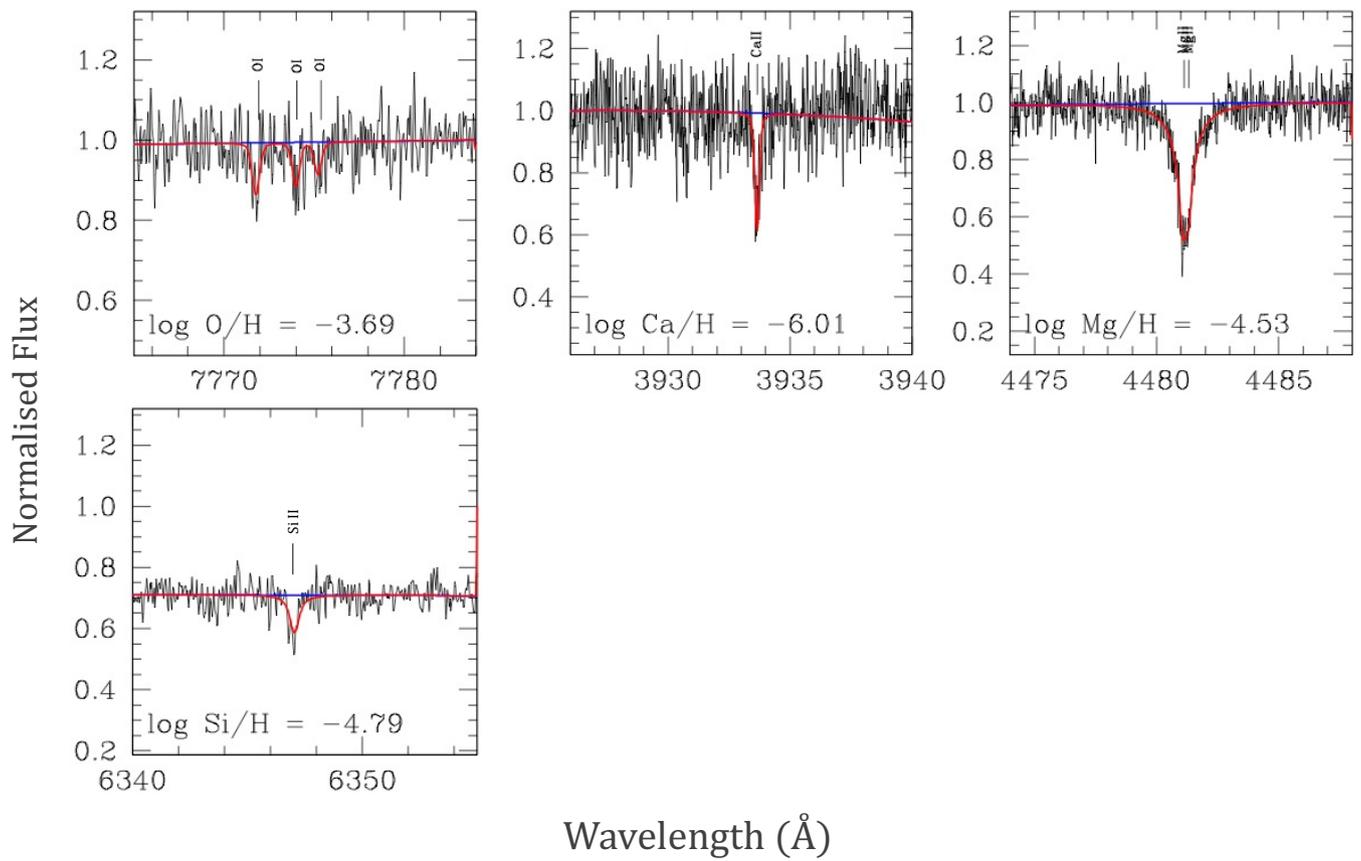

**Figure D7.** Model fits (red lines) to individual spectral lines in the MIKE optical spectra for Gaia J0611−6931. Each panel correlates to a different element measured in the spectra (O, Ca, Mg, and Si), and the lines fitted are denoted with a black dash and labelled. The blue line shows the model without the photospheric contribution of that element to the model.

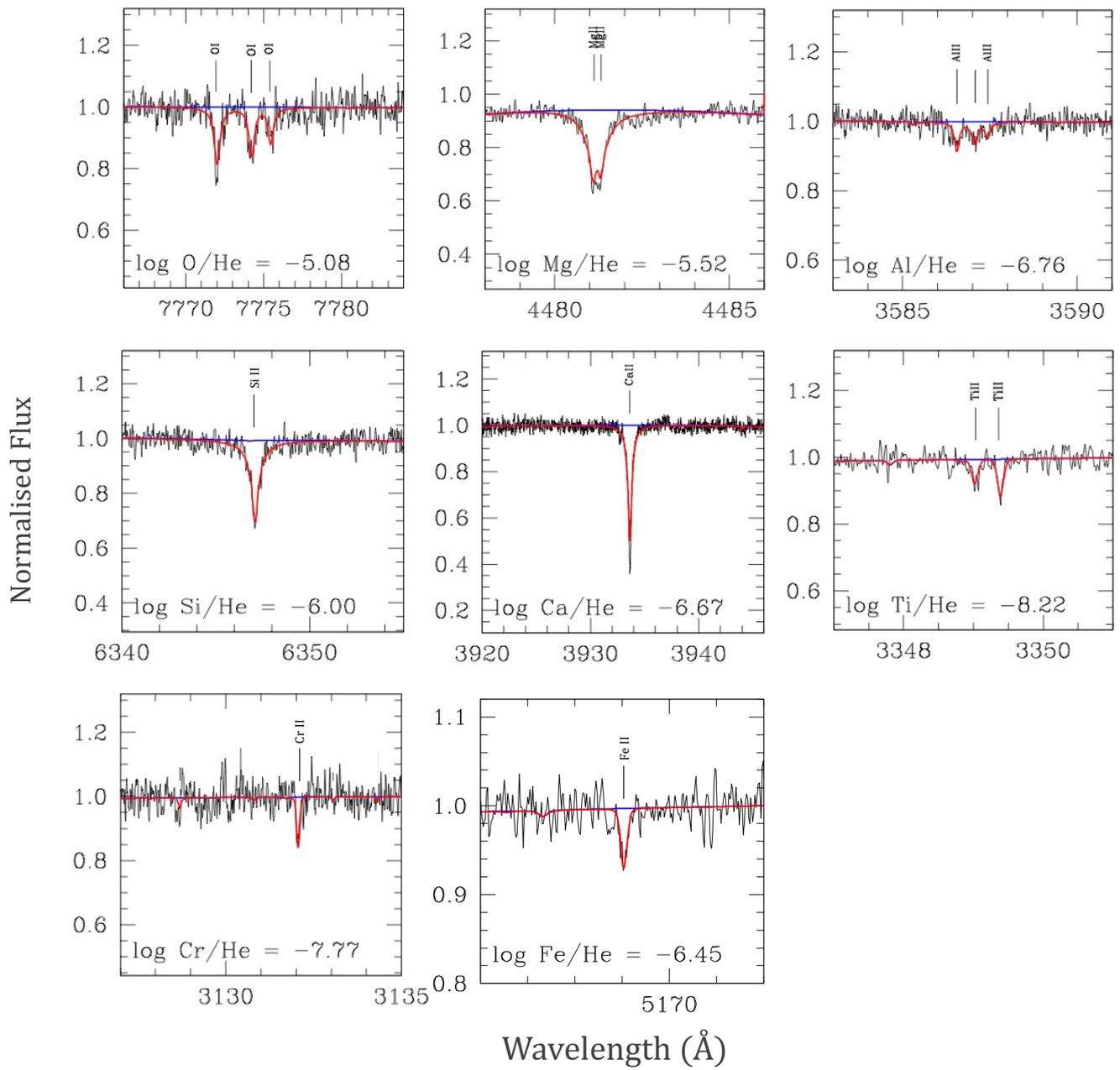

**Figure D8.** Model fits (red lines) to individual spectral lines in the HIRES optical spectra for Gaia J0644−0352, the spectra have a five point smoothing applied for clarity. Each panel correlates to a different element measured in the spectra (O, Mg, Al, Si, Ca, Ti, Cr, and Fe), and the lines fitted are denoted with a black dash and labelled. The blue line shows the model without the photospheric contribution of that element to the model.

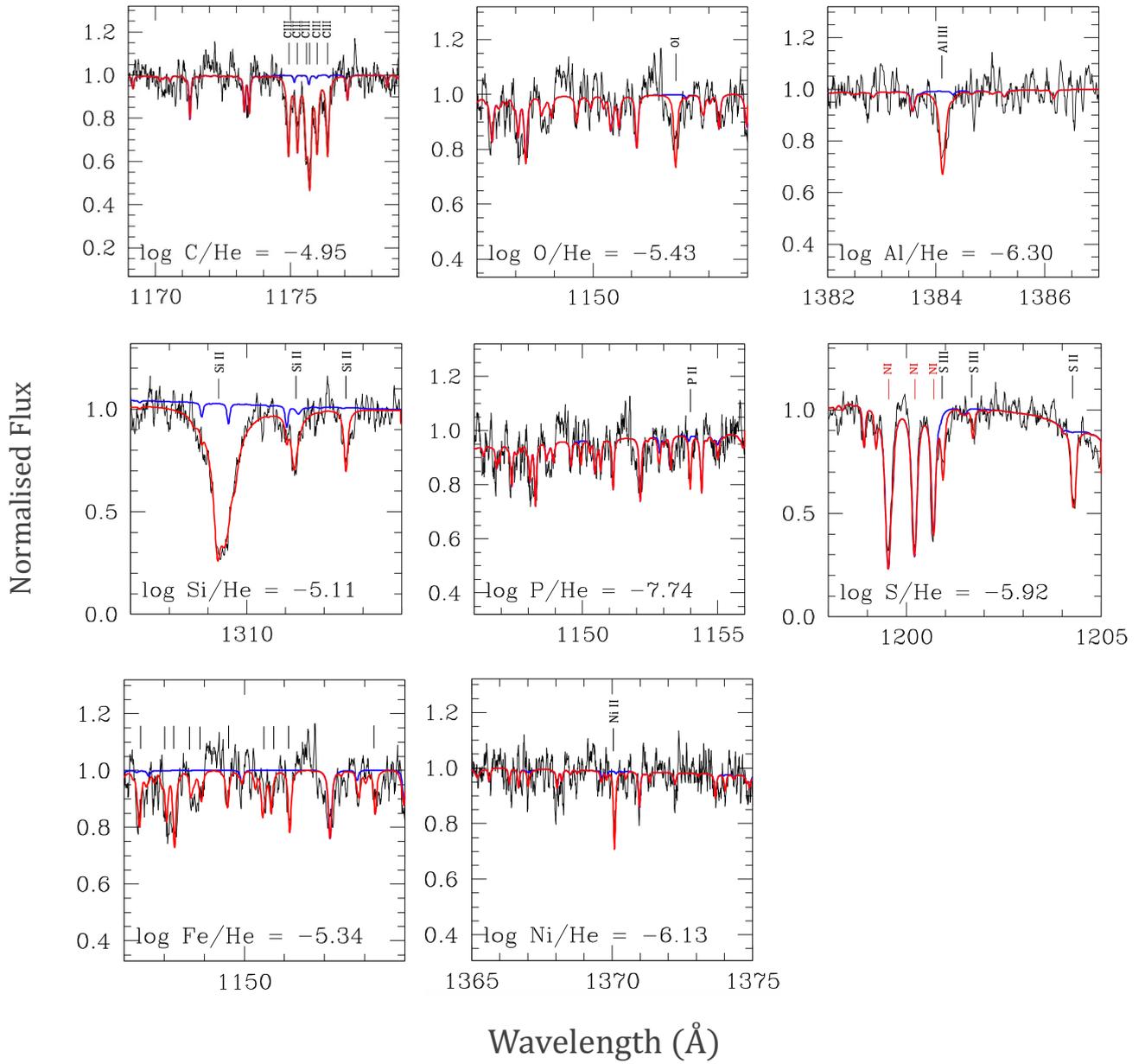

**Figure D9.** Model fits (red lines) to individual spectral lines in the *HST*/COS FUV optical spectra for WD 1622+587, the spectra have a five point smoothing applied for clarity. Each panel correlates to a different element measured in the spectra (C, O, Al, Si, P, S, Fe, and Ni), and the lines fitted are denoted with a black dash and labelled. The blue line shows the model without the photospheric contribution of that element to the model. The blue model therefore encapsulates the photospheric components from other elements, as well as a model which uses a Voigt profile to match the non-photospheric absorption contribution to the spectral lines, these lines are marked with a red dash and are blueshifted in comparison to the photospheric component. It should be noted that the strong N I lines shown in the S panel are non-photospheric.

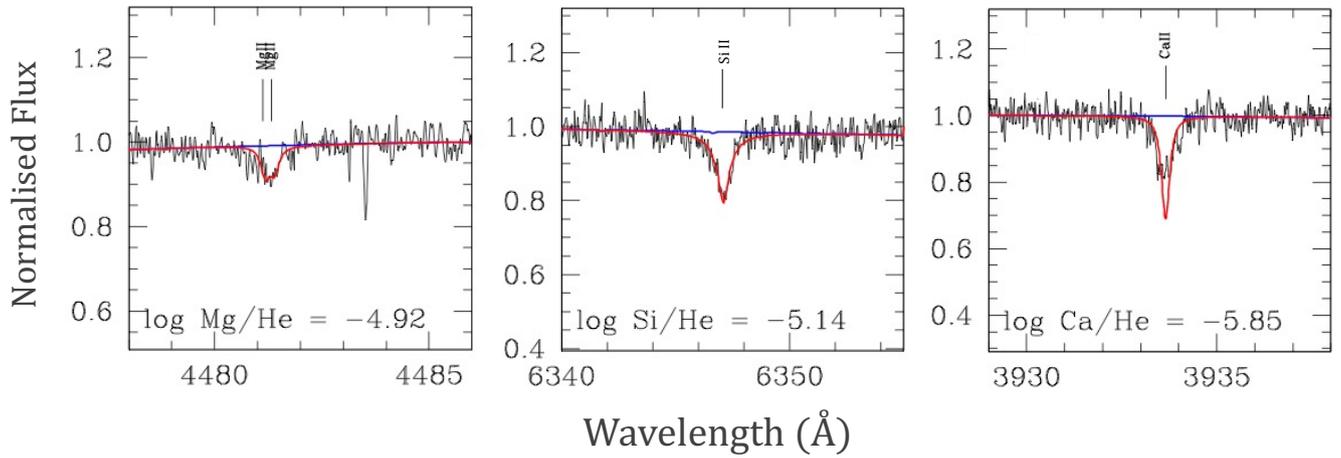

**Figure D10.** Model fits (red lines) to individual spectral lines in the HIRES optical spectra for WD 1622+587. Each panel correlates to a different element measured in the spectra (Mg, Si, and Ca), and the lines fitted are denoted with a black dash and labelled. The blue line shows the model without the photospheric contribution of that element to the model.

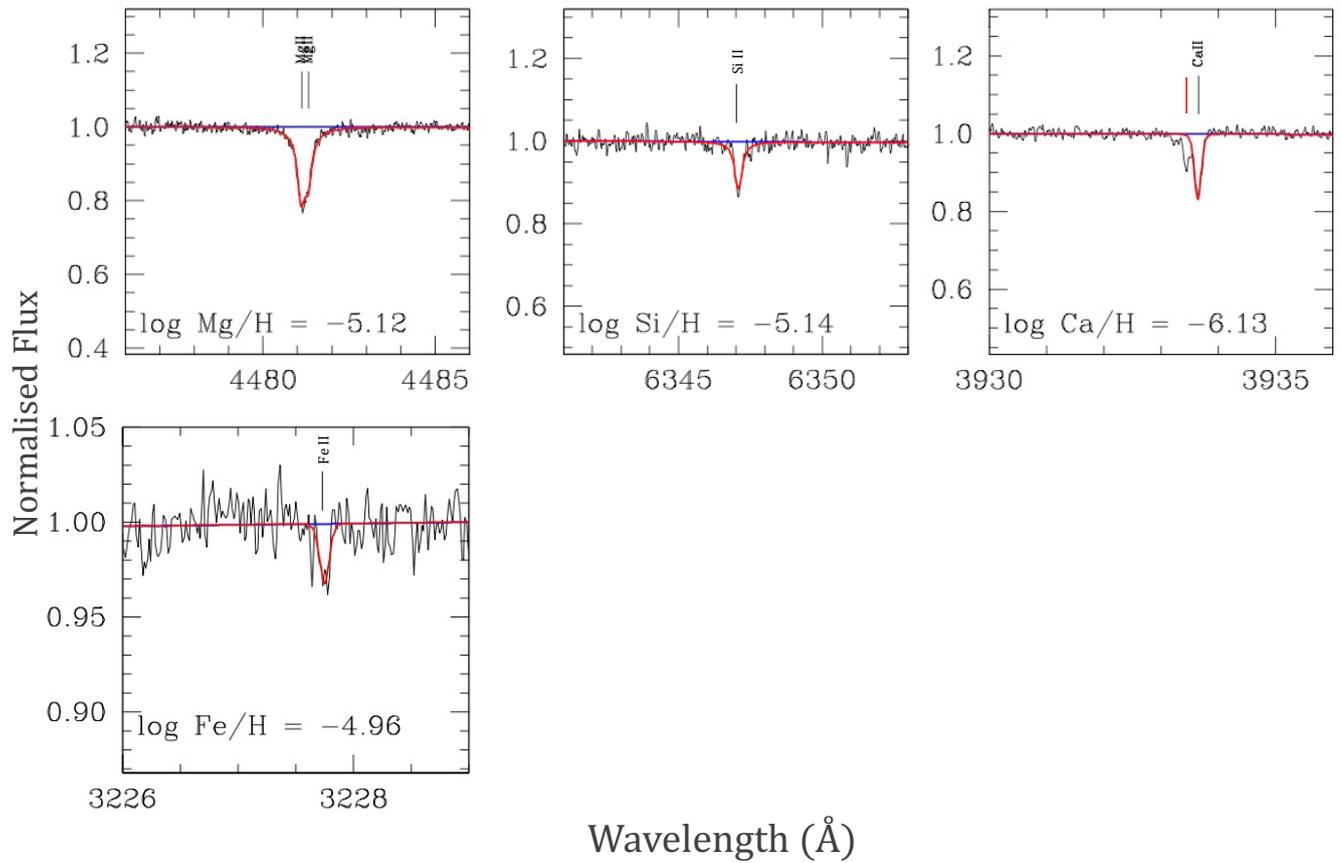

**Figure D11.** Model fits (red lines) to individual spectral lines in the HIRES optical spectra for Gaia J2100+2122. Each panel correlates to a different element measured in the spectra (Mg, Si, Ca, and Fe), and the lines fitted are denoted with a black dash and labelled. The blue line shows the model without the photospheric contribution of that element to the model. There is additional non-photospheric absorption bluewards of the Ca II K line and is marked with a red dash, this is excluded from the fit.



\end{document}